\documentclass[preprint,final,3p]{elsarticle}

\biboptions{sort&compress}

\usepackage{hyperref}
\usepackage{fancyvrb}
\usepackage{tabularx}
\usepackage{mdframed}
\usepackage{booktabs}
\usepackage{slashed}
\usepackage{units}
\usepackage{caption}
\usepackage{subcaption}
\usepackage{textcomp}
\usepackage[normalem]{ulem}
\usepackage[utf8]{inputenc}

\usepackage{url}

\usepackage[usenames,dvipsnames]{xcolor}

\usepackage[titletoc]{appendix}

\usepackage{amssymb}
\usepackage{amsmath,mathrsfs}

\usepackage{cleveref}
\allowdisplaybreaks

\journal{Computer Physics Communications}

\def \TeV{\text{Te\hspace{-0.05cm}V}}
\def \GeV{\text{Ge\hspace{-0.05cm}V}}

\def \etmiss{$\slashed{E}_T$}
\def \Root{ROOT}
\def \ATLAS{ATLAS}
\def \Atlas{ATLAS}
\def \CMS{CMS}
\def \Cms{CMS}
\def \Checkmate{\texttt{CheckMATE}}
\def \CheckMATE{\texttt{CheckMATE}}
\def \CheckMATETwo{\texttt{CheckMATE~2}}

\def \cmate{\texttt{CheckMATE}}

\def \Prospino{\texttt{Prospino}}
\def \Nllfast{\texttt{NLLFast}}
\def \Madanalysis{\texttt{MadAnalysis5}}
\def \Madgraph{\texttt{MadGraph5\_aMC@NLO}}
\def \Ufo{\texttt{UFO}}
\def \Slha{\texttt{SLHA}}
\def \Delphes{\texttt{Delphes}}
\def \Pythia{\texttt{Pythia}}
\def \Pythiaeight{\texttt{Pythia\,8}}
\def \Python{\texttt{Python}}
\def \Fritz{\texttt{FRITZ}}
\def \Cls{CL$_{\text{S}}$}

\newcommand{\userinputcolor}{\color{gray}}

\newcommand{\badcolor}{\color{red}}

\newcommand{\CLs}{CL$_\text{S}$\ }

\newcommand{\pT}{p_\text{T}}

\begin{document}

\fvset{samepage=true, fontsize=\footnotesize}
\begin{frontmatter}

\begin{center}
\includegraphics[width=0.7\textwidth]{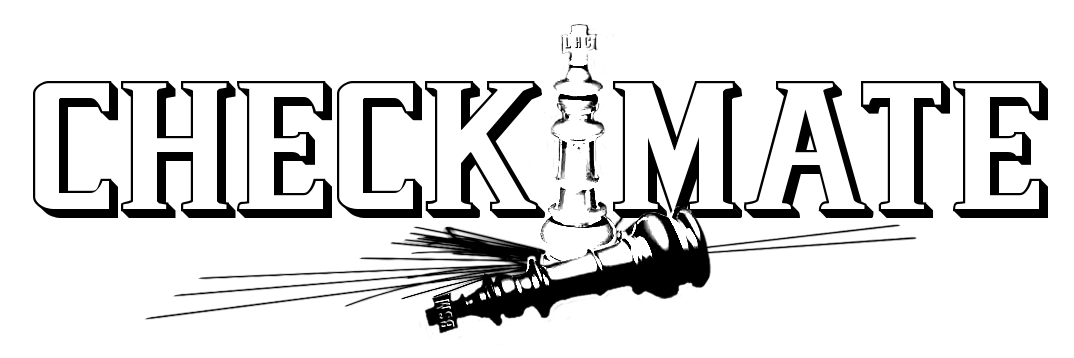}
\end{center}


{\tiny CTPU-16-36}
{\tiny CSIC-16-116}
{\tiny TTK-16-47}
\title{CheckMATE 2: From the model to the limit}


\author[label1,label7]{Daniel Dercks\fnref{aut1}}
\author[label2]{Nishita Desai\fnref{aut2}}
\author[label3,label4]{Jong Soo Kim\fnref{aut3}}
\author[label5]{Krzysztof Rolbiecki\fnref{aut4}}
\author[label6]{Jamie Tattersall\fnref{aut5}}
\author[label6]{Torsten~Weber\fnref{aut6}}
\address[label1]{II. Institute for Theoretical Physics, University of Hamburg, Luruper Chaussee 149, D-22761 Hamburg,
Germany}
\address[label7]{Bethe Center for Theoretical Physics \& Physikalisches Institut der Universit\"at Bonn, Nussallee 12, D-53115 Bonn, Germany}
\address[label2]{Laboratoire Charles Coulomb (L2C) UMR 5221 \& Laboratoire Univers et Particules de Montpellier (LUPM) UMR 5299, CNRS-Universit\'e de Montpellier, 34090 Montpellier, France
}
\address[label3]{Center for Theoretical Physics of the Universe, Institute for Basic Science (IBS), Daejeon, 34051, Korea}
\address[label4]{Universidad Aut\'onoma de Madrid, Instituto de F\'isica Te\'orica, Calle Nicol\'as Cabrera 13--15, Cantoblanco, 28049 Madrid, Spain}
\address[label5]{Faculty of Physics, University of Warsaw, Pasteura 5, 02-093 Warsaw, Poland}
\address[label6]{Institute for Theoretical Particle Physics and Cosmology, RWTH Aachen University, D-52056 Aachen, Germany}

\fntext[aut1]{daniel.dercks@desy.de}
\fntext[aut2]{nishita.desai@umontpellier.fr}
\fntext[aut3]{jongsoo.kim@tu-dortmund.de}
\fntext[aut4]{krzysztof.rolbiecki@fuw.edu.pl}
\fntext[aut5]{tattersall@physik.rwth-aachen.de}
\fntext[aut6]{torsten.weber@rwth-aachen.de}
\begin{abstract}
We present the latest developments to the \CheckMATE{} program that allows models of new physics to be 
easily tested against the recent LHC data. To achieve this goal, the core of \CheckMATE{} now contains 
over 60 LHC analyses of which 12 are from the 13~TeV run. The main new feature is that 
\CheckMATETwo{} now integrates the Monte Carlo event generation via \Madgraph{} and \Pythiaeight{}. This 
allows users to go directly from a \Slha{} file or \Ufo{} model to the result of whether a model 
is allowed or not. In addition, the integration of the event generation leads to a significant 
increase in the speed of the program. Many other improvements have also been made, including the 
possibility to now combine signal regions to give a total likelihood for a model.

\end{abstract}


\end{frontmatter}
{\newpage 
   \noindent
{\bf PROGRAM SUMMARY}

\captionsetup[subfigure]{aboveskip=-15pt}

\begin{small}
\noindent                         \\
{\em Program Title:} CheckMATE                                          \\
{\em Journal Reference:}                                      \\
{\em Catalogue identifier:}                                   \\
{\em Licensing provisions:} none                                   \\
{\em Programming language:} C++, Python                                   \\
{\em Computer:} PC, Mac                                               \\
{\em Operating system:} Linux, Mac OS                                       
{\em Classification:} 11.9                                         \\
{\em External routines/libraries:} ROOT, Python, HepMC (optional)                \\
{\em Subprograms used:} Delphes                                \\
{\em Nature of problem:}
The LHC experiments have performed a huge number of searches for new physics in the past few years. 
However the results can only be given for a few benchmark models out of the huge number that exist in 
the literature.
\\
{\em Solution method:}
CheckMATE is a program that automatically calculates limits for new physics models. The original 
version required the user to generate Monte Carlo events themselves before CheckMATE could be run 
but the new version now integrates this step. The simplest output of CheckMATE is whether the model 
is ruled out at 95\% CLs or not. However, more complicated statistical metrics are also available, 
including the combination of many signal regions.
\\
{\em Restrictions:} Only a subset of available experimental results have been implemented. \\ 
{\em Running time:} The running time scales about linearly with the number of input events provided 
by the user. The event generation, detector simulation and analysis of 10000 events needs about 245~s for a single core 
calculation on an Intel(R) Xeon(R) CPU E5-2650 v2 @ 2.60\,GHz with 32\,GB RAM.\\
\end{small}



\subsection*{Important Note}

\begin{itemize}
 \item \Checkmate{} is built upon the tools and hard work of many people. If \Checkmate{} is used in your 
    publication it is extremely important that all of the following citations are included,
\begin{itemize}
  \item \texttt{Delphes 3} \cite{deFavereau:2013fsa}.  \\
  \texttt{https://cp3.irmp.ucl.ac.be/projects/delphes}
  \item \texttt{FastJet} \cite{Cacciari:2011ma,Cacciari:2005hq}.  \\
  \texttt{http://fastjet.fr/}
  \item Anti-$k_t$ jet algorithm \cite{Cacciari:2008gp}.
  \item \CLs prescription \cite{Read:2002hq}.
    \item All experimental analyses that were used to set limits in the study and if the analysis
    was implemented by non-\Checkmate{} authors, the relevant implementation reference.
  \item \Madgraph~\cite{Alwall:2014hca} if \Madgraph{} is used to calculate the hard matrix element events
  from within \CheckMATE. \\
  \texttt{https://launchpad.net/mg5amcnlo}
  \item \Pythia{} \texttt{8.2} \cite{Sjostrand:2014zea}  if showering or matching  done from within \Checkmate{}. \\
  \texttt{http://home.thep.lu.se/~torbjorn/Pythia.html}
  \item The Monte Carlo event generator that was used if \texttt{.hepmc} or \texttt{.lhe} files were generated externally.
  \item In analyses that use the $m_{T2}$ kinematical discriminant \cite{Lester:1999tx,Barr:2003rg} we use 
  the \texttt{mt2\_bisect} library \cite{Cheng:2008hk}. We also include the $M_{T2}^{b\ell}$ and $M_{T2}^W$ derivatives 
  \cite{Bai:2012gs}.  \\
  \texttt{http://particle.physics.ucdavis.edu/hefti/projects/doku.php?id=wimpmass} \\
  \texttt{https://sites.google.com/a/ucdavis.edu/mass/}
  \item In analyses that use the $M_{CT}$ family of kinematical discriminants we use the \texttt{MctLib} library
  that includes the following variables, $M_{CT}$ \cite{Tovey:2008ui}, $M_{CT}$ corrected \cite{Polesello:2009rn}, 
  $M_{CT}$ parallel and perpendicular \cite{Matchev:2009ad}. \\
  \texttt{https://mctlib.hepforge.org/}
  \item In analyses that use topness variable we use the topness library \cite{Graesser:2012qy}. \\
  \texttt{https://github.com/michaelgraesser/topness}
  \item In analyses that use Super-Razor \cite{Buckley:2013kua}.
\end{itemize}  

\end{itemize}  
}
\newenvironment{verbtextfile}[1]
  { 
\VerbatimEnvironment
\scriptsize \mdfsetup{
    frametitle={\colorbox{white}{\space \texttt{#1}\space}},
    innertopmargin=0pt,
    leftmargin=1cm,
    frametitleaboveskip=-\ht\strutbox,
    nobreak=false,
    roundcorner=3pt,
    frametitlealignment={\hspace*{0.010\linewidth}}
    }
  \begin{mdframed}%
\begin{Verbatim}
  }
  {
\end{Verbatim}\end{mdframed}
}

\newenvironment{bigtextfilestart}[1]
  {
\mdfsetup{
    frametitle={\colorbox{white}{\space \scriptsize \texttt{#1}\space}},
    innertopmargin=0pt,
    leftmargin=1cm,
    nobreak=false,
    roundcorner=3pt,
    bottomline=false,
    frametitleaboveskip=-\ht\strutbox,
    frametitlealignment={\hspace*{0.010\linewidth}}
    }
  \begin{mdframed} \vspace{-0.5\ht\strutbox} \colorbox{white}{\space \scriptsize \texttt{#1}\space}\scriptsize
  } 
  {\end{mdframed}
\normalsize
}
\newenvironment{bigtextfilemiddle}
  { 
\mdfsetup{
    innertopmargin=0pt,
    leftmargin=1cm,
    nobreak=false,
    roundcorner=3pt,
    topline=false,
    bottomline=false
    }
  \begin{mdframed} \scriptsize
  } 
  {\end{mdframed}
\normalsize
}

\newenvironment{bigtextfileend}
  { 
\mdfsetup{
    innertopmargin=0pt,
    leftmargin=1cm,
    nobreak=false,
    roundcorner=3pt,
    topline=false
    }
  \begin{mdframed}\scriptsize
  }  
  {\end{mdframed}
\normalsize
}

\newenvironment{bigtextfile}[1]
  { 
\mdfsetup{
    innertopmargin=1pt,
    leftmargin=1cm,
    nobreak=false,
    roundcorner=3pt,
    skipabove=3pt,
    frametitleaboveskip=-\ht\strutbox,
    frametitlealignment={\hspace*{0.010\linewidth}}
    }
  \begin{mdframed} \vspace{-0.5\ht\strutbox} \colorbox{white}{\space \scriptsize \texttt{#1}\space}\scriptsize
  } 
  {\end{mdframed}
\normalsize
}

\newenvironment{textfile}[1]
  { 
\scriptsize \mdfsetup{
    frametitle={\colorbox{white}{\space \texttt{#1}\space}},
    innertopmargin=0pt,
    leftmargin=1cm,
    frametitleaboveskip=-\ht\strutbox,
    nobreak=true,
    roundcorner=3pt,
    frametitlealignment={\hspace*{0.010\linewidth}},
    }
  \begin{mdframed}%
  }
  {\end{mdframed}

\normalsize
}

\newpage
{\small\tableofcontents}
\newpage

\section{Introduction}
\label{sec:intro}

The first years of the Large Hadron Collider (LHC) running have been a triumph with the notable discovery of the Higgs boson being
a particular highlight \cite{Aad:2015zhl,Khachatryan:2016vau}. Whilst no other new states have been found yet, a large number of searches 
for physics beyond the Standard Model (BSM) have been performed. This dataset has profound implications
for many of these BSM theories but the experimental collaborations themselves only have limited resources
to investigate the many models on the market. Therefore it is imperative that theorists
take up the task of testing their models against the wealth of data available.

To help with such studies, a number of tools have now been made public that allow for easy and fast
model testing. One class of programs are based on the so-called `simplified models' \cite{Alves:2011wf} 
approach. Here the tools make use of the actual LHC limits on particular topologies that are commonly
seen in BSM. The limits are then adjusted to account for the correct branching ratios in the actual 
model under test. The advantage of such an approach is that these programs are very quick to return an
answer. However the big disadvantage is that if the new physics model contains final state topologies not originally
tested, the limit will be severely weakened compared to the true result. Unfortunately this is very common
in realistic models that may have longer or asymmetric decay chains. Examples of tools that use
simplified models are \texttt{SModelS} \cite{Kraml:2013mwa,Kraml:2014sna} and \texttt{FastLim} \cite{Papucci:2014rja}
for supersymmetry and \texttt{XQCUT} \cite{Barducci:2014gna} for models with extended quark sectors.  Futhermore, 
if limits from a simplified model analysis are used for a different model (e.g.\ one predicting different angular 
correlations or extra intermediate particles) the results may not be accurate. This severely limits the applicability 
to classes of models with particles of exactly the same spin and an identical decay chain to
the one originally tested.

The second approach\footnote{Another approach, that we do not discuss here, is training a machine
learning algorithm on already tested models which has been pioneered by \texttt{SUSY-AI}~\cite{Caron:2016hib}.}  
for theorists to test models against the LHC data is to essentially
follow the same procedure that the experimental collaborations
perform themselves, or to ``recast''.\footnote{``To Recast'' is a a verb coined from Ref.~\cite{Cranmer:2010hk} and is now increasingly 
used as shorthand to describe the full process of reproducing an experimental analysis.}
Here, Monte Carlo (MC) events are simulated for the particular
model under test and these events are then passed through a detector simulation that returns
reconstructed final state objects. The same experimental
cuts used by the experiment are then placed on these objects and limits are placed on the number of
events seen in pre-defined signal regions. This chain is validated against benchmark models provided by the experiments.  Once validated, the recast analysis can be used to test any model by changing the MC events supplied.  Whilst this method has the disadvantage of being slower than
the simplified model approach, the big advantage is the generality that allows for a large range of 
different theories to be tested. Examples of tools that use recast analyses are
\CheckMATE{} \cite{Drees:2013wra,Kim:2015wza} and \texttt{MadAnalysis} \cite{Conte:2012fm,Dumont:2014tja}
while the newly released \texttt{Contur} \cite{Butterworth:2016sqg} uses the \texttt{RIVET} \cite{Buckley:2010ar}
library of Standard Model measurements.

So far all the recast based tools require the user to provide externally generated events that have been showered, 
hadronised and already have any required underlying event modelling. This has severe
performance drawbacks aside from the obvious extra effort for the end user to generate these events. Firstly,
the Monte Carlo events must be stored to disk and this process
involves writing and re-reading several 10's or even 100's GB for the required 
statistics.\footnote{We find e.g.\ that 10,000 hadronised events require between $\sim$1 and $\sim$10 GB.}
Consequently, even when we are just testing a single model point the Monte Carlo events already require
significant amounts of free space. However, a greater
issue is if we want to test many model points on a large cluster. In a cluster architechture where
many jobs run from a single hard drive, the reading and writing of events is already the limiting factor for just 2--3
simultaneous processes.

To solve this problem, we present \CheckMATETwo{} which integrates both \Madgraph{} and \Pythiaeight{}
 into a complete model testing loop. As a consequence, the user only has to provide 
an \Slha{} file \cite{Skands:2003cj,Allanach:2008qq} in the case of supersymmetry or a  
\texttt{UFO} \cite{Degrande:2011ua} model file
that can be produced by a variety of tools such as \texttt{FeynRules} \cite{Christensen:2008py,Alloul:2013bka},
\texttt{SARAH} \cite{Staub:2015kfa,Staub:2013tta} or \texttt{LanHEP} \cite{Semenov:2014rea} for other models and the 
event generation is then taken care of internally by \CheckMATE{}. We also note that the improved efficiency due to skipping the event file storage
results in a single process running up to $\sim$40\% faster. More importantly however, is the far improved
cluster processing. That means running many simultaneous jobs from a single hard disk is now made possible
without any reduction in performance.

The incorporation of the event generation is not the only improvement in \CheckMATETwo{} though. Over 60
experimental analyses from the LHC collaborations are now available covering the 7, 8 and 13~TeV runs 
and more are continually being added. In addition, \CheckMATETwo{} now contains 14~TeV high luminosity
LHC analyses as well. Consequently the user can now go beyond simply setting limits on models with the 
current data and investigate what is the ultimate LHC reach to the model of interest. Further 
improvements have also been made to the \texttt{AnalysisManager} introduced in Ref.~\cite{Kim:2015wza} so that users can
more easily investigate new ideas for LHC searches. These include more LHC kinematical observables 
being included and additional tools that make the statistical analysis of the results easier.

We begin this paper in \Cref{sec:cm:overview} by giving an overview of the individual building blocks 
that make up the \CheckMATE{} program. We follow in \Cref{sec:app:fulllist} by giving the full list of available \CheckMATE{} parameters that
may be useful to more advanced users. To more easily explain how \CheckMATE{} is used, we then provide
an example run in \Cref{sec:cm:tutorial} that also discusses in more detail some of the most commonly used options.
A brief summary of the currently available analyses is provided in \Cref{sec:cm:analyses}.
\Cref{sec:performance} investigates the performance improvements 
of the new version of \CheckMATE{} over the old in more detail while \Cref{sec:analyismanager} explains
the improvements to the \texttt{AnalysisManager}. Finally, we summarise in \Cref{sec:summary}. 
Appendix~\ref{sec:Install} gives
installation instructions, Appendix~\ref{sec:app:statistics} covers the statistical methods used 
in \CheckMATE{} in more detail and Appendix~\ref{sec:app:tuning} describes the updated lepton identification performance and
$b$-tagging efficiencies now used in \CheckMATE{}.

\section{General Program Flow}
\label{sec:cm:overview}
%

\begin{figure}
\centering
\includegraphics[width=\textwidth]{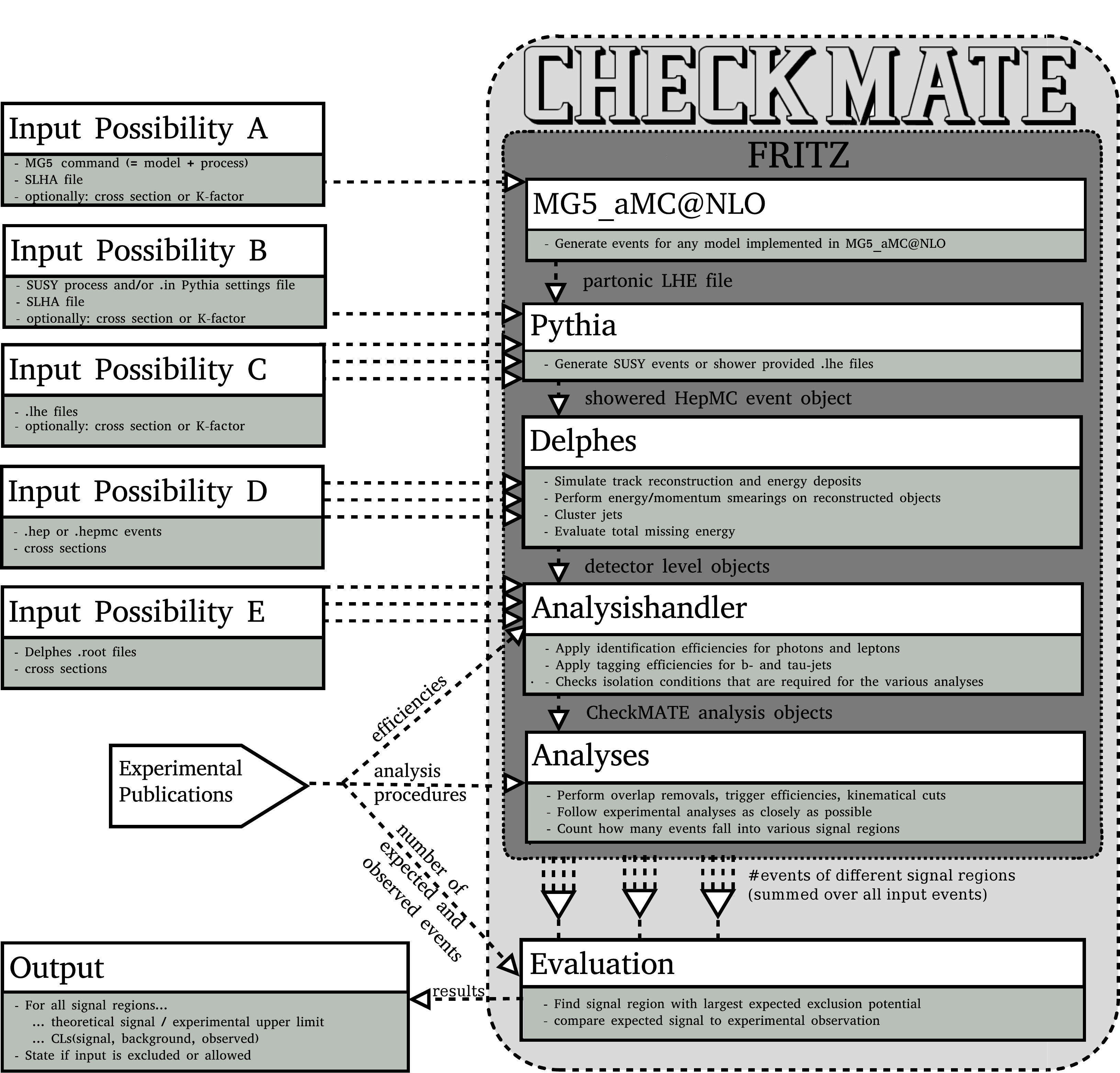}
\caption{Flow chart to demonstrate the chain of data processing within \Checkmate{}. }
\label{fig:flowdiagram}
\end{figure}
\Checkmate{}{} incorporates many individual modules which cover the steps necessary for model 
testing. A flowchart illustrating the modules and the data passed internally between them 
 is given in \Cref{fig:flowdiagram}. The modules are embedded in 
a \Python{} scaffolding which handles the user prompts, the file I/O and the setup 
of the core modules which we describe in more detail below.

\subsection{The \Python{} Scaffolding}
\label{sec:cm:init}
The initialisation is performed by a \cmate{} \Python{} script which reads in user input (either via a file or command line), and writes the configuration files required by the \Fritz{} and \texttt{AnalysisHandler} modules to set up event generation, showering and detector simulation followed by recasting of the experimental analysis.  The computation-heavy parts are taken over by the individual modules coded in \texttt{C++} described below, which call as required external libraries or programs (viz.\ \Madgraph{}, \Pythiaeight{}, \texttt{HepMC}, \Delphes{} and \Root{}).  After the recast analyses are run, the results are again collected and processed by the \Python{} script to check against published 95\% confidence level (CL) upper limits.  The ratio of expected signal to the 95\% CL upper limit and whether the point is still allowed is reported as the default primary result.  Further options for statistical evaluation are discussed in \Cref{sec:cm:eval}. 

\subsection{\Fritz}
\label{sec:cm:fritz}
\Fritz\footnote{The name \emph{Fritz} is derived from a German chess program of the same 
name, see Ref.~\cite{fritz1,fritz2}, and the very first chess computer program one of the authors (DD) played.}   
(Flexible Rapid Interactive Tool Zipper) denotes the core \texttt{C++} 
 program of \Checkmate{}{}. Depending on the provided data and settings, it connects to and runs \Madgraph, \Pythiaeight{}  
 and \Delphes{}, followed by the \texttt{AnalysisHandler} and all the analyses requested by the user.
 Except for the LHE files produced by \Madgraph{}, intermediate data, e.g.\ the simulated Monte Carlo events generated 
by \Pythia{} and/or the detector level objects produced by \Delphes{}, are passed on-the-fly between the individual modules.  
This is a great improvement on the original \Checkmate{}{} version \texttt{1} in which the generated events as well as the detector level objects had to be stored and then re-read from hard disk. Given that a typical BSM Monte Carlo event file including hadronised final states and a sufficiently high statistical sample easily reaches file sizes of several GB, significant time is saved in I/O by this improvement\footnote{We show later that we can gain a factor of 3 in speed between \Checkmate{} version \texttt{1} and \Checkmate{} version \texttt{2} depending on the details of the benchmark model and the number of parallel runs. Details can be found in \Cref{sec:performance}.} besides removing the requirement for large storage.  However, if the user requires so, the intermediate objects can be written to the disk using the switches \texttt{WritePythiaEvents} and \texttt{WriteDelphesEvents}, cf.\ \Cref{sec:app:fulllist}.

\subsection{Event Use or Generation:}

One of the core parts of Monte Carlo based collider phenomenology is the simulation of 
final state configurations that would be produced in a collider experiment if a particular 
model of BSM physics was true. In the first version of \Checkmate{}, the 
event generation had to be done externally by the user. MC event files and the 
corresponding cross section --- either from the same event generator or from an external 
cross section calculator like \texttt{Prospino} \cite{Beenakker:1996ch,Beenakker:1997ut,Beenakker:1999xh,Spira:2002rd,Plehn:2004rp} 
or \Nllfast~\cite{Beenakker:1996ch,Beenakker:1997ut,Kulesza:2008jb,Kulesza:2009kq,Beenakker:2009ha,Beenakker:2010nq,Beenakker:2011fu} 
--- were mandatory input parameters which were then processed via \Delphes{} within \Checkmate{}. Besides the practical inconvenience 
that every \Checkmate{}{} user had to use an external event generator, the forced split between event generation 
and detector simulation/analysis also yields a computational disadvantage as already explained above. Consequently,  
the new \Checkmate{}{} version now provides an automatic link to both \Madgraph{}  \cite{Alwall:2014hca} and
the \Pythiaeight{}~\cite{Sjostrand:2014zea} event 
generation. With this new functionality, \Checkmate{} provides different types of modes to either run 
\Madgraph{}  and \Pythiaeight{} or simply use already generated event files:
\begin{description}

\item[Provide an externally produced \texttt{.hepmc} or \texttt{.hep} event file:]
We first emphasise that if the user wishes to provide Monte Carlo events in either \texttt{.hepmc} or \texttt{.hep}
format to \Checkmate{}, this option is still supported. \CheckMATE{} will then pass these events directly to
\Delphes{} for detector simulation.

\item[Generate events entirely using \Pythiaeight:] \Pythiaeight{} is capable of generating events for 
BSM models followed by parton showering and hadronisation of the final state. This functionality 
can be accessed by \Checkmate{} in two different ways. 

The first possibility is to provide the \Pythiaeight{} setup via an \texttt{.in} file which uses the \Pythiaeight{} 
internal syntax, see refs.~\cite{Sjostrand:2007gs,Sjostrand:2014zea,pythiawebpage}, to set the internal 
parameters. This mode allows for the full flexibility of the \Pythiaeight{} program as all parameters can be 
changed via this input file method. Most importantly, the \texttt{.in} file is used to define the model and 
the list of processes which should be generated. All model parameters (e.g.\ couplings, masses, widths, branching ratios etc.) must be provided in the input file. If a supersymmetric (SUSY) process is desired, the \Slha{} file \cite{Skands:2003cj,Allanach:2008qq} which specifies the parameter point must be provided within the input file in the standard \Pythiaeight{} syntax.

Given the popularity of SUSY models, we provide an additional shortcut to generate SUSY processes without the need to provide a full input card.  All showering and hadronisation parameters will then be taken from the default \texttt{Pythia} settings.  The process can be set directly using a simplified \Madgraph-like syntax
e.g.\ \texttt{Pythia8Process: p p > go go} to initiate gluino pair production. A further shorthand to refer to classes of SUSY particles is also available.  The full list of available processes using this running mode is given in \Cref{sec:app:fulllist}. 

In both cases, \Checkmate{} will use \Pythiaeight{} to simulate the given processes and will directly  
perform detector simulation followed by applying analysis routines. As explained above the events are 
not stored unless explicitly demanded by the user. 

By default, \Checkmate{} uses the cross section and statistical uncertainty provided by \Pythiaeight. A user 
can also provide a cross section or a $K$-factor, i.e.\ $\sigma_{\text{NLO}}/\sigma_\text{LO}$, 
calculated via an external code (see \Cref{sec:app:fulllist}).

\item[Generate events with \Madgraph, shower and hadronise with \Pythiaeight:] This option allows a 
user to both generate the hard matrix element using \Madgraph{}  by providing a process card 
(and parameter and run cards if necessary) followed by showering and hadronisation of the 
resultant \texttt{LHE} files via \Pythia. This allows  users to generate Monte Carlo events for a 
huge range of BSM models using the \Ufo{} file format~\cite{Degrande:2011ua}.

\item[Shower and hadronise externally provided \texttt{.lhe} files:] 
If a user wishes to calculate the matrix element and generate events with a different generator than \Madgraph{}, e.g.\ \texttt{WHIZARD} \cite{Kilian:2007gr,Moretti:2001zz} or \texttt{CalcHep} \cite{Belyaev:2012qa}, this is also possible as long as the program is able to output event files in the \texttt{.lhe} format.
The advantage of this approach is that the hard process simulation and the parton showering and 
hadronisation steps can be performed rather 
independently and it is usually the latter two which take the most computational effort and require most disk space. \Checkmate{} can take lightweight \texttt{.lhe} files 
produced by an external tool and use \Pythiaeight{} to shower/hadronise on-the-fly as described above.  A default  
\texttt{.in} file is then used to set the collision energy and to retain default \Pythiaeight{} settings but if necessary users can also 
provide their own file instead (if e.g.\ a user wishes to turn off multi-particle interactions etc.)
\end{description}

\Pythiaeight{} internally generates random numbers starting from a default seed. The user can further provide the random number seed in two ways --- first by setting an integer seed via \texttt{RandomSeed: X} and second by providing a binary state for the \Pythiaeight{} run using the \texttt{Pythia8Rndm} key in the \CheckMATE{} input file.  Each \Pythiaeight{} run writes out the random state at the beginning and end of the run in the files \texttt{rndm-init.dat} and \texttt{rndm-end.dat} respectively.  These can then be used either to reproduce a run or while adding events to ensure there is statistical independence of the two runs.

\subsection{\Delphes{}}
\label{sec:cm:expl:delphes}
The events that were either generated internally or provided by the user are then passed to the detector simulation program \Delphes{} 
\cite{deFavereau:2013fsa}. In \Checkmate{} \texttt{2.0}, 
contrary to the original \Checkmate{} described in Ref.~\cite{Drees:2013wra}, \Delphes{} is now only
used to simulate the calorimeter and tracking using the standard detector settings for \Atlas{} and \Cms{} along
with jet reconstruction.  The identification efficiencies for final state particles are now performed internally by \CheckMATE{}.
In addition, the output event object is extended to include generator level particles which are required for external $b$- and 
$\tau$-tagging.

The results of the detector simulation are passed as \Root{} objects and are typically not saved on 
disk (unless explicitly requested) but instead immediately processed by the analysis 
framework described below. By setting \texttt{WriteDelphesEvents} to \texttt{True} (see also \Cref{sec:app:fulllist}), 
a \texttt{.root} output file can optionally be created and used as an event 
file in a future \Checkmate{} run, see also \Cref{fig:flowdiagram}. 

If a user decides to test only \Atlas{} or only \Cms{} analyses, \Delphes{} needs to run only once per event. Otherwise, 
each input event is processed by two independent \Delphes{} runs and independent detector level objects 
for \Atlas{} and \Cms{} analyses are respectively created. 
\subsection{\texttt{AnalysisHandler}}
\label{sec:delphes:introah}
The detector level objects created in the previous step contain reconstructed electrons, muons, photons, 
jets, tracks, clustered calorimeter cells and the missing transverse momentum vector. These are now further processed 
by a so-called \texttt{AnalysisHandler} before being passed to the actual data analysis codes. 

Depending on the list of analyses selected by a user, final state objects are tested against a list of isolation, 
identification and tagging conditions which set individual tags 
on those candidates which pass the respective constraints. For that purpose, \Checkmate{} first determines 
which constraints have to be considered in order to provide all analyses with the 
required information on the final state objects. As an example, let us assume a user chose three 
analyses out of which two require tight leptons (usually this means electrons or muons at the LHC) and the 
third one vetoes events with medium leptons. Then the \texttt{AnalysisHandler} using a lepton list passed 
from \Delphes{} will create three new lepton lists, corresponding to different identification working 
points: \texttt{tight} $\subset$ \texttt{medium} $\subset$ \texttt{loose}. The decision about the 
assignment for a particular lepton is based on its momentum and takes into account identification 
efficiencies reported by the experiments, see also  Appendix~\ref{sec:app:tuning}. Finally, 
isolation criteria will be checked for each of the leptons. The \texttt{AnalysisHandler} will 
also apply simplified $b$- and $\tau$-tagging algorithms, if required.\footnote{In the first \Checkmate{} version,
all these steps have been performed within the \Delphes{} framework by generating a run-specific \Delphes{}
detector card. That made it impossible to re-use \Delphes{} output files, either within \Checkmate{} 
to \emph{a posteriori} 
test additional analyses or outside \Checkmate{} with a different \Delphes{}-based analysis framework 
like e.g.\ \Madanalysis~\cite{Conte:2012fm}.  Therefore, the \Checkmate{} and \Madanalysis{} teams mutually agreed to 
switch to a final state post-processing outside \Delphes{}, see also~\cite{miniworkshop}.}

\Checkmate{} uses a set of independent \texttt{AnalysisHandler}s with individual tagging efficiencies depending on the list of analyses chosen 
by the user. Due to the evolution of the reconstruction and identification algorithms, there exist 
separate \texttt{AnalysisHandler}s for analyses performed  on 7, 8 and \unit[13]{TeV} data and 
for projective studies at \unit[14]{TeV} centre-of-mass energy. For each of those there is an 
independent \Atlas{} and \Cms{} version. The tunings for 7 and \unit[8]{TeV} are described in more detail in 
Ref.~\cite{Drees:2013wra} while the updates for \unit[13]{TeV} are given in Appendix~\ref{sec:app:tuning}.
\subsection{Analyses}
After all detector level objects have been properly prepared by the \texttt{AnalysisHandler}(s), these are processed event by event by each analysis 
selected by a user when starting \Checkmate{}. They are internally coded in a framework that allows for an easy extension to new upcoming experimental 
results and allows users to easily update given analyses or implement their own, see also the original \Checkmate{} publication in Ref.~\cite{Drees:2013wra}. For a detailed description of the \texttt{AnalysisManager} framework see \cite{Kim:2015wza}.

Each event is processed by checking isolation criteria, removing overlapping objects and implementing the cuts that define the signal regions. The analysis program then determines how many events in total satisfy certain signal region criteria and stores this information in a human-readable output for each separate input event file or alternatively each separate \Pythia{} run. In addition we also store the actual number of Monte Carlo events (i.e.\ the efficiency times acceptance $\mathcal{A} \times \epsilon$) and the number of signal events $S = \mathcal{L}_{\text{int}} \times \sigma \times \mathcal{A} \times \epsilon$, where $\mathcal{L}_{\text{int}}$ is the integrated luminosity, and the cross section $\sigma$ is provided either internally by \Pythiaeight{} or externally by the user. In case of weighted events, the event weights are taken into account while calculating the efficiency.  
\subsection{Evaluation and Output}
\label{sec:cm:eval}
The final step of the program consists of a statistical evaluation of the results. For 
each individual signal region of every chosen analyses, the total number of expected 
signal events $S$ is determined by summing up the results from each input event 
file (or event generation run) as explained later in \Cref{sec:cm:example:eval}. The total $1 \sigma$ uncertainty $\Delta S$ on this number 
is determined from both the statistical uncertainty, given by the number of 
Monte Carlo events, and the systematic uncertainty, which is estimated from the 
total uncertainty on the signal cross section given by the user as an optional parameter. These numbers are compared to the 
results published in the respective experimental search. There are two possible ways to perform the comparison 
(see also Ref.~\cite{Drees:2013wra}): in the standard approach, the number $S$ is tested against a
pre-calculated model independent 95~\% CL limit $S95$. Alternatively, a user can choose to calculate the proper 
\Cls{} value by folding in the uncertainty on the model prediction. This is not only more accurate but 
allows for testing against limits other than 95~\% CL. In both cases, \CheckMATE{} calculates the observed
\Cls{} value using the actual data recorded and the expected \Cls{} value which assumes
the observed data equals the background expectation.
In contrast to the earlier \Checkmate{} version, the current version uses its own routines to 
perform the statistical calculations using the algorithm explained in 
Appendix~\ref{sec:app:statistics}.  Besides those two means of model testing, \Checkmate{} 
optionally allows for the calculation of the likelihood of the final result. This allows 
for model parameter fits and corresponding confidence limit evaluations. The calculation of the likelihood is explained in detail in Appendix~\ref{sec:app:statistics}.

The results of the evaluation are then output for every signal region of every analysis.  If multiple signal regions are considered, the model point is determined to be ``excluded'' or ``allowed'' based on the signal region which has the best sensitivity assuming background-only hypothesis evaluated by the experiment.  
This is done to avoid erroneously ruling out a model point due a downward fluctuation of the observed events in a signal region that is not expected to be sensitive in the first place.

%


\section{Full List of \Checkmate{} Parameters}
\label{sec:app:fulllist}

There exist many optional parameters within \Checkmate{} which can change the standard behaviour of the 
code and here we describe their usage. These parameters can either be provided via the input file 
or alternatively, \Checkmate{} can be set up directly within 
the command line by adding $\texttt{-parameter value}$  pairs after the \texttt{./CheckMATE} command. 
The second alternative is unfortunately only possible for a setup with a single process as 
only one $\texttt{-p}$ command can be provided. If more than one process needs to be analysed 
in the same run, one either has to use the input file or use the \texttt{add} feature described below.

\begin{table}
\begin{tabularx}{\textwidth}{llX}
\toprule \midrule
Parameter card & Terminal & Description \\
\midrule
\multicolumn{3}{c}{General options} \\
\midrule
\texttt{Name: X} &  \texttt{-n X} & Gives name \texttt{X} to the run and specifies output directory.  \\
\texttt{Analyses: X} &  \texttt{-a X} & States which analysis/es \texttt{X} should be applied to the event files; see the text for more details. \\
\texttt{SLHAFile: X} &  \texttt{-slha X} & Use \Slha{} file \texttt{X}. Mandatory if event generation using \Pythiaeight{} is requested. \\
\texttt{InvisiblePIDs: X} &  \texttt{-invpids X} & BSM Monte Carlo Particle IDs \cite{Agashe:2014kda} which are 
invisible for the detector; see the text. \\
\texttt{QuietMode: True} &  \texttt{-q} & No terminal output is produced. Automatically sets \texttt{-sp}. \\
\texttt{SkipParamCheck: True} &  \texttt{-sp} & Skips startup parameter check. \\
\texttt{SkipAnalysis: True} &  \texttt{-sa } & Skips analysis step. Requires \texttt{-wp8} or \texttt{-wd}. \\
\texttt{SkipPythia: True} &  \texttt{-spy } &  Only if \texttt{.lhe} files are provided. These are not showered by \Pythiaeight{} but instead directly processed via \Delphes{}.$^*$ \\
\texttt{SkipEvaluation: True} &  \texttt{-se} & Skips evaluation step. \\
\texttt{RandomSeed: X} &  \texttt{-rs X} &  Chooses fixed seed \texttt{X} for the random number generator to render output deterministic. \\
\midrule
\multicolumn{3}{c}{Options related to output} \\
\midrule
\texttt{WritePythiaEvents: True} &  \texttt{-wp8} &  Writes \texttt{.hepmc} files produced by  \Pythiaeight{} on disk. \\
\texttt{WriteDelphesEvents: True} &  \texttt{-wd} &  Writes \texttt{.root} files produced by \Delphes{} on disk. \\
\parbox[t]{4cm}{\texttt{EventResult-}\\\phantom{X}\texttt{FileColumns: X}} &  \texttt{-erfc X} &
                        Sets columns which are stored in event-wise result files; see the text for more details. \\
\parbox[t]{4cm}{\texttt{ProcessResult-}\\ \phantom{X}\texttt{FileColumns: X}} &  \texttt{-prfc X} &
                        Sets columns which are stored in process-wise result files; see the text for more details. \\
\parbox[t]{4cm}{\texttt{TotalResult-}\\\phantom{X}\texttt{FileColumns: X}} &  \texttt{-tefc X} &
                        Sets columns in \texttt{TotalResults.txt} after evaluation; see the text for more details. \\
\parbox[t]{4cm}{\texttt{BestPerAnalysisResult-}\\ \phantom{X}\texttt{FileColumns: X}} &  \texttt{-bpaefc X} &
                        Sets columns in \texttt{BestPerAnalysis.txt} after evaluation; see the text for more details. \\
\texttt{OutputDirectory: X} &  \texttt{-od X} &
                        Specifies directory in which the results should be stored. \\
\texttt{OutputExists: X} &  \texttt{-oe X} &
                        Specifies what to do if output directory already exists. \texttt{overwrite} will
                        delete existing output and overwrite it with the new
                        results. \texttt{add} will add the current results to the old
                        ones, see text. \\
\midrule \bottomrule
\end{tabularx}
\caption{Summary of all parameters which can be set within \Checkmate{}, either via the parameter card introduced 
in \Cref{sec:cm:tutorial} or as command line input $\texttt{./CheckMATE -X -Y ...}$.   Occasional dash symbols (\texttt{-}) in the 
first column only indicate that a command is split in two lines to reduce the column width and are \emph{not} part of the keyword. 
$^*${\it The use of LHE events that have only been generated at the parton level or showered events that have been pre-clustered is not 
recommended and may lead to substantial efficiency and acceptance errors.}}
\label{tbl:cm:params1}
\end{table}

\begin{table}
\begin{tabularx}{\textwidth}{llX}
\toprule \midrule
Parameter card & Terminal & Description \\
\midrule
\multicolumn{3}{c}{Options related to statistical evaluation} \\
\midrule
\texttt{FullCLs: True} &  \texttt{-cls} &   Evaluate full observed and expected \CLs for all signal regions and use it for exclusion test.  \\
\texttt{BestCLs: X} &  \texttt{-bcls X} &  As above but only for the $X$ signal regions with highest \texttt{rexpcons} value. \\
\texttt{Likelihood: True} &  \texttt{-likeli} &  Evaluate likelihood for all signal regions and sum the individual contributions; see the text. \\
\texttt{EffTab: B} &  \texttt{-eff\_tab} &   Creates efficiency tables for every signal region in
                        each analysis run. \\      
\texttt{No\_MC\_Stat\_Err: True} &  \texttt{-mcstats\_off} & Uncertainty associated with limited Monte Carlo signal statistics is not included
                                                          in any statistical evaluation. \\                        

\midrule
\multicolumn{3}{c}{Process dependent options} \\
\midrule
\texttt{[X]} &  \texttt{-p X} &   Sets up a process with name \texttt{X}. \\
\texttt{MaxEvents: X} &  \texttt{-maxev X} & Defines the  number of generated events if \Pythiaeight{} or \Madgraph{} is used for event generation. If events are provided, simulation stops when either the end of the event file or \texttt{MaxEvents} is reached. \\
\texttt{XSect: X} &  \texttt{-xs X} & Sets the cross section for the given process including unit (e.g.\ '\texttt{10.7 fb}'. Note that the space between the value and the unit is required). \emph{Must} be provided for \texttt{.hepmc} and \texttt{.root} input and \emph{can} be provided to override the \Pythiaeight{} calculated cross section. \\
\texttt{XSectErr: X} &  \texttt{-xse X} & Sets the systematic cross section error for the given process including unit which can be given with a unit or provided as a percentage (e.g.\ '\texttt{2.2 fb}' or '\texttt{20.0 \%}'. Note that the space between the value and the unit is required). \\
\texttt{KFactor: X} &  \texttt{-kf X} & Sets the $K$-factor for the cross section. \\
\texttt{Events: X} &  \texttt{-ev X} &  Sets the \texttt{.hepmc, .hep, .lhe} or \texttt{.root} event files which are to be analysed for the given process. \\
\texttt{Pythia8Process: X} &  \texttt{-pyp X} &  Specifies the SUSY process to be generated by \Pythiaeight{}; see the text for more details. \\
\texttt{Pythia8Card: X} &  \texttt{-pyc X} &  Specifies the \texttt{.in} input card used by \Pythiaeight{}.  It can also be used to provide settings for showering and hadronisation of \texttt{LHE} files provided externally or generated with \Madgraph{}.  \\
\texttt{Pythia8Rndm: X} & \texttt{-pyr X} & Provides the binary random number generator state (output during previous run) for the \Pythiaeight{} run. \\
\texttt{MGprocess: X} & \texttt{-mgproc X} &  Specifies the process card for generating events with \Madgraph{}.\\
\texttt{MGcommand: X} & \texttt{-mgcommand X} &  Specifies the content of the \Madgraph{} process card excplicitly.\\
\texttt{MGparam: X} & \texttt{-mgparam X} &  Optional parameter card for \Madgraph{} (e.g.\ \texttt{SLHA} card).\\
\texttt{MGrun: X} & \texttt{-mgrun X} &  Optional run card for \Madgraph{}. \\
\texttt{MGconfig: X} & \texttt{-mgconfig X} &  Optional configuration file for \Madgraph{} (e.g.\ \texttt{mg5\_configuration.txt}). \\
\midrule \bottomrule
\end{tabularx} 
\caption{Continuation of \Cref{tbl:cm:params1}. Occasional dash symbols (\texttt{-}) in the 
first column only indicate that a command is split in two lines to reduce the column width and are \emph{not} part of the keyword.}
\label{tbl:cm:params2}
\end{table}

\begin{description}
\item[Analyses:] The full list of currently implemented analyses is given in \Cref{sec:cm:analyses}. The following examples show how to specify
which of these \Checkmate{} should take into account using an input file (command line),
\begin{itemize}
\item \texttt{Analyses: atlas\_1404\_2500 (-a atlas\_1404\_2500)} tests only the analysis \texttt{atlas\_1404\_2500}.
\item \texttt{Analyses: 8TeV (-a 8TeV)} tests all implemented analyses which correspond 
to  $\sqrt{s} = \unit[8]{TeV}$. Alternative values are \texttt{7TeV}, \texttt{13TeV} and \texttt{14TeV}. 
\item \texttt{Analyses: atlas8TeV (-a atlas8TeV)} tests all implemented \Atlas{} analyses of the given centre-of-mass energy. 
Similarly for \texttt{cms}. 
\end{itemize}
The above specifiers can be combined via a simple separation with commas,  \\
e.g.\ \texttt{-a cms8TeV, atlas\_1404\_2500} tests all \unit[8]{TeV} \Cms{} analyses and the 
single \Atlas{} analysis  \texttt{atlas\_1404\_2500}. All analyses combined this way \emph{must} 
correspond to the same center-of-mass energy, otherwise \Checkmate{} will abort.
\item[Invisible PIDs:] Physics beyond the Standard Model which addresses the dark matter problem often 
predicts the existence of one or more stable, light particles that only interact weakly 
with ordinary matter. Whilst \Delphes{} automatically identifies all neutral particles in the Minimal Supersymmetric Standard Model (MSSM),   
other BSM particles have to be explicitly declared as invisible as they are otherwise 
considered as exotic hadrons which deposit their energy in the hadronic calorimeter. As an 
example, a Higgs portal  model with a stable scalar would require placing in the
parameter file the following setting: \texttt{Invisible PIDs: 35}.
\item[Result file columns:] \Checkmate{} stores the results of its analyses in various files to allow for a detailed investigation how the final result was determined. 
  The standard content of these result files --- we describe them in more detail in \cref{sec:cm:tutorial} --- is 
adaptable such that more intermediate results may be stored. By setting the corresponding 
\texttt{ResultFileColumns} parameter to \texttt{a,b,c,...} the corresponding file(s) are set to contain respective information $a, b, c$. \texttt{EventResultFileColumns} and \texttt{ProcessResultFileColumns} can be taken out of the following set:
\begin{sloppypar}
\texttt{analysis, sr, totalmcevents, totalnormevents, totalsumofweights, totalsumofweights2, signalsumofweights, signalsumofweights2, signalnormevents, signal\_err\_stat,  signal\_err\_sys, signal\_err\_tot}.
\end{sloppypar}
The names are mostly self-explanatory; the prefix \texttt{total-} refers to the full input sample whereas \texttt{signal-} corresponds to the subset of events which pass the respective signal region cuts. \texttt{sumofweights2} corresponds to sum of squared weights which is an important quantity to calculate the statistical uncertainty properly in case of weighted events. \texttt{normevents} correspond to the physical number of events after normalising to the provided cross section and the analysis' respective integrated luminosity.

\texttt{TotalResultFileColumns} and \texttt{BestSignalRegionResultFileColumns} can in addition 
use the following columns:
\begin{sloppypar}
\texttt{obs, bkg, bkgerr, eff, eff\_err\_stat, eff\_err\_sys, eff\_err\_tot, s95obs, s95exp, 
robs, robscons, robsconssysonly, rexp, rexpcons, rexpconssysonly, clsobs, clsobs\_err, 
clsexp, clsexp\_err, likelihood},
\end{sloppypar}
which are mostly self-explanatory. The suffix \texttt{-cons} refers to the $r$ value in \cref{eq:rvalue} and columns without this suffix do not include the conservative subtraction term of $1.64 \cdot \Delta S$ in the numerator. Note that the output of \Cls{} and 
likelihood related columns are set to $-1$ unless the calculation of the respective quantity is enabled via the corresponding flag; see \Cref{tbl:cm:params2}.

\item[\texttt{add} mode:] After a \Checkmate{} run is completed, the user might realise that the 
events which were processed were insufficient. For example, the size of the tested Monte Carlo samples 
might be too small to find the number $S$ with desired statistical precision.  
It might also happen that \emph{a posteriori} it becomes apparent that other processes need to be taken 
into account which were expected to be negligible before the first run. For those cases, 
\Checkmate{} allows for new results to be added to old ones. To do so, \Checkmate{} has to be set up 
with the same name and the same output directory as the original run. During the initialisation step 
the user is then explicitly asked if the new results are supposed to replace or to be added to the 
existing ones. If the second option is chosen, the original \Checkmate{} settings are restored 
from the earlier run, all events in the current setup file are processed and properly added to 
the ones of the first run. This procedure can be repeated as many times as the user wishes. This behaviour can be controlled by the \texttt{OutputExists} parameter or using \texttt{-oe add} from the command prompt.    
\item[Likelihood:] Instead of exclusion tests, \Checkmate{} can also be used for model fits; see Ref.~\cite{Kim:2014eva} for an 
example. With this option \Checkmate{} can calculate 
the likelihood ratios for all signal regions to test the compatibility of a given model with the experimental 
results. A formulae for a simplified LHC profile likelihood ratios without nuisance parameters is 
given by Eq.~\eqref{eq:coll:likelihood} in Appendix~\ref{sec:app:1bin_like}. \Checkmate{} uses the full version 
including $\Delta B$ and $\Delta S$ as nuisance parameters given by Eq.~\eqref{eq:coll:likelihood2} 
of \Cref{sec:app:1bin_like}. In the final results, \Checkmate{} also returns the sum of the likelihood ratios over all signal 
regions but this value should be used with care since no checks for kinematically overlapping regions are considered. The user
is advised to check the signal regions of interest and only sum those that are independent.
Analysing the dependence of this quantity on model 
parameters allows one to find best fit points and the corresponding confidence intervals, see 
e.g.\ Ref.~\cite{Agashe:2014kda}. 
\item[Pythia8Process:] 
Due to the popularity of the MSSM, \Checkmate{} also allows 
one to easily set up generation of SUSY production processes using the \texttt{Pythia8Process} 
keyword. Possible values for this parameter are \texttt{p p > X}, with \texttt{X} being any of the following:
\begin{description}
\item[\texttt{go go}:] gluino pair production;
\item[\texttt{go sq}:] gluino-squark and gluino-antisquark associated production;
\item[\texttt{sq sq~}:] squark-antisquark production;
\item[\texttt{t1 t1~}:] pair production of the lightest stop;
\item[\texttt{3gen}:] pair production of stops and sbottoms;
\item[\texttt{sq sq}:] squark pair production;
\item[\texttt{colsusy}:] all coloured SUSY pair-production;
\item[\texttt{ewsusy}:] pair production of neutralinos, charginos and neutralino-chargino production;
\item[\texttt{allsusy}:] all of the above.
\end{description}
Note that here ``squark'' always corresponds to the squarks of the first two generations. To simulate 
any other process or any combination of the above, an explicit \Pythiaeight{} \texttt{.in} file has 
to be provided. For all processes the default parton distribution function \cite{Ball:2012cx} is used and 
the underlying event is switched off.
\end{description}
\color{black}

	\section{Example: Running \Checkmate{} and Understanding the Results}
\label{sec:cm:tutorial}

To illustrate how the individual steps explained in \Cref{sec:cm:overview} work in practice, we discuss an 
example \Checkmate{} run. It is designed in such a way that it covers the most common scenarios to provide 
input data within the current \Checkmate{} version. It also attempts to apply some optional settings to 
illustrate their meaning.  After the example run is completed, we take a closer look at the auxiliary 
files which are created along the way and what additional information a user can find in these. A tarball containing all files which are used as input in this example run can be downloaded from \url{http://www.hepforge.org/archive/checkmate/ExampleFilesForCheckMATE2Manual.tar.gz}
\subsection{Benchmark Model}
Within this section we test a simplified supersymmetric model where the only kinematically 
accessible particles are the gluino with mass \unit[1.5]{TeV}, the eight mass-degenerate squarks of the 
first two generations with mass \unit[1.5]{TeV} and a \unit[100]{GeV} stable bino-like neutralino lightest supersymmetric particle (LSP). Here, the 
gluino is expected to always decay democratically\footnote{For a bino-like LSP, a gluino would actually decay with different branching-ratios into up-like and 
down-like quarks due to their different quantum numbers. However, phenomenologically these quarks are almost indistinguishable at the LHC so we can safely set all branching ratios equal.} into same-flavour quark-antiquark pairs of the first two fermion generations and the stable lightest neutralino. Squarks always decay into the associated Standard Model quark and the neutralino. The masses of squarks and gluino are chosen such that neither of the two can directly decay into the other.
The important parts of the \Slha{} file for this model are as follows: 
\begin{bigtextfile}{point.slha}
\begin{Verbatim}[commandchars=\\\@\@]
[...]
Block MASS   # Scalar and gaugino mass spectrum
#  PDG code    mass    particle
      1000001 1.500000e+03 # downL squark
      1000002 1.500000e+03 # upL squark
      1000003 1.500000e+03 # strangeL squark
      1000004 1.500000e+03 # charmL squark
      1000005 5.000000e+03 # bottom1 squark
      1000006 5.000000e+03 # top1 squark
      2000001 1.500000e+03 # downR squark
      2000002 1.500000e+03 # upR squark
      2000003 1.500000e+03 # strangeR squark
      2000004 1.500000e+03 # charmR squark
      2000005 5.000000e+03 # bottom2 squark
      2000006 5.000000e+03 # top2 squark
      1000011 5.000000e+03 # electronL slepton (and other sfermions)
      [...]
      1000021 1.500000e+03 # gluino
      1000022 1.000000e+02 # neutralino 1
      1000023 5.000000e+03 # neutralino 2 (and other neutralino/charginos)
      [...]
 [...]
#         PDG         Width
DECAY   1000021  3.154880e-02
      2.500000e-01 3 -1 1 1000022 # gluino -> up anti-up neutralino1
      2.500000e-01 3 -2 2 1000022 # gluino -> down anti-down neutralino1
      2.500000e-01 3 -3 3 1000022 # gluino -> strange anti-strange neutralino1
      2.500000e-01 3 -4 4 1000022 # gluino -> charm anti-charm neutralino1
DECAY   1000001  2.105978e-01        # downL decays
      1    2      1000022         1    # downL -> down neutralino1
DECAY   1000002  2.105978e-01        # upL   decays
      1    2      1000022         2    # upL -> up neutralino1
DECAY   1000003  2.105978e-01        # strangeL   decays
      1    2      1000022         3    # strangeL -> strange neutralino1
      \end{Verbatim}
\begin{Verbatim}[commandchars=\\\@\@] 
DECAY   1000004  2.105978e-01        # charmL   decays
      1    2      1000022         4    # charmL -> charm neutralino1                       
DECAY   2000001  8.423913e-01        # downR   decays
      1    2      1000022         1    # downR -> down neutralino1                        
DECAY   2000002  8.423913e-01        # upR   decays
      1    2      1000022         2    # upR -> up neutralino1                       
DECAY   2000003  8.423913e-01        # strangeR   decays
      1    2      1000022         3    # strangeR -> strange neutralino1                       
DECAY   2000004  8.423913e-01        # charmR   decays
      1    2      1000022         4    # charmR -> charm neutralino1                       
\end{Verbatim}
\end{bigtextfile}
The most relevant production modes for such a model are the $2$-body final states 
$p p \to \tilde g \tilde g, \tilde g \tilde q, \tilde q \tilde q$ and $\tilde q \tilde q^*$. In our 
example run, we use different approaches\footnote{The different input modes are only combined for 
illustrative purpose here. In practice, one would normally use the same tool setup for all the different 
hadronic SUSY final states.} to generate the events for these processes:
\begin{itemize}
\item For squark-antisquark production, $p p \to \tilde q \tilde q^*$, we call \Madgraph{} internally to do the parton level event generation and subsequently let \Pythiaeight{} do the parton showering. This mode 
is one example of how to perform the event generation entirely on the fly. In our example we explicitly give the commands for \Madgraph{} to simulate the correct final state.
\item Similarly, for the final state $\tilde q \tilde q$ of squark pair production, we also generate events within \Checkmate{}, however this time we entirely rely on \Pythiaeight{} to do both the partonic event generation and parton showering. Here, we setup \Pythiaeight{} via its \texttt{.in} settings file. 

\item For events of type $\tilde g \tilde g$, we provide partonic \texttt{.lhe} files generated with \Madgraph{} beforehand and perform the parton showering and hadronisation with \Pythiaeight{} directly within \Checkmate{}.
\item Associated gluino-squark production has been performed completely externally and we provide two fully showered \texttt{.hepmc} files. The two files contain the same physics process generated with different random seeds such that they contain statistically independent samples. Such a setup with multiple files per process can for example be required when event generation was parallelised on a computing cluster. 
\end{itemize}
\subsection{Initialising and Starting \Checkmate{}}
\newcommand{\cmdir}{\texttt{\$CMDIR}}
We assume that \Checkmate{} has already been properly installed in folder \texttt{\cmdir} including \Pythia{} and \Madgraph{} functionalities; see also Appendix~\ref{sec:Install}. The pre-generated \texttt{.lhe} file for the $\tilde g \tilde g$ and the \texttt{.hepmc} file for the $\tilde g\tilde q$ process are located in \texttt{/scratch/files}. To run \Checkmate{}, some mandatory information has to be provided, either via a command line input or via a text-based parameter card. As the former only works for runs with single process, we have to choose the second approach. A (not minimal) working example for the above setup reads as follows:

\begin{bigtextfile}{checkmate\_example.in}
\begin{Verbatim}
[Parameters]
Name: ExampleRun
SLHAFile: /scratch/files/point.slha
Analyses: 8TeV
RandomSeed: 10

[squ_asq]
MGCommand: import model mssm;
           define sq  = ul ur sl sr dl dr cl cr;
           define sq~ = ul~ ur~ sl~ sr~ dl~ dr~ cl~ cr~;
           generate p p > sq sq~;
KFactor: 1.96
      \end{Verbatim}
\begin{Verbatim}[commandchars=\\\@\@] 
[squ_squ]
Pythia8Card: /scratch/files/pythiasqusqu.in
MaxEvents: 1000

[glu_glu]
Events: /scratch/events/glu_glu.lhe
XSectErr: 20 %

[glu_sq]
Events: /scratch/events/glu_squ_1.hepmc, /scratch/events/glu_squ_2.hepmc
XSect: 1.90 fb 
\end{Verbatim}
\end{bigtextfile}
The general structure of such a file consists of blocks separated by \verb@[]@ expressions which contain one or more \verb@Key: Value@ pairs. 

The first such block, \verb@[Parameters]@, is a special block type which lists  general settings 
for the \Checkmate{} run common to all processes. In our example, we first give our run a specific 
name \texttt{ExampleRun} which specifies the name of the output directory.  With a nonzero \texttt{RandomSeed} we make the results deterministic as this parameter fixes the sequence of random numbers which is used to e.g.\ simulate kinematic configurations in the event generation and apply finite efficiencies on final state objects in the detector simulation phase. 

We then provide the already
explained \texttt{.slha} spectrum file which informs  \Madgraph{} and  \Pythiaeight{} about the masses and decay tables of all SUSY particles. This file is common to all processes since obviously the 
same physics scenario should be considered within one \Checkmate{} run. In our case, 
the \texttt{SLHAFile} is a mandatory parameter as we ask \Checkmate{} to simulate the events internally 
and therefore not providing this parameter would result in an immediate abort.\footnote{If we instead only 
provided showered \texttt{.hepmc} files, this parameter would not be mandatory as no model-dependent information 
would be required. Note that providing partonic \texttt{.lhe} \emph{may} require an \Slha{} file including the full decay table of all BSM particles, namely if the BSM final state 
particles are not yet fully decayed and if the decay table is not included in the header part of the \texttt{.lhe} event file already.} Additional 
possible settings which can be changed via the \verb@[Parameters]@ block are summarised in \Cref{sec:app:fulllist}.

Besides the special \verb@[Parameters]@ block, any other \verb@[X]@ block combines the information for a 
particular production process \verb@X@, where \verb@X@ is a freely chosen identifier. In our particular case, we need 
four such blocks for all the different production modes we wish to take into account. Within each such process 
block we have to provide the information that describes the form of the Monte Carlo events for the particular process given.
\begin{itemize}
\item We start with the block \texttt{[squ\_asq]} responsible for $\tilde q \tilde q^*$ production. With 
the \texttt{MGCommand} keyword, we specify the set of commands to internally call \Madgraph{}. For our 
example, we load the Minimal Supersymmetric Standard Model via the \texttt{import model mssm} command, 
combine all squarks and all antisquarks into single identifiers \texttt{sq, sq\~}, respectively and 
conveniently setup pair production of all squark-antisquark pairs 
via \texttt{generate p p > sq sq\~}.\footnote{Note that this call will form all possible final 
state combinations of the product \texttt{\texttt{(ul ur  dl ...)} $\times$ \texttt{(ul\~\ ur\~\ dl\~\ ...)}}, 
including flavour-nondiagonal pairs, because of different combinations of initial state quarks in 
protons. 
} Since we do not specify otherwise, \Madgraph{} will be set up so that it simulates 5,000 partonic events for the given process.
Because no explicit cross section is provided, \Checkmate{} uses the result from \Madgraph{} determined during 
the generation of the events. We use the optional \texttt{KFactor} parameter to specify a $K$-factor 
which we determined using \Nllfast. It multiplies the \Madgraph{} leading order cross section with a 
fixed quantitiy to estimate the cross section at the next-to-leading order plus next-to-leading log accuracy in QCD. 

\item For the simulation of $\tilde q \tilde q$ pair production, we set up a second block 
called \texttt{[squ\_squ]}. Here, we also want the event generation to be done entirely 
internally via \Pythiaeight{}, however this time we explicitly provide the \texttt{.in} 
setting file for \Pythia{}.
\begin{bigtextfile}{pythiasqusqu.in}
\begin{Verbatim}
PDF:pSet = 8 !(CTEQ6L1)

Beams:idA = 2212   ! first beam, p = 2212, pbar = -2212
Beams:idB = 2212   ! second beam, p = 2212, pbar = -2212
Beams:eCM = 8000.

SLHA:file = /scratch/files/point.slha
SUSY:qq2squarksquark = on
SUSY:idVecA =  1000001,1000002,1000003,1000004,2000001,2000002,2000003,2000004
\end{Verbatim}
\end{bigtextfile}
The meaning of the individual lines should be self explanatory. The last row specifies the set of squarks 
which should be taken into account and which we set to all left- and right-chiral squarks of the first 
two generations. In principle, any of \Pythiaeight's parameters listed in Ref.~\cite{pythiawebpage} can be 
changed via this file. 
Note that for this block, we explicitly specify the number of generated events to be 1,000 via the optional parameter \texttt{MaxEvents}. 

\item The third process block \texttt{[glu\_glu]} sets up the gluino pair production process where we
have already generated 1,000 events using \Madgraph{}. Here, we simply have to provide a reference to this 
file via the \texttt{Events} keyword. There is no other mandatory parameter in this case. Most 
importantly, \Checkmate{} uses standard \Pythia{} settings for showering and hadronisation, see below, and 
takes the cross section from the \texttt{.lhe} file itself. In this example, we provide the optional 
parameter \texttt{XSectErr} to inform \Checkmate{} about the systematic error it should consider for 
this process, which we assume to be \unit[20]{\%} of the signal cross section. If no such parameter 
is provided, as we do for the other three processes, the systematic error associated
with the signal is set to zero. 

\item Finally, we provide two fully hadronic \texttt{.hepmc} files for associated gluino squark production 
in the \texttt{[glu\_squ]} block. We can simply list all available files for a given process in 
one \texttt{Events} command and in the end the results of all files are properly averaged as 
explained below. If \CheckMATE{} is run with showered \texttt{.hep} or \texttt{.hepmc} files the cross-section \emph{must} be provided
explicitly by the user as contrarily to the \texttt{.lhe} format, these event formats do not store this information. In 
our case, the provided cross section is taken from \Pythiaeight{} which we used to simulate the events 
but we could have instead used \Nllfast{} or \Prospino{}.
\end{itemize}

With the above files ready, we can start \Checkmate{} with the following command:\footnote{Within this chapter, gray text denotes input to be entered by a user.}

\begin{textfile}{Terminal}
\begin{Verbatim}[commandchars=\\\@\@]
   $CMDIR/bin: \userinputcolor ./CheckMATE checkmate\_example.in
\end{Verbatim}
\end{textfile}
\Checkmate{} then responds with a summary of the used settings for the given run and asks a user for confirmation.
\begin{bigtextfile}{Terminal}
\begin{Verbatim}
  ____ _               _    __  __    _  _____ _____ ____  
 / ___| |__   ___  ___| | _|  \/  |  / \|_   _| ____|___ \ 
| |   | '_ \ / _ \/ __| |/ / |\/| | / _ \ | | |  _|   __) |
| |___| | | |  __/ (__|   <| |  | |/ ___ \| | | |___ / __/ 
 \____|_| |_|\___|\___|_|\_\_|  |_/_/   \_\_| |_____|_____|
\end{Verbatim}
\begin{Verbatim}[commandchars=\\\@\@]
The following settings are used:
Analyses: 
	cms_1301_4698_WW (WW production only, 8 TeV, 3.5 fb-1)
	cms_1303_2985 (CMS, alpha_T + b-jets)
        [...]
	atlas_conf_2014_056 (Constraint on stop production from ttbar spin correlations)
	atlas_conf_2015_004 (Search for an invisibly decaying Higgs boson produced via vector ...
E_CM: 8.0
Processes: 
	Process Name: squ_asq
	Input KFactor: 1.96
	Associated event files and/or Monte-Carlo generation runs:
		MG5_aMC\@NLO Events
			 - internal identifier:  'squ_asq'
			 - command: import model mssm;
			            define sq  = ul ur sl sr dl dr cl cr;
			            define sq~ = ul~ ur~ sl~ sr~ dl~ dr~ cl~ cr~;
			            generate p p > sq sq~;

\end{Verbatim}
\begin{Verbatim}[commandchars=\\\@\@]

	Process Name: squ_squ
	Associated event files and/or Monte-Carlo generation runs:
		 Pythia8 Events
			 - internal identifier:  'squ_squ'
			 - .in settings file: /scratch/files/pythiasqusqu.in
			 - at most 1000 events are generated and analysed



	Process Name: glu_glu
	Input cross section error: 20.0 %
	Associated event files and/or Monte-Carlo generation runs:
		LHE Events
			 - internal identifier: 'glu_glu'
			 - path to .lhe file: /scratch/events/glu_glu.lhe



	Process Name: glu_sq
	Input Cross section: 1.9 fb
	Associated event files and/or Monte-Carlo generation runs:
		 HepMC events
			 - internal identifier:  'glu_sq_event1'
			 - path to eventfile: /scratch/events/glu_squ_1.hepmc

		 HepMC events
			 - internal identifier:  'glu_sq_event2'
			 - path to eventfile: /scratch/events/glu_squ_2.hepmc


Output Directory: 
	$CMDIR/results/ExampleRun
Additional Settings: 
	 - SLHA file /scratch/files/point.slha will be used for event generation
	 - Fixed random seed of 10
Is this correct? (y/n) 
\end{Verbatim}
\end{bigtextfile}
Here we chose that \Checkmate{} generates events at $\sqrt{s} = \unit[8]{TeV}$ centre-of-mass energy and 
tests against all implemented \Atlas{} and \Cms{} analyses for that particular energy. Note how a unique \texttt{internal identifier} is given to each process which will help us to associate output files to the corresponding input events later. In the following, when we use an expression \texttt{\textit{event}}, it is to be understood as a placeholder for one such internal identifier.

As soon as we start \Checkmate{} by answering \texttt{\userinputcolor y}, it informs us that --- since we did not specify otherwise --- \Madgraph{} is set up to generate 5,000 events for the \texttt{squ\_asq}  process.
\begin{bigtextfile}{Terminal}
\begin{Verbatim}[commandchars=\\\@\@]
 	 squ_asq:prepare(): Setting number of to-be-generated MC events to 5000. 
	                             Use the 'maxEvents' Parameter to change this default behaviour.
\end{Verbatim}
\end{bigtextfile}
After about five to ten minutes, depending on user's CPU, during which \Checkmate{} continuously updates us with the current 
status of the analysis chain, it returns the following result:
\begin{bigtextfile}{Terminal}
\begin{Verbatim}[commandchars=\\\@\@]
Evaluating Results
Test: Calculation of r = signal/(95%CL limit on signal)
Result: \badcolor Excluded
Result for r: 2.24273341815
Analysis: atlas_1405_7875
SR: SR02_3j
\end{Verbatim}
\end{bigtextfile}
We find that after simulating all events, passing them through a detector simulation and 
performing 40 different analyses, \Checkmate{} concludes that the input parameter point is excluded 
because in signal region \verb@SR02_3j@ of analysis \verb@atlas_1405_7875@, see Ref.~\cite{Aad:2014wea}, the 
number of predicted signal events $S$ exceeds the \unit[95]{\%} upper limit \texttt{S95} when testing the 
conservative value,
\begin{align}
  r = \frac{(S - 1.64 \cdot \Delta S)}{S95}\,.  \label{eq:rvalue}
\end{align}
This agrees with the result of the experimental collaboration, see Figure~9 of Ref.~\cite{Aad:2014wea}. 

For most users, this information would be sufficient for checking exclusion status of a given 
model. Simply by changing and testing different values of $m_{\tilde q}$ and $m_{\tilde g}$ given in the \Slha{} 
file, one easily finds the allowed and excluded regions in parameter space.



%

\subsection{Structure of the Results Folder}
\label{sec:resultfolder}
We now take a closer look at the additional information that is stored in many files in the 
results folder. These files may be ignored by a user who simply performs a 
test for a given parameter point. However, knowing which information can be found in these files can be very helpful  if for instance a more detailed breakdown of intermediate results is required or if \Checkmate{} behaves in an unexpected way. Furthermore, analysing these files aids us in understanding how \Checkmate{} internally works. 

\begin{sloppypar}
For our example case, the results folder would be located under \texttt{\$CMDIR/}\texttt{results/}\texttt{ExampleRun}. 
It contains the 
following files and directories:
\end{sloppypar}
\begin{bigtextfile}{Terminal}
\begin{Verbatim}[commandchars=\\\@\@]
$CMDIR/results/ExampleRun: \userinputcolor ls
   analysis  delphes  evaluation  fritz  internal  mg5amcatnlo  pythia  result.txt
\end{Verbatim}
\end{bigtextfile}
The file \verb@result.txt@ stores exactly the same information as printed on screen at the end of the \Checkmate{} run.  The other folders store the respective individual information of the different modules as explained in the introductory section and we discuss them in the same order.

\subsubsection{Folder \texttt{internal/}}
The \texttt{internal} folder stores all internally set \Checkmate{} parameters in a \texttt{Python} 
readable format such that \emph{a posteriori} one is capable of reproducing the exact \texttt{Python} 
instance of \Checkmate{} and is required if \Checkmate{} is run in the \texttt{add} mode 
explained in \Cref{sec:app:fulllist}. In normal use they are of no relevance to a user and therefore we do 
not further discuss them here. Note that files in this folder are only created if \Checkmate{} finished successfully and did not abort due to an internal error.
\subsubsection{Folder \texttt{fritz/}}
As \Fritz{} is the steering code which runs and calls the respective submodules, we continue our discussion with a content of the folder \texttt{fritz}. After running our  example it should contain the following files:

\begin{textfile}{Terminal}
\begin{Verbatim}[commandchars=\\\@\@]
$CMDIR/results/ExampleRun/fritz: \userinputcolor ls
fritz_error.log          fritz_glu_sq_event2.log  glu_glu.ini        squ_asq.ini
fritz_glu_glu.log        fritz_squ_asq.log        glu_sq_event1.ini  squ_squ.ini
fritz_glu_sq_event1.log  fritz_squ_squ.log        glu_sq_event2.ini
\end{Verbatim}
\end{textfile}
Here, the \texttt{fritz\_\textit{event}.log} files contain the runtime output of \Fritz{} which was also 
printed on-screen while \Checkmate{} was analysing \textit{event}. \texttt{fritz\_error.log} 
combines the standard error output of all runs and should hopefully be empty at all times. Note that --- 
besides the \texttt{fritz\_error.log} file --- there exists one \texttt{.log} file for each of the 
individually tested event files. Therefore, in our case, we have \emph{five} files as we 
have \emph{four processes} including {one which uses two separate \texttt{.hepmc} event files}. The 
respective \texttt{.log} files informs a user about the order in which individual modules were initialised, 
combined and finalised in the end, for example:
\vspace{0.2cm}
\begin{bigtextfile}{fritz\_squ\_squ.log}
\begin{Verbatim}[commandchars=\\\@\@]
Fritz: Initialising handlers from file $CMDIR/results/ExampleRun/fritz/squ_squ.ini
Fritz: Set random seed to 10
PythiaHandler: Output redirected to $CMDIR/results/ExampleRun/pythia/pythia_squ_squ.log
PythiaHandler: Initializing Pythia8 with /scratch/files/pythiasqusqu.in
PythiaHandler 'squ_squ': Pythia8 initialized successfully!
PythiaHandler 'squ_squ': Pythia8 will generate 1000 events
DelphesHandler 'atlas8tev': Initialising Delphes via linking to PythiaHandler 'squ_squ'
DelphesHandler 'atlas8tev': Initialising settings from $CMDIR/results/ExampleRun/delphes/modified_...
DelphesHandler 'atlas8tev': Delphes successfully initialised!
DelphesHandler 'cms8tev': Initialising Delphes via linking to PythiaHandler 'squ_squ'
[...]
AnalysisHandler 'atlas8tev': Initialising AnalysisHandler
AnalysisHandler 'atlas8tev': Loading Analysis atlas_1308_1841
AnalysisHandler 'atlas8tev': Successfully loaded analysis atlas_1308_1841
AnalysisHandler 'atlas8tev': Successfully loaded analysis atlas_conf_2015_004
AnalysisHandler 'atlas8tev': Linking to DelphesHandler 'atlas8tev' tree
AnalysisHandler 'atlas8tev': AnalysisHandler successfully linked to DelphesHandler 'atlas8tev'
AnalysisHandler 'cms8tev': Initialising AnalysisHandler
[...]
Fritz: Fritz successfully loaded command line parameters!
Fritz:  >> Successfully initialized and linked all handlers! <<
Fritz: Starting event loop!
Fritz: Progress: 10 %
[...]
Fritz: Progress: 100 %
Fritz:  >> Finalising after 1000 events. <<
AnalysisHandler 'atlas8tev': Asking DelphesHandler 'atlas8tev' for cross section information
DelphesHandler 'atlas8tev': Asking PythiaHandler 'squ_squ' for cross section information
PythiaHandler 'squ_squ': Pythia8 returned cross section of 2.43366 fb
PythiaHandler 'squ_squ': Pythia8 returned cross section error of 0 fb
AnalysisHandler 'atlas8tev': Analyses updated with sigma = 2.43366 fb and dSigma = 0 fb
AnalysisHandler 'atlas8tev': Analyses successfully finished!
AnalysisHandler 'cms8tev': Asking DelphesHandler 'cms8tev' for cross section information
DelphesHandler 'cms8tev': Asking PythiaHandler 'squ_squ' for cross section information
PythiaHandler 'squ_squ': Pythia8 returned cross section of 2.43366 fb
PythiaHandler 'squ_squ': Pythia8 returned cross section error of 0 fb
AnalysisHandler 'cms8tev': Analyses updated with sigma = 2.43366 fb and dSigma = 0 fb
AnalysisHandler 'cms8tev': Analyses successfully finished!
DelphesHandler 'atlas8tev': Delphes finished successfully!
DelphesHandler 'cms8tev': Delphes finished successfully!
PythiaHandler 'squ_squ': Pythia8 finished successfully!
Fritz:  >> Done <<
\end{Verbatim}
\end{bigtextfile}
This file enables us to trace exactly which modules have been loaded using which settings and how they were respectively linked. In our particular example, for the squark pair production process we need a \texttt{PythiaHandler} which takes care of the event generation within \Pythiaeight{}.\footnote{Note that for practical purposes, each process starts a separate \Fritz{} run. This is why only one \Pythiaeight{} instance appears in the above example logfile for the gluino run.} Then, since we test against all \unit[8]{TeV} analyses, we require two separate \Delphes{} instances, one for the \Atlas{} and another one for the \Cms{} detector description. Each of these \texttt{DelphesHandler}s takes its event information from the same single \texttt{PythiaHandler}, i.e.\ the same generated events are used for both \Atlas{} and \Cms{}. The detector level objects are then respectively passed to two individual \texttt{AnalysisHandler}s which perform flavour tagging and final state isolation checks according to the requirements of the respective analyses. These then pass the information to all the loaded analyses which independently perform the signal region categorisation. In the end, the cross section information --- which is needed by the individual analyses to properly normalise their final results --- in this particular case is taken from \Pythiaeight{} itself. Note that the cross section error from the event generator is set to zero, even though this value --- as can be seen in the respective output files showed below --- has a numerical uncertainty. This \emph{statistical} uncertainty is however already accounted for by \Checkmate{} internally. Any additional \emph{systematic} uncertainty has to be provided by the user via the \texttt{XSectErr} keyword.

The required set of handlers, their properties and how they are linked is determined by the \texttt{Python} part of \Checkmate{} and for each \textit{event} is passed via an \texttt{\textit{event}.ini} file, for example:
\begin{bigtextfile}{glu\_glu.ini}
\begin{Verbatim}[commandchars=\\\@\@]
[Global]
randomseed = 10

[...]

[PythiaHandler: glu_glu]
pythiapath = $CMDIR/results/ExampleRun/pythia
logfile = $CMDIR/results/ExampleRun/pythia/pythia_glu_glu.log
usemg5 = false
settings = $CMDIR/results/ExampleRun/pythia/glu_glucard_0.in
xsecterrfactor = 0.2

[DelphesHandler: cms8tev]
settings = $CMDIR/results/ExampleRun/delphes/modified_cms8tev_card.tcl
logfile = $CMDIR/results/ExampleRun/delphes/delphes_glu_glu.log
pythiahandler = glu_glu

[DelphesHandler: atlas8tev]
settings = $CMDIR/results/ExampleRun/delphes/modified_atlas8tev_card.tcl
logfile = $CMDIR/results/ExampleRun/delphes/delphes_glu_glu.log
pythiahandler = glu_glu

[AnalysisHandler: atlas8tev]
analysistype = atlas8tev
outputprefix = gluinos
outputdirectory = $CMDIR/results/ExampleRun/analysis
logfile = $CMDIR/results/ExampleRun/analysis/output
delpheshandler = atlas8tev

[AnalysisHandler: cms8tev]
analysistype = cms8tev
outputprefix = gluinos
outputdirectory = $CMDIR/results/ExampleRun/analysis
logfile = $CMDIR/results/ExampleRun/analysis/output
delpheshandler = cms8tev
\end{Verbatim}
\end{bigtextfile}

The two \texttt{AnalysisHandler}s responsible for \Atlas{} and \Cms{} analyses need to apply flavour tagging and isolation conditions after the detector simulation step. As explained before, \Checkmate{} first analyses the respective analysis implementation in order to create a list of all required settings for each analysis. All this is stored in the same \texttt{.ini} file:

\begin{bigtextfile}{glu\_glu.ini}
\begin{Verbatim}[commandchars=\\\@\@]
[ANALYSIS: atlas_1308_1841]
analysishandler = atlas8tev
jet_btags = atlas8tev7 atlas8tev8 atlas8tev9
electron_isolation = atlas8tev0 atlas8tev3 atlas8tev4
muon_isolation = atlas8tev0 atlas8tev11 atlas8tev12
photon_isolation = atlas8tev0

[...]

[ANALYSIS: atlas_conf_2015_004]
analysishandler = atlas8tev
jet_btags = atlas8tev3
electron_isolation = atlas8tev0 atlas8tev6
muon_isolation = atlas8tev0 atlas8tev6
photon_isolation = atlas8tev0


[BTAG: atlas8tev0]
eff = 70.
analysishandler = atlas8tev

[...]

[BTAG: atlas8tev13]
eff = 75.
analysishandler = atlas8tev

[TAUTAG: atlas8tev0]
analysishandler = atlas8tev

[ELECTRONISO: atlas8tev0]
source = c
analysishandler = atlas8tev
dr = 0.2
ptmin = 0.1
absorrel = r
maxval = 0.2

[ELECTRONISO: atlas8tev1]
source = t
analysishandler = atlas8tev
dr = 0.2
ptmin = 1.
absorrel = r
maxval = 0.1

[..]

[ELECTRONISO: atlas8tev30]
source = c
analysishandler = atlas8tev
dr = 0.3
ptmin = 0.1
absorrel = r
maxval = 0.14

\end{Verbatim}
\begin{Verbatim}[commandchars=\\\@\@]

[MUONISO: atlas8tev0]
source = t
analysishandler = atlas8tev
dr = 0.05
ptmin = 0.5
absorrel = r
maxval = 0.2

[MUONISO: atlas8tev1]
source = t
analysishandler = atlas8tev
dr = 0.2
ptmin = 0.5
absorrel = a
maxval = 1.8

[...]

[MUONISO: atlas8tev27]
source = t
analysishandler = atlas8tev
dr = 0.2
ptmin = 0.4
absorrel = r
maxval = 0.10

[PHOTONISO: atlas8tev0]
source = c
analysishandler = atlas8tev
dr = 0.2
ptmin = 0.1
absorrel = r
maxval = 0.2

[PHOTONISO: atlas8tev1]
source = c
analysishandler = atlas8tev
dr = 0.4
ptmin = 0.1
absorrel = a
maxval = 4.0

[ANALYSIS: cms_1303_2985]
analysishandler = cms8tev
jet_btags = cms8tev3
electron_isolation = cms8tev0 cms8tev8
muon_isolation = cms8tev0 cms8tev9
photon_isolation = cms8tev0 cms8tev1

[...]

[PHOTONISO: cms8tev1]
source = c
analysishandler = cms8tev
dr = 0.3
ptmin = 0.1
absorrel = r
maxval = 0.2
[...]
\end{Verbatim}
\end{bigtextfile}
We find, for example, that  the 31 \Atlas{} analyses (including those that are only partially validated, cf.\ \Cref{sec:cm:analyses})
of our general \unit[8]{TeV} run require in total 30 distinct isolation tests for electrons, 
27 different muon isolation tests and 13 different $b$-tagging working points.

The flexibility of \Fritz{} becomes apparent when inspecting the corresponding \texttt{.log} file for e.g.\ the $\tilde{g} \tilde{q}$ process for which we provided the fully hadronised \texttt{.hepmc} files:

\begin{bigtextfile}{fritz\_glu\_sq\_event1.log}
\begin{Verbatim}
Fritz: Initialising handlers from file $CMDIR/results/ExampleRun/fritz/glu_sq_event1.ini
Fritz: Set random seed to 10
DelphesHandler 'atlas8tev': Initialising Delphes via input event /scratch/events/glu_squ_1.hepmc
DelphesHandler 'atlas8tev': Input File determined to be HepMC.
DelphesHandler 'atlas8tev': Initialising settings from $CMDIR/results/ExampleRun/delphes/modified_...
DelphesHandler 'atlas8tev': Delphes successfully initialised!
DelphesHandler 'atlas8tev': Input file successfully opened!
[...]
\end{Verbatim}
\begin{Verbatim}[commandchars=\\\@\@]
Fritz:  >> Finalising after 1000 events. <<
AnalysisHandler 'atlas8tev': Asking DelphesHandler 'atlas8tev' for cross section information
DelphesHandler 'atlas8tev': Asking EventFile 'glu_sq_event1' for cross section information
EventFile 'glu_sq_event1':  Returning cross section of 1.9 fb
EventFile 'glu_sq_event1':  Returning cross section error of 0 fb
AnalysisHandler 'atlas8tev': Analyses successfully Finished!
[...]
Fritz:  >> Done <<
\end{Verbatim}
\end{bigtextfile}
One finds that no \texttt{PythiaHandler} is loaded in this case. The \texttt{.hepmc} files are directly loaded into \Delphes{} and the cross section is taken from a user which is passed to \Fritz{} and then to the analyses via the corresponding \texttt{.ini} file:
\begin{bigtextfile}{glu\_sq\_event1.ini}
\begin{Verbatim}
[...]

[EventFile: glu_sq_event1]
file = /scratch/events/glu_squ_1.hepmc
xsect = 1.9
xsecterr = 0

[...]
\end{Verbatim}
\end{bigtextfile}
The analysis part, however, is the same for all processes. 
\subsubsection{Folder \texttt{mg5amcatnlo/}}
Results of the event generation by \Madgraph{} are stored in this folder. We get the following files and directories:

\begin{textfile}{Terminal}
\begin{Verbatim}[commandchars=\\\@\@]
$CMDIR/results/ExampleRun/mg5amcatnlo/: \userinputcolor ls
mg5amcatnlo_squ_asq.log  squ_asq  squ_asq_proc_card.dat  squ_asq_run_card.dat
\end{Verbatim}
\end{textfile}
The names of all files and directories can be associated to the process that was generated via the respective unique identifier. In our case, we only encounter the identifier \texttt{squ\_asq} as we enabled interal event generation with \Madgraph{} only for this process.

A typical \Madgraph{} run requires three files: 
\begin{itemize} 
\item A \texttt{proc} card lists the commands which should be given to \Madgraph{} to simulate the correct events. In our case, the file \texttt{squ\_asq\_proc\_card.dat} contains the exact command we provided via the \texttt{MGCommand} parameter in our initial \Checkmate{} input file. 
\item A \texttt{param} card typically defines the parameters of the to-be-analysed BSM model in the form of an \texttt{SLHA} file. In our case, this file is already existing and was defined by the \texttt{SLHA} command. It, therefore, does not appear as a separate file in this folder.
\item A \texttt{run} card specifies the details of the partonic Monte Carlo simulation. This file is created by \Checkmate{}, taking a standardised \texttt{run} card and filling it with information given by a user, e.g.\ the number of requested events and the centre-of-mass energy.

\begin{textfile}{squ\_asq\_run\_card.dat}
\begin{Verbatim}[commandchars=\\\@\@]
#*********************************************************************
#                       MadGraph5_aMC\@NLO                            *
#                                                                    *
#                     run_card.dat MadEvent                          *
#                                                                    *
#  This file is used to set the parameters of the run.               *
#                                                                    *
#  Some notation/conventions:                                        *
#                                                                    *
#   Lines starting with a '# ' are info or comments                  *
#                                                                    *
#   mind the format:   value    = variable     ! comment             *
#*********************************************************************
#
#*******************                                                 
# Running parameters
#*******************                                                 
#                                                                    
[...]
  5000 = nevents ! Number of unweighted events requested 
  10   = iseed   ! rnd seed (0=assigned automatically=default))
[...]
     1        = lpp1    ! beam 1 type 
     1        = lpp2    ! beam 2 type
     4000.0     = ebeam1  ! beam 1 total energy in GeV
     4000.0     = ebeam2  ! beam 2 total energy in GeV
[...]
\end{Verbatim}
\end{textfile}
\end{itemize}
The \texttt{squ\_asq} folder is the working directory of \Madgraph{} which contains all input files, executables and output files. For more information regarding the meaning of these we refer to the original publication, Ref.~\cite{Alwall:2014hca}.

\subsubsection{Folder \texttt{pythia/}}
The \verb@pythia@ folder contains all files that have been used to simulate the events with \Pythiaeight{}. In our particular case, these are

\begin{textfile}{Terminal}
\begin{Verbatim}[commandchars=\\\@\@]
$CMDIR/results/ExampleRun/pythia: \userinputcolor ls
glu_glucard_0.in    pythia_squ_asq.log  rndm-end.dat   squ_asq_showercard.in
pythia_glu_glu.log  pythia_squ_squ.log  rndm-init.dat
\end{Verbatim}
\end{textfile}
The \verb@.in@ files correspond to setup files for \Pythiaeight{} which have been created by \Checkmate{} and which contain the commands to simulate the respective final state with the correct centre-of-mass energy using the SUSY parameter point from the \verb@.slha@ file. We have such a card for the \texttt{[squ\_asq]} process, for which we used the \texttt{MGCommand} parameter. Here, \Checkmate{} generates the \texttt{.in} file used for showering the parton events from \Madgraph{} automatically. 
\begin{bigtextfile}{squ\_asq\_showercard.in}
\begin{Verbatim}[commandchars=\\\@\@]
Init:showChangedSettings = on      ! list changed settings
Init:showChangedParticleData = on ! list changed particle data
Main:timesAllowErrors = 300          ! how many aborts before run stops
PartonLevel:MPI = off              ! no multiparton interactions
SLHA:file = /scratch/files/point.slha
\end{Verbatim}
\end{bigtextfile}
For the $\tilde{g} \tilde{g}$ production, we also have a card which sets up \Pythiaeight{} for hadronisation of \texttt{.lhe} events:

\begin{textfile}{glu\_glucard.in}
\begin{Verbatim}
[...]
Next:numberShowEvent = 0           ! print event record n times

Beams:frameType = 4
Beams:LHEF = /scratch/events/glu_glu.lhe
\end{Verbatim}
\end{textfile}
The \verb@.log@ files contain the verbatim output produced by \Pythiaeight{} before, during and after the event generation. If the simulation finished successfully, this file concludes with a summary of the numerically evaluated cross sections, for example:
\begin{bigtextfile}{pythia\_squ\_squ.log}
\begin{Verbatim}[commandchars=\\\@\@]
[...]
 *-------  PYTHIA Event and Cross Section Statistics  -----------------------------------------*
 |                                                                                             |
 | Subprocess                  de |            Number of events       |      sigma +- delta    |
 |                                |       Tried   Selected   Accepted |     (estimated) (mb)   |
 |                                |                                   |                        |
 |---------------------------------------------------------------------------------------------|
 |                                |                                   |                        |
 | q q' -> ~d_L ~d_L + c.c.  1351 |         625         12         12 |   4.536e-14  4.978e-15 |
 | q q' -> ~d_L ~s_L + c.c.  1352 |          22          0          0 |   0.000e+00  0.000e+00 |
 | q q' -> ~d_L ~b_1 + c.c.  1353 |           0          0          0 |   0.000e+00  0.000e+00 |
 [...]
 | q q' -> ~t_2 ~d_L + c.c.  1423 |           0          0          0 |   0.000e+00  0.000e+00 |
 | q q' -> ~t_2 ~s_L + c.c.  1424 |           0          0          0 |   0.000e+00  0.000e+00 |
 |                                |                                   |                        |
 | sum                            |       27950       1000       1000 |    2.367e-12  3.489e-14|
 |                                                                                             |
 *-------  End PYTHIA Event and Cross Section Statistics --------------------------------------*
[...]
\end{Verbatim}
\end{bigtextfile}
Note that, as expected, the final numbers coincide with the cross section values quoted in the \texttt{fritz\_squ\_squ.log} file that have been used for the analysis normalisation. If, for some reason, the event generation has to be aborted, this file can contain more information about the cause. 

The run also produces two binary files \texttt{rndm-init.dat} and \texttt{rndm-end.dat} which contain the state of the random number generator.  The random number sequence can be reproduced by providing \texttt{Pythia8Rndm: <path>/rndm-init.dat} in the parameter file. Alternatively, providing \texttt{rndm-end.dat} ensures that the new random number sequence is independent.  If a user requires the directory to be overwritten, these files should be first copied to another location.
\subsubsection{Folder \texttt{delphes/}}
Analogously to the \verb@pythia@ folder, intermediate results of the detector simulation step are stored in the \verb@delphes@ folder. 

\begin{textfile}{Terminal}
\begin{Verbatim}[commandchars=\\\@\@]
$CMDIR/results/ExampleRun/delphes: \userinputcolor ls
delphes_glu_glu.log        delphes_squ_asq.log          modified_cms8tev_card.tcl
delphes_glu_sq_event1.log  delphes_squ_squ.log
delphes_glu_sq_event2.log  modified_atlas8tev_card.tcl
\end{Verbatim}
\end{textfile}
For our example, this folder simply contains the log files produced by \Delphes{} for each of the five independent event files. These do not contain any interesting information if a run succeeded but can be of assistance if the detector simulation encountered an unexpected problem. Typically, \Checkmate{} uses standardised \texttt{.tcl} detector cards which can be found in \texttt{data/cards} within the \Checkmate{} installation directory. Only in our specific case, since we fix the random seed which can only be done in \Delphes{} in the \texttt{.tcl} files, \Checkmate{} produces \texttt{modified\_\**.tcl} cards. The only difference to the standard cards is the appearance of the line \texttt{set RandomSeed 10} at the very end.
For more information about the fast detector simulation \Delphes{}, we refer to Ref.~\cite{deFavereau:2013fsa}.\color{black}

\subsubsection{Folder \texttt{analysis/}}
A closer look into the \verb@analysis@ folder reveals a plethora of files.
\begin{textfile}{terminal}
\begin{Verbatim}[commandchars=\\\@\@]
$CMDIR/results/ExampleRun/analysis: \userinputcolor ls
glu_glu_atlas_1308_1841_cutflow.dat            glu_sq_event2_atlas_conf_2013_024_signal.dat
glu_glu_atlas_1308_1841_signal.dat             glu_sq_event2_atlas_conf_2013_031_cutflow.dat
[...]
glu_sq_event2_atlas_conf_2012_147_signal.dat   squ_squ_cms_sus_13_016_cutflow.dat
glu_sq_event2_atlas_conf_2013_021_signal.dat   squ_squ_cms_sus_13_016_signal.dat
glu_sq_event2_atlas_conf_2013_024_cutflow.dat
\end{Verbatim}
\end{textfile}
To be precise, there are two files for each of the analysis for each process event file which in our case sums up to almost 400 different files. 
 The content of these files has not changed since the original \Checkmate{} publication in Ref.~\cite{Drees:2013wra}. For completeness, we discuss the logic and content of these files here.

Each analysis in \Checkmate{} produces two types of output: \verb@cutflow@-files show the absolute and relative 
numbers of events that pass the individual selection cuts of the 
corresponding analysis step-by-step, whereas \verb@signal@-files give the final number of 
events that pass all signal region cuts defined within the analysis. As shown  below, both files have a common structure. For a detailed discussion we choose the analysis \texttt{atlas\_1405\_7875} which \Checkmate{} determined to be responsible for the signal exclusion:\vspace{0.1cm}
\begin{bigtextfile}{glu\_glu\_atlas\_1405\_7875\_cutflow.dat}
\begin{Verbatim}[commandchars=\\\@\@]
# ATLAS
# ATLAS-1405-7875
# 0 lepton, 2-6 jets, etmiss
# sqrt(s) = 8 TeV
# int(L) = 20.3 fb^-1

Inputfile:
XSect:           0.0591437 fb
 Error:          0.0118287 fb
MCEvents:        1000
 SumOfWeights:   1000
 SumOfWeights2:  1000
 NormEvents:     1.19582
 
Cut                       Sum_W  Sum_W2  Acc    N_Norm      
a.2jl_CR01_all            1000   1000    1      1.19582     
a.2jl_CR02_missETjetsPT   940    940     0.94   1.12407     
a.2jl_CR07_dphiMin2J3J    762    762     0.762  0.911212    
a.2jl_CR11_RHT            616    616     0.616  0.736622    
[...]
o.6jt+_CR12_Rmeff         146    146     0.146  0.174589    
o.6jt+_CR13_meffIncl      139    139     0.139  0.166218    
\end{Verbatim}
\end{bigtextfile}
\begin{bigtextfile}{glu\_glu\_atlas\_1405\_7875\_signal.dat}
\begin{Verbatim}[commandchars=\\\@\@]
[...]
# ATLAS
# ATLAS-1405-7875
# 0 lepton, 2-6 jets, etmiss
# sqrt(s) = 8 TeV
# int(L) = 20.3 fb^-1

Inputfile:
XSect:           0.0591437 fb
 Error:          0.0118287 fb
MCEvents:        1000
 SumOfWeights:   1000
 SumOfWeights2:  1000
 NormEvents:     1.19582
\end{Verbatim}
\begin{Verbatim}[commandchars=\\\@\@]
SR           Sum_W  Sum_W2  Acc    N_Norm      
SR01_a.2jl   616    616     0.616  0.736622    
SR01_b.2jm   326    326     0.326  0.389836    
SR01_c.2jt   314    314     0.314  0.375486    
SR01_d.2jW   27     27      0.027  0.032287    
SR02_3j      237    237     0.237  0.283408    
SR03_a.4jl-  432    432     0.432  0.516592    
SR03_b.4jl   432    432     0.432  0.516592    
SR03_c.4jm   50     50      0.05   0.0597908   
SR03_d.4jt   232    232     0.232  0.277429    
SR03_e.4jW   7      7       0.007  0.00837071  
SR04_5j      269    269     0.269  0.321674    
SR05_a.6jl   119    119     0.119  0.142302    
SR05_b.6jm   119    119     0.119  0.142302    
SR05_c.6jt   82     82      0.082  0.0980569   
SR05_d.6jt+  139    139     0.139  0.166218    
\end{Verbatim}
\end{bigtextfile}
These files start with some general information about the analysis and the analysed events. Note that the cross section error corresponds to \unit[20]{\%} of the total cross section as specified in our \Checkmate{} input file for the $\tilde{g}\tilde{g}$ process. 

After this, a list of all individual 
cutflow milestones/signal regions follows. For each of these, \Checkmate{} lists 
the sum of weights and sum of squared weights of all events that passed the corresponding cut(s) (\verb@Sum_W, Sum_W2@), the 
relative efficiency times acceptance factor (\verb@Acc@) as well as the 
  expected number of events  after normalising 
to the given total cross section and the luminosity of the respective 
analysis (\verb@N_Norm@). In case of unweighted events, \verb@Sum_W@ and \verb@Sum_W2@ corresponds to 
the number of Monte Carlo events in the respective region and this is true for the 
example above. However, if weighted events are used, they are properly taken 
into account and both \texttt{Sum\_W} and \texttt{Sum\_W2} are required by \Checkmate's 
evaluation routines to properly calculate the statistical error in the upcoming evaluation step.

The cutflow information, similarly to all the files discussed in the previous paragraph, can 
be used e.g.\ for validation purposes. It is, however,  currently not further processed 
by \Checkmate{}. The \verb@signal@ files, on the other hand, contain crucial information used for 
the subsequent evaluation step explained below. 

Any output or warning/error messages generated during the analysis runs are stored in \texttt{analysisstdout} \texttt{\_\textit{analysisname}.log} files. If after a successful \Checkmate{} run these files are empty --- as usually expected --- they are removed automatically.
\subsubsection{Folder \texttt{evaluation/}}
\label{sec:cm:example:eval}
The \texttt{evaluation} folder of our example run contains the following files:
\begin{textfile}{Terminal}
\begin{Verbatim}[commandchars=\\\@\@]
$CMDIR/results/ExampleRun/evaluation: \userinputcolor ls
best_signal_regions.txt          glu_sq_event2_eventsResults.txt  squ_squ_eventsResults.txt
glu_glu_eventsResults.txt        glu_sq_processResults.txt        squ_squ_processResults.txt
glu_glu_processResults.txt       squ_asq_eventsResults.txt        total_results.txt
glu_sq_event1_eventsResults.txt  squ_asq_processResults.txt
\end{Verbatim}
\end{textfile}
Files with name \texttt{X\_eventResults.txt} collect the results returned by all signal regions 
in all analyses for a given process \texttt{X}. By default, the data stored are the normalised number of 
predicted signal events (\texttt{signal\_normevents}) and the total error on this number (\texttt{signal\_err\_tot}).
\vspace{0.1cm}
\begin{bigtextfile}{glu\_glu\_eventResults.txt}
\begin{Verbatim}
analysis             sr                        signal_normevents  signal_err_tot     
atlas_1308_1841      SR01_8j50_a.0b            0.0263608          0.00770593751115   
atlas_1308_1841      SR01_8j50_b.1b            0.00838752         0.00358665176569   
[...]
\end{Verbatim}
\begin{Verbatim}[commandchars=\\\@\@]
atlas_1404_2500      SR3Llow                   0.0                0                  
atlas_1404_2500      SR3b                      0.0                0                  
atlas_1405_7875      SR01_a.2jl                0.736622           0.150283717126     
atlas_1405_7875      SR01_b.2jm                0.389836           0.080901268657     
[...]
cms_sus_13_016       SR1                       0.0                0  
\end{Verbatim}
\end{bigtextfile}
The error is defined as the quadratic sum of the statistical error, calculated internally 
from the size of the Monte Carlo sample, and the systematic error provided by the user. These 
individual error sources and additional columns can be requested 
by setting the correct options in the \texttt{[Parameters]} block in the \Checkmate{} setup file; 
see \Cref{sec:app:fulllist}. 

During the evaluation phase, results from all individual  processes are 
combined. First, results  that correspond 
to the same process will be \textit{averaged} by taking the corresponding weights properly into account. The statistical error 
is then calculated from the combined sum of weights and combined sum of squared weights. The statistical error 
for all signal regions with 0 Monte Carlo events at this stage is set to the corresponding statistical error of 
1 Monte Carlo event.\footnote{This prescription ensures that Monte Carlo samples for processes with very large cross sections but with an insufficiently small number of 
events contribute with large statistical uncertainty to the final number, even if no signal event passed the cuts.} 
The combined results of this procedure are stored in \verb@X_processResults.txt@. 
\begin{bigtextfile}{glu\_sq\_event1\_eventResults.txt}
\begin{Verbatim}
[...]
atlas_1405_7875      SR02_3j                   10.9686            0.459030034105 
[...]
\end{Verbatim}
\end{bigtextfile}
\begin{bigtextfile}{glu\_sq\_event2\_eventResults.txt}
\begin{Verbatim}
[...]
atlas_1405_7875      SR02_3j                   11.2566            0.657620946569 
[...]
\end{Verbatim}
\end{bigtextfile}
\begin{bigtextfile}{glu\_sq\_processResults.txt}
\begin{Verbatim}
[...]
atlas_1405_7875      SR02_3j                   11.0645722039      0.376429600571 
[...]
\end{Verbatim}
\end{bigtextfile}   
For processes with only one event file, the corresponding \texttt{eventResults} and \texttt{processResults} files are almost identical, except for the statistical error of signal regions with 0 events which is only set process-wise, not event-wise. 

For the next step, results from different processes are \textit{added} to determine the total expected number of signal events for each signal region. All 
errors are considered independent and hence added in quadrature. This is done for each signal region in each selected analysis separately. These results are then compared to the experimental limits using the chosen method, in our case the conservative $r$-limit since we did not specify anything else. The results for each analysis and each signal region are then stored in \verb@total_results.txt@. Here, the standard columns are the number of experimentally observed, \texttt{o}, and expected Standard Model events, \texttt{b} $\pm$ \texttt{db}, quoted by the experiments, the \Checkmate{}  predicted number of signal events, \texttt{s}, and the corresponding error, \texttt{ds}, the model independent \unit[95]{\%} observed and expected limits, \texttt{s95obs} and \texttt{s95exp}, and the conservative $r$ value as defined in Eq.~\eqref{eq:rvalue}.
\pagebreak
\begin{bigtextfile}{total\_results.txt}
\begin{Verbatim}
analysis        sr               o    b    db   s       ds    s95obs s95exp robscons rexpcons
atlas_1308_1841 SR01_8j50_a.0b   40.0 35.0 4.0  0.155   0.063 20.0   16.0   0.002543 0.003178
atlas_1308_1841 SR01_8j50_b.1b   44.0 40.0 10.0 0.021   0.050 23.0   23.0   0        0       
atlas_1308_1841 SR01_8j50_c.GE2b 44.0 50.0 10.0 0.028   0.051 22.0   26.0   0        0       
atlas_1308_1841 SR02_9j50_a.0b   5.0  3.3  0.7  0.048   0.053 7.0    5.0    0        0       
[...]
atlas_1405_7875 SR01_d.2jW       0.0  2.3  1.4  2.482   1.286 4.8    4.0    0.420647 0.504776
atlas_1405_7875 SR02_3j          7.0  5.0  1.2  19.58   1.222 8.2    6.4    2.242733 2.873502
atlas_1405_7875 SR03_a.4jl-      2169.2120 110. 22.54   0.282 270.0  240.0  00.07914 0.089036
[...]
cms_sus_13_016  SR1              1.0  1.2  1.04 0.0     0.048 4.0    3.9    0        0    
\end{Verbatim}
\end{bigtextfile}
(Note that we rounded the numbers in the table compared to the actual file content to fit the page width.) 

In the last step of the evaluation procedure, \Checkmate{} will search for the signal region with 
the largest expected sensitivity. For the $r$-limits this corresponds to 
the signal region with the largest \verb@rexpcons@. The results of the most 
sensitive signal region of each analysis is written in the file \verb@best_signal_regions.txt@. 
\vspace{0.1cm}
\begin{bigtextfile}{best\_signal\_regions.txt}
\begin{Verbatim}
analysis           sr                     b    db     s    ds       s95obs s95exp robscons rexpcons 
atlas_1308_1841    SR04_7j80_a.0b         11.0 2.2    0.21 0.0685   10.0   10.0   0.010426 0.0104267
atlas_1308_2631    SRA3                   15.8 2.8    0.0  0.0492   9.0    10.2   0        0        
atlas_1402_7029    SR0taua20              0.29 0.18   0.0  0.0497   2.9    2.9    0        0        
atlas_1403_4853    L110                   9.3  3.5    0.0  0.0497   9.0    9.4    0        0        
atlas_1403_5222    SR2B                   2.4  0.9    0.0  0.0497   3.4    4.5    0        0        
atlas_1403_5294    WWa_DF                 73.6 7.9    0.0  0.0497   20.3   22.533 0        0        
atlas_1403_5294_CR CRmT2_top              789. 126.0  0.0  0.0497   253.0  250.0  0        0        
atlas_1404_2500    SR1b                   4.7  2.1    0.0  0.0497   13.3   8.0    0        0        
atlas_1405_7875    SR02_3j                5.0  1.2    19.5 0.7296   8.2    6.4    2.242733 2.8735021
atlas_1407_0583    bCd_high1              11.0 1.5    0.0  0.0497   13.2   8.5    0        0        
atlas_1407_0600    SR0l7jA                21.2 4.6    0.01 0.0492   13.9   13.8   0        0        
atlas_1407_0608    M3                     1770 81.0   24.4 1.0220   195.0  190.0  0.116903 0.1199802
atlas_1411_1559    SRTotal                557. 45.0   0.21 0.0880   70.0   91.0   0.000980 0.0007540
atlas_1501_07110   SRmm-1                 3.8  0.9    0.0  0.0497   7.9    6.0    0        0        
atlas_1502_01518   SR9                    97.0 14.0   24.4 0.9640   58.0   36.0   0.394622 0.6357800
atlas_1503_03290   SR-Z                   10.6 3.2    0.0  0.0497   29.6   12.0   0        0        
atlas_1506_08616   SRinB                  14.1 2.8    0.0  0.0497   16.1   11.2   0        0        
atlas_conf_2012_10 el                     9.0  2.8    0.0  0.0142   9.9    9.3    0        0        
atlas_conf_2012_14 4                      380. 73.40  4.47 0.3204   210.0  210.0  0.018805 0.0188058
atlas_conf_2013_02 emumu                  287. 19.0   0.0  0.0318   58.2   49.7   0        0        
atlas_conf_2013_02 SR1                    17.5 3.2    0.07 0.0511   10.0   10.6   0        0        
atlas_conf_2013_03 Higgs                  3450 180.0  0.0  0.0507   484.0  363.0  0        0        
atlas_conf_2013_03 SR1Z                   1.3  1.0    0.0  0.0507   6.5    4.5    0        0        
atlas_conf_2013_04 SR_mT2_110_elmu        4.4  2.0    0.0  0.0497   7.105  6.699  0        0        
atlas_conf_2013_06 SR0L7JA                22.5 6.9    0.01 0.0492   15.3   14.6   0        0        
atlas_conf_2013_06 SoftLep1BHigh          4.0  1.1    0.02 0.0508   7.9    6.3    0        0        
atlas_conf_2013_08 SR1OF                  103. 15.0   0.0  0.0497   24.0   31.0   0        0        
atlas_conf_2014_01 SRa                    53.0 10.0   0.0  0.0497   30.2   27.0   0        0        
atlas_conf_2014_03 emu                    4376 281.2  0.0  0.0497   1176.0 566.0  0        0        
atlas_conf_2014_05 sig                    6000 3600.  0.0  0.0497   6902.0 6717.0 0        0        
atlas_conf_2015_00 M1                     578. 48.41  0.04 0.0497   73.0   96.0   0        0        
cms_1301_4698_WW   combined               1000 60.0   0.0  0.00855  240.4  135.7  0        0        
cms_1303_2985      23j_0b_875             16.1 1.7    7.75 0.4298   18.545 10.11550.379938 0.6965528
cms_1405_7570      Zjj_030                2136 859.0  0.0  0.0477   1378.8 1595.450        0        
cms_1408_3583      550                    509. 66.0   9.28 0.6481   129.0  123.0  0.063758 0.0668690
cms_1502_06031     SR01_GE2jets_c.highMET 12.8 4.3    0.0  0.0475   7.6    7.6    0        0        
cms_1504_03198     SR1                    16.4 3.640  0.0  0.0482   12.9   11.4   0        0        
cms_smp_12_006     0e                     487. 40.0   0.0  0.0480   151.62 88.98  0        0        
cms_sus_12_019     For_OF                 155. 16.40  0.0  0.0475   31.8   31.8   0        0        
cms_sus_13_016     SR1                    1.2  1.048  0.0  0.0477   4.0    3.9    0        0        
\end{Verbatim}
\end{bigtextfile}
This file is helpful in getting a good overview  of which analyses yield a non-vanishing $r$ value and 
hence  show sensitivity to the tested model. In our example, one would expect that the most sensitive 
analyses are those targeting final states with 
a large jet multiplicity and missing transverse 
energy. Indeed, one can identify three such analyses with sizeable $r$-values: 
\texttt{atlas\_1405\_7875} \cite{Aad:2014wea}, the zero lepton multijet search, the \Atlas{} 
monojet\footnote{Despite the description, this analysis allows for events with up to three hard jets 
in the final state and hence is also sensitive to our expected multijet signature.} 
search \texttt{atlas\_1502\_01518} \cite{Aad:2015zva}  and the \Cms{} 
search \texttt{cms\_1303\_2985} \cite{Chatrchyan:2013lya} which uses the $\alpha_T$ variable 
to identify BSM events with a large hadronic activity.

\Checkmate{} then again chooses the most sensitive signal region among these. The corresponding 
\emph{observed} result will be used to finally conclude whether the input can be 
considered excluded or not, i.e.\ in the case of the $r$-limit if \verb@robscons@ is 
larger than 1. In the above example, the best signal region would be \texttt{SR02\_3j} in the analysis \texttt{atlas\_1405\_7875} which with the \texttt{robscons} value of about 2.2 excludes the tested model. This is exactly the result which was printed on screen at the end of our \Checkmate{} run.

With that, we have illustrated how \Checkmate{} can be used to test various BSM models with a range of input methods and which content can be found in all the produced output files. This knowledge should be sufficient for standard users to test their models of interest without much effort.

\section{Available Analyses \label{sec:cm:analyses}}

A large number of \ATLAS{} and \CMS{} studies have been implemented covering a wide range of 
final state configurations. The vast majority of studies are dedicated searches for 
supersymmetry and generally require significant missing transverse momentum. They can be divided 
into four main categories.
\begin{itemize}
\item The powerful inclusive SUSY searches in final states 
with large jet multiplicities and large missing transverse momentum target the production of 
gluino and first/second generation squark pairs which subsequently cascade or directly decay 
into the lightest stable supersymmetric particle. 

\item Third generation searches are of 
particular interests due to the emerging little hierarchy problem and are sensitive to 
direct production of stop and sbottom pairs. In addition, searches for gluino induced stop 
and sbottom production (in cascade decays) are implemented in \CheckMATE{}. 
The decays of the third generation sparticles yield top or bottom quarks in 
the final state.

\item Electroweak searches target the direct production of electroweakinos and 
sleptons and generally focus on final states with at least two leptons (electrons or muons) or taus.

\item Monojet (monophoton) searches which are sensitive to compressed spectra, simplified dark matter models, large extra dimensions, 
 Higgs portals and other scenarios predicting a high momentum jet (photon) recoiling against missing transverse momentum. 
\end{itemize}

\newcolumntype{H}{>{\setbox0=\hbox\bgroup}c<{\egroup}@{}}
\begin{table}
{\small
\begin{tabularx}{\textwidth}{lXcclll}
\multicolumn{2}{l}{\ATLAS{} 8~TeV searches} \\
\toprule \midrule
Name & Search  & $\sqrt{s}$ &  $\mathcal{L}$ & $N_{\text{SR}}$  & Ref. & Cite \\
 &  &  [TeV] &  [fb$^{-1}$] &   &  &  \\
\midrule
\texttt{atlas\_1308\_1841} &  New phenomena in final states with large jet multiplicities and \etmiss{}  &  8 &  20.3 &  19 & \cite{Aad:2013wta} & \cite{Cao:2015ara} \\
\texttt{atlas\_1308\_2631} &  Third-generation squark pair production in final states with \etmiss{}  and two $b$-jets  &  8 &  20.1 &  6 & \cite{Aad:2013ija} \\
\texttt{atlas\_1402\_7029} &  Production of charginos and neutralinos in events with three leptons and \etmiss{}   &  8 &  20.3 &  24 & \cite{Aad:2014nua}\\
\texttt{atlas\_1403\_4853} &  Top-squark pair production in final states with two leptons &  8 &  20.3 &  12 & \cite{Aad:2014qaa}\\
\texttt{atlas\_1403\_5222} &  Top squark pair production in events with a $Z$ boson, $b$-jets and \etmiss{}  &  8 &  20.3 &  5 & \cite{Aad:2014mha}\\
\texttt{atlas\_1404\_2500} & Supersymmetry in final states with jets and two same-sign leptons or three leptons  &  8 &  20.3 &  5 & \cite{Aad:2014pda} \\
\texttt{atlas\_1405\_7875} & Squarks and gluinos in final states with jets and \etmiss{}  &  8 &  20.3 &  15 & \cite{Aad:2014wea}  & \cite{Cao:2015ara} \\
\texttt{atlas\_1407\_0583} &  Top squark pair production in final states with one isolated lepton, jets, and \etmiss{}  &  8 &  20.3 &  27 & \cite{Aad:2014kra}\\
\texttt{atlas\_1407\_0608} &  Pair-produced third-generation squarks decaying via charm quarks or in compressed supersymmetric scenarios &  8 &  20.3 &  3 & \cite{Aad:2014nra}\\
\texttt{atlas\_1411\_1559} &  Monophoton search with one energetic photon and large \etmiss{} &  8 &  20.3 &  1 & \cite{Aad:2014tda}\\
\texttt{atlas\_1501\_07110} &  Search for direct pair production of a chargino and a neutralino decaying to the 125 GeV Higgs boson  &  8 &  20.3 &  12 & \cite{Aad:2015jqa} \\
\texttt{atlas\_1502\_01518} &  New phenomena in final states with an energetic jet and large \etmiss{}  &  8 &  20.3 &  9 & \cite{Aad:2015zva} \\
\texttt{atlas\_1503\_03290} &  Supersymmetry in events containing a same-flavour opposite-sign dilepton pair, jets, and large \etmiss{}  &  8 &  20.3 &  1 & \cite{Aad:2015wqa} & \cite{Cao:2015ara} \\
\texttt{atlas\_1506\_08616} &  Pair production of third-generation squarks &  8 &  20.3 &  11 & \cite{Aad:2015pfx} \\
\texttt{atlas\_conf\_2012\_104} &  Supersymmetry in final states with jets, \etmiss{}  and one isolated lepton &  8 &  5.8 &  2 & \cite{ATLAS-CONF-2012-104} \\
\texttt{atlas\_conf\_2012\_147} & New phenomena in monojet plus \etmiss{}  final states &  8 &  10 &  4 & \cite{ATLAS-CONF-2012-147} \\
\texttt{atlas\_conf\_2013\_024} & Production of the top squark in the all-hadronic $t \bar t$ and \etmiss{} final state &  8 &  20.5 &  3 & \cite{ATLAS-CONF-2013-024}  \\
\texttt{atlas\_conf\_2013\_049} &  Direct-slepton and direct-chargino production in final states with two opposite-sign leptons, \etmiss{}  and no jets &  8 &  20.3 &  9 & \cite{ATLAS-CONF-2013-049} \\
\texttt{atlas\_conf\_2013\_061} &  Strong production of supersymmetric particles in final states with \etmiss{}  and at least three $b$-jets &  8 &  20.1 &  9 & \cite{ATLAS-CONF-2013-061} \\
\texttt{atlas\_conf\_2013\_089} &  Strongly produced supersymmetric particles in decays with two leptons &  8 &  20.3 &  12 & \cite{ATLAS-CONF-2013-089} \\
\texttt{atlas\_conf\_2015\_004} &  Invisibly decaying Higgs boson produced via vector boson fusion &  8 &  20.3 &  1 & \cite{ATLAS-CONF-2015-004} \\
\midrule
\bottomrule
\end{tabularx}
\caption{List of 8~TeV \ATLAS{} analyses which are available in the public alpha version of \Checkmate{} and which have been validated against published experimental results. The ``Cite'' column refers to an original paper by external authors who implemented a search and should be cited along with \Checkmate. 
\label{tab:cm:analysis_atlas8}}
}
\end{table}

\begin{table}
\begin{tabularx}{\textwidth}{lXcclll}
\multicolumn{2}{l}{\CMS{} 8~TeV searches} \\
\toprule \midrule
Name & Search  & $\sqrt{s}$ &  $\mathcal{L}$ & $N_{\text{SR}}$  & Ref. & Cite \\
 &  &  [TeV] &  [fb$^{-1}$] &   &  &  \\
\midrule
\texttt{cms\_1303\_2985} &  Supersymmetry in hadronic final states with missing transverse energy using the variables $\alpha_T$ and $b$-quark multiplicity &  8 &  11.7 &  59 & \cite{Chatrchyan:2013lya} \\
\texttt{cms\_1408\_3583} & Dark matter, extra dimensions, and unparticles in monojet events &  8 &  19.7 &  7 & \cite{Khachatryan:2014rra} \\
\texttt{cms\_1502\_06031} &  Physics beyond the Standard Model in events with two Leptons, jets, and \etmiss{}  &  8 &  19.4 &  6 & \cite{Khachatryan:2015lwa} & \cite{Cao:2015ara} \\
\texttt{cms\_1504\_03198} &  Production of dark matter in association with top-quark pairs in the single-lepton final state &  8 &  19.7 &  1 & \cite{Khachatryan:2015nua} & \cite{Baek:2016lnv} \\
\texttt{cms\_sus\_13\_016} &  New physics in events with same-sign dileptons and jets &  8 &  19.5 &  1 & \cite{CMS:2013jea} \\
\midrule
\bottomrule
\end{tabularx}
\caption{List of 8~TeV \CMS{} analyses which are available in the public alpha version of \Checkmate{} and which have been validated against published experimental results. The ``Cite'' column refers to an original paper by external authors who implemented a search. }
\label{tab:cm:analysis_cms8}
\end{table}

\begin{table}
\begin{tabularx}{\textwidth}{lXcclll}
\multicolumn{2}{l}{\ATLAS{} 13~TeV searches} \\
\toprule \midrule
Name & Search  & $\sqrt{s}$ &  $\mathcal{L}$ & $N_{\text{SR}}$  & Ref. & Cite \\
 &  &  [TeV] &  [fb$^{-1}$] &   &  &  \\
\midrule
\texttt{atlas\_1602\_09058} &  Supersymmetry in final states with jets and 2 same sign leptons or 3 leptons  &  13 &  3.2 &  4 & \cite{Aad:2016tuk} \\
\texttt{atlas\_1604\_01306} &  Search for new phenomena in events with a photon and missing transverse momentum   &  13 &  3.2 &  1 & \cite{Aaboud:2016uro} \\
\texttt{atlas\_1604\_07773} &  Search for new phenomena in monojet events   &  13 &  3.2 &  13 & \cite{Aaboud:2016tnv} \\
\texttt{atlas\_1605\_03814} &  Squarks and gluinos in final states with 2 - 6 jets + \etmiss{} &  13 &  3.2 &  7 & \cite{Aaboud:2016zdn} \\
\texttt{atlas\_1605\_04285} &  Gluino search in final states with an isolated lepton + jets + \etmiss{}  &  13 &  3.2 &  7 & \cite{Aad:2016qqk} \\
\texttt{atlas\_1605\_09318} &  Gluino pair productions in final states with $b$ jets and \etmiss{}   &  13 &  3.2 &  3 & \cite{Aad:2016eki} \\
\texttt{atlas\_1606\_03903} &  Stop search in final states with 1 lepton, jets and \etmiss{}  &  13 &  3.2 &  3 & \cite{Aaboud:2016lwz} \\
\texttt{atlas\_conf\_2015\_082} &  Supersymmetry search in final states with leptonic Z + jets + \etmiss{}  &  13 &  3.2 &  1 & \cite{ATLAS-CONF-2015-082} \\
\texttt{atlas\_conf\_2016\_013} &  Vector like top quark production and 4 top quark production in lepton plus jets final state & 13 &  3.2 &  10 & \cite{ATLAS-CONF-2016-013} \\
\texttt{atlas\_conf\_2016\_050} &  Search for top squarks in final states with one isolated lepton, jets, and \etmiss{} & 13 &  13.2 &  5 & \cite{ATLAS-CONF-2016-050} \\
\texttt{atlas\_conf\_2016\_076} &  Search for direct top squark pair production and dark matter production in final states with two leptons & 13 &  13.3 &  6 & \cite{ATLAS-CONF-2016-076} \\
\midrule
\bottomrule
\end{tabularx} 
\caption{List of 13 TeV \ATLAS{} analyses which are available in the public alpha version of \Checkmate{} and which have been validated against published experimental results. }
\label{tab:cm:analysis_atlas13}
\end{table}

\begin{table}
\begin{tabularx}{\textwidth}{lXcclll}
\multicolumn{2}{l}{\CMS{} 13~TeV searches} \\
\toprule \midrule
Name & Search  & $\sqrt{s}$ &  $\mathcal{L}$ & $N_{\text{SR}}$  & Ref. & Cite \\
 &  &  [TeV] &  [fb$^{-1}$] &   &  &  \\
\midrule
\texttt{cms\_pas\_sus\_15\_011} &  Search for new physics in final states with two opposite-sign same-flavor leptons, jets and  &  13 &  2.2 &  47 & \cite{CMS-PAS-SUS-15-011} \\
\midrule
\bottomrule
\end{tabularx} 
\caption{List of 13 TeV \CMS{} analyses which are available in the public alpha version of \Checkmate{} and which have been validated against published experimental results. }
\label{tab:cm:analysis_cms13}
\end{table}

\begin{table}
\begin{tabularx}{\textwidth}{lXcclll}
\multicolumn{2}{l}{\ATLAS{} 14~TeV high luminosity studies}  \\
\toprule \midrule
Name & Search  & $\sqrt{s}$ &  $\mathcal{L}$ & $N_{\text{SR}}$  & Ref. & Cite \\
 &  &  [TeV] &  [fb$^{-1}$] &   &  &  \\
\midrule
\texttt{atlas\_phys\_2014\_010\_sq\_hl} & Gluino and squark production in final states with large jet multiplicities and \etmiss{} and with no leptons &  14 &  3000 &  10 & \cite{ATL-PHYS-PUB-2014-010} \\
\texttt{atlas\_phys\_2014\_010\_300} & Gluino and squark production in final states with large jet multiplicities and \etmiss{} and with no leptons &  14 &  300 &  10 & \cite{ATL-PHYS-PUB-2014-010} \\
\texttt{atlas\_phys\_2014\_010\_hl\_3l} & Direct production of charginos and neutralinos in final states with three leptons and \etmiss{} &  14 &  3000 &  1 & \cite{ATL-PHYS-PUB-2014-010} \\
\texttt{atlas\_phys\_2014\_010\_sbottom} & Direct production of sbottom pairs in final states with 2$b$ jets and \etmiss{} &  14 &  3000 &  6 & \cite{ATL-PHYS-PUB-2014-010} \\
\texttt{atlas\_phys\_pub\_013\_011} & Top squark pair production in final states with ($b$) jets, 0-1 lepton and \etmiss{} &  14 &  3000 &  4 & \cite{ATL-PHYS-PUB-2013-011} \\
\midrule
\bottomrule
\end{tabularx}
\caption{List of official \ATLAS{} 14 TeV high luminosity analyses which are available in the public alpha version of \Checkmate{} and which have been validated against ATLAS MC results.}
\label{tab:cm:analysis_atlas14}
\end{table}

A search for production of vector like top quarks as well as the rapidity gap signature in the vector 
boson fusion have also been implemented. Several analyses focused on SM measurements, like cross sections, are available as well.  
Searches for long-lived particles as well as the heavy Higgs boson 
searches are not included in the current \CheckMATE{} version. Even though most of the currently implemented searches focus on SUSY,
they can be applied to {\it any} non-supersymmetric BSM scenarios. In many cases some missing transverse momentum is expected, e.g.\ due to neutrinos 
from the SM gauge bosons decays which arise in a cascade decays of vector like quarks, for example. The analyses measuring SM cross sections typically also require rather small missing transverse momentum.

Most searches contain multiple signal regions, e.g.\ the stop pair production search with 
one isolated lepton, jets and missing transverse momentum~\cite{Aad:2014kra} has 27 signal regions. It is, therefore, sensitive 
to a large class of mass hierarchies between the stop, 
chargino NLSP and neutralino LSP. Counting all signal regions of the implemented searches, 
190 signal regions are employed just for the ATLAS searches at 8 TeV. We would like to caution users that different signal regions across different analyses are in many cases not statistically independent. The correlations can be particularly strong for the signal regions with similar final states and kinematic cuts. Any statistical  combination should take this into account.

All analyses listed in the above tables are fully validated against published cut-flows, distributions and/or exclusion limit plots from 
both experimental collaborations. The validation notes can be found on the official \CheckMATE{} webpage. The analyses 
are grouped into \ATLAS{} and \CMS{} searches at the center of mass energies of 8 and 13 TeV which are listed in 
Tables~\ref{tab:cm:analysis_atlas8}, \ref{tab:cm:analysis_cms8}, \ref{tab:cm:analysis_atlas13} and \ref{tab:cm:analysis_cms13}. It is clear 
from the tables that \ATLAS{} outnumbers \CMS{} in the number of implemented analyses. This is due to the 
fact that the efficiencies of the final state particles have only been optimized and fully validated for the
\ATLAS{} detector. However, some \CMS{} searches are also included, especially when there is no ATLAS equivalent or are of particular
interest for our own phenomenological studies. The number of signal regions and the total integrated luminosity 
are given for each analysis in the corresponding tables. More details of the implemented searches can be found 
in the respective references provided in the tables.

In addition to the fully validated analyses, \Checkmate{} also includes some analyses which have not been completely 
validated. This can be, for example, due to insufficient information from the collaborations. Therefore some of these 
analyses are only partially validated and the full list can be found on the \Checkmate{} web page. If available, 
partial validation notes may also be provided and these often contain details regarding the outstanding issues 
that are still to be solved. Obviously, some caution is necessary when using these analyses for physics 
studies, especially if the particular signal regions of interest have not been completely tested. Altogether 
there are currently about 60 analyses available in \Checkmate{} and the list is expanding
rapidly.   

Current searches already push the exclusion limits of gluinos and first generation squarks well beyond the 
TeV scale. In the high luminosity phase, the limits will significantly improve and are of general interest. Therefore 
official ATLAS SUSY high-luminosity studies at a centre-of-mass energy of 14~TeV for a total integrated luminosity 
of 300 and 3000~$\mathrm{fb}^{-1}$ have now been included. Here, the high luminosity studies cover squark and gluino 
pair production, stop and sbottom pair production as well as chargino and neutralino production which are summarised 
in Table~\ref{tab:cm:analysis_atlas14}.

\section{Performance Studies}
\label{sec:performance}
\Fritz{} allows for a significant gain in performance by bypassing the generation of \texttt{HepMC} or \texttt{STDHEP} 
event files, as well as not storing the detector level objects in a \Root{} file. This 
new module interfaces \Pythia{}, \Delphes{} and the \texttt{AnalysisHandler} without the necessity to write and 
read information on the hard disk as described in detail in section~\ref{sec:cm:overview}. Here, 
we compare the performance between \CheckMATE{}~1 and \CheckMATE{}~2. In both frameworks, MC 
events are generated with \Pythia~8 and the events are stored in a \texttt{HepMC} file for the \CheckMATE{}~1 setup whereas 
for the \CheckMATE{}~2 case the MC events are directly passed to the \Delphes{} module with \Fritz{}. 
The computations are performed on an Intel(R) Xeon(R) CPU E5-2650 v2 @ 2.60\,GHz with 32\,GB RAM. 
The performance has been quantified by generating 10000 gluino pairs at the LHC at the 
centre-of-mass energy of 8 TeV with underlying event, initial and final state radiation and hadronisation switched on. 
The current version {\tt Pythia~8.2.1.9} was employed for this purpose. The events were passed 
to {\tt HepMC~2.0.6.09} and the resulting event file had a size of 2.3~GB. The truth level MC event 
generation and the write operation to hard disk took 255 seconds in total. The \texttt{HepMC} file was then passed to {\tt CheckMATE~1.2.2} 
and tested against a single inclusive supersymmetry search~\cite{Aad:2014wea} and it 
took 85~seconds for \CheckMATE~1 to process the \texttt{HepMC} file. Thus the total processing time was
340~seconds. This time was compared to the computing time of {\tt CheckMATE~2.0.0}. The same process with 
the same sparticle spectrum and the identical \Pythia{} settings were employed. After generating 10000 events on the
fly, the total computational time was 245~seconds which clearly shows the improved performance.

For the next comparison, several instances were run at the same time. Here, it becomes 
clear that \CheckMATE{}~2 has a big advantage over \CheckMATE{}~1 since simultaneous read 
and write operations significantly affect the performance. Ten \Pythia{} instances linked with \CheckMATE{}~1 were simultaneously run.
20000 events were generated with the same settings as before. The event generation with \Pythia{} took about 564~seconds. The truth
level events were passed to \CheckMATE{}~1 and writing the root files for the detector level objects took 1237 seconds on average. Contrary to \CheckMATE{}~2
the root file must have been created before the analysis step could have been performed. The total processing time
was 1802 seconds. We repeated the test with \CheckMATE{}~2 and the average running time was 544 seconds. The \CheckMATE{}~2 run is much faster since the reconstructed detector level objects are stored as \Root{} objects which are immediately processed by the analysis module. It is evident that the bottleneck are
the simultaneous write operations of \Root{} files and this example clearly demonstrates the performance gain of \CheckMATE{}~2.
	
\section{Analysis Manager}
\label{sec:analyismanager}

Along with the main \CheckMATE{} program, several improvements have also been made to the 
\texttt{AnalysisManager}. Since these are only minor updates to the 
already existing program we refer the reader to original manual~\cite{Kim:2015wza} and 
here we only list the changes.

\subsection{Prototyping New Analyses}
\label{sec:proto_new_analyses}

The most significant addition to the \texttt{AnalysisManager} is the improved support for analysis 
prototyping i.e.\ developing new LHC analyses. Analyses can now be added (using the 
usual interactive procedure) without accompanying signal region
data (numbers of observed/expected events etc.) to allow a user to first simulate various SM backgrounds. The various
background contributions can then be added for each defined signal region and for the development
of new analyses we assume the expected background and observed data are equal.

After all the Standard Model contributions have been calculated, the \texttt{AnalysisManager} can then 
be rerun with a newly added option to edit the analysis information. Here a user should enter
the total Standard Model background for all signal regions defined in the analysis. At this
step the \texttt{AnalysisManager} will internally calculate $S95$ limits so that 
new physics models can be quickly tested.\footnote{By default the \texttt{AnalysisManager} also now internally
calculates all $S95$ values rather than using those quoted by the experiments. The internally
calculated values only show small differences compared to the numbers obtained the experiments. This approach ensures
that the \CheckMATE{} results are statistically consistent across all included analyses.}

\subsection{Kinematical Variables}
\label{sec:kinvar}

To enable a quick and easy implementation of analyses, \Checkmate{} contains a large 
library of kinematical variables that are often used by the LHC collaborations.
Firstly, \Checkmate{} is interfaced with the \Delphes{} and \Root{} libraries and thus a large number of 
kinematical variables such as transverse momentum, energy, pseudorapidity, boosts, etc.\
are immediately available. A list of \Delphes{} objects and their methods 
can be found at \url{https://cp3.irmp.ucl.ac.be/projects/delphes/wiki/WorkBook/RootTreeDescription}. 
In addition, the \texttt{TLorentzVector} class from \Root{} is included; see \url{https://root.cern.ch/root/html/TLorentzVector.html} for more details. 

\Checkmate{} also 
contains a number of LHC mass-reconstruction variables which are directly implemented or included from 
other libraries, e.g.\ implementation of $M_{T2}$ and derivatives \cite{Cheng:2008hk,Bai:2012gs} and 
\texttt{MctLib}~\cite{Tovey:2008ui,Polesello:2009rn}. A full list of these kinematical variables is
given in \Cref{tab:kinematic_variables} and further details can be found 
in our \texttt{doxygen} documentation \url{http://checkmate.hepforge.org/documentation/index.html}. 
See Ref.~\cite{Barr:2010zj} for a review of the kinematical variables 
proposed for mass reconstruction and BSM searches at the LHC.

\begin{table}
\begin{tabularx}{\textwidth}{lXl}
\multicolumn{2}{l}{Kinematical variables} \\
\toprule \midrule
Name & Description/Example application & Ref.\\
\midrule
\texttt{$M_T$} &  Transverse Mass; reconstruction of $W$ mass in $W\rightarrow \ell\nu$ decay &  \cite{vanNeerven:1982mz,Arnison:1983rp,Banner:1983jy,Smith:1983aa,Barger:1987du}  \\
\texttt{$M_{T2}$} &  Stransverse Mass; generalisation of \texttt{$M_T$} to events with more than one invisible particle &  \cite{Lester:1999tx,Barr:2003rg,Cheng:2008hk}  \\
\texttt{$M_{T2}^{b\ell}$} &  Asymmetric Stransverse Mass; suppression of SM top background &  \cite{Cheng:2008hk,Bai:2012gs}  \\
\texttt{$M_{T2}^{W}$} & Asymmetric Stransverse Mass including $W$ mass condition; suppression of SM top background   & \cite{Cheng:2008hk,Bai:2012gs}\\
\texttt{$M_{CT}$} & Cotransverse Mass; invariant under contra-linear transverse boosts &  \cite{Tovey:2008ui}  \\
\texttt{$M_{CT}$ corrected} & Cotransverse Mass corrected; takes initial state radiation into account  &   \cite{Polesello:2009rn}  \\
\texttt{$M_{CT_{\bot}}$ and $M_{CT_{\lVert}}$} & Decomposed Cotransverse Mass &    \cite{Matchev:2009ad}  \\
\texttt{$\alpha_T$} & Suppression of fake \etmiss{} in QCD events  &   \cite{Randall:2008rw,Khachatryan:2011tk}  \\
\texttt{Razor} & {\it Mega}-dijet kinematic variable without relying on \etmiss{}  & \cite{Rogan:2010kb,Chatrchyan:2011ek,Khachatryan:2015pwa}  \\
\texttt{`Super'-Razor} &  Improved Razor that more accurately determines production and centre-of-mass frames   & \cite{Buckley:2013kua} \\
\texttt{Topness} &  Suppression of SM top background   & \cite{Graesser:2012qy} \\
\texttt{Aplanarity} & Suppression of QCD events & \cite{Chen:2011ah} \\ 
\midrule
\bottomrule
\end{tabularx}
\caption{List of kinematical observables which are available in the public alpha version of \Checkmate{}.}
\label{tab:kinematic_variables}
\end{table}

	
	%
\section{Summary}
\label{sec:summary}

We have introduced the second version of the program \CheckMATE{} which greatly improves the
ease and speed with which models can be tested against the latest LHC data compared to its 
predecessor. The major improvement in this release is the integration of Monte Carlo event
generation that allows a user to go directly from a model defined in the UFO format
to the LHC results. In addition, this integration significantly reduces the CPU load required to
investigate models.
 
Further improvements are the inclusion of over 50 analyses which now cover the vast majority
of LHC searches that include missing energy. Moreover, high luminosity studies at 14~TeV are 
also included for the first time and these allow a user to understand the ultimate LHC
reach for their model.

For users who wish to include their own analyses or develop new LHC searches, the \texttt{AnalysisManager}
also has a number of improvements to aid this process. As an example, backgrounds can be far 
more easily included and then used to test the reach of an analysis. In addition, the library
of kinematical variables has grown significantly.

We should emphasise that the release of \CheckMATETwo{} is simply a snapshot of a continuously
evolving program. New analyses will regularly be included in updated versions available on
the website. Besides, many new developments are also planned for the \CheckMATE{} including
a fast parameter scanning technique that does not require Monte Carlo events, automatic
merging of matrix elements containing different jet multiplicities and the inclusion
of systematic correlations between signal regions to allow proper combinations
of many different analyses.

\section*{Acknowledgements}
\noindent 
We would like to thank all of the following people for their contribution to \CheckMATE. 

\begin{itemize}
 \item Especially Liangliang Shang for contibuting analyses, performing validation studies and finding many bugs in \CheckMATE{}. 
 \item Junjie Cao, Jin Min Yang, Peiwen Wu, Jinmin Yang and Yang Zhang for contributing analyses to the \CheckMATE{} database.
 \item Swasti Beswal, Anke Biek\"{o}tter, Tim Keller and Jan Sch\"{u}tte-Engel for contributing analyses to the \CheckMATE{} database.
 \item Sebastian Belkner for improving the statistical tools within \CheckMATE{}.
 \item Florian Jetter for providing an updated muon resolution tuning.
 \item Daniel Antrim, Philip Bechtle, Jamie Boyd, Sascha Caron,
Geraldine Conti, Carolina Deluca, Klaus Desch, Monica D'Onofrio, Till Eifert, Frank Filthaut, Eva Halkiadakis, Nicolai Hartman, Andreas Hoecker,
Emma Kuwertz, Tommaso Lari, Zachary Marshall,
Antoine Marzin, Federico Meloni, Alaettin Serhan Mete, Marija Vranjes Milosavljevic, Maurizio Pierini, 
Tina Potter, George Redlinger, Iacopo Vivarelli, Steven Worm, Frank Wuerthwein and Takashi Yamanaka for help with interpreting the experimental analyses.
\end{itemize}

The work has been supported
by the BMBF grant 00160200. ND acknowledges partial support of the OCEVU Labex (ANR-11-LABX-0060), the A*MIDEX project (ANR-11-IDEX-0001-02) 
funded by the French Government programme ``Investissements d'Avenir'' and the German Research Foundation (DFG) through the Forschergruppe
\textsl{New Physics at the Large Hadron Collider} (FOR 2239).  The work of JSK was supported by IBS under the project code, 
IBS-R018-D1 and was partially supported by the MINECO, Spain, under contract FPA2013-44773-P; Consolider-Ingenio CPAN CSD2007-00042 
and the Spanish MINECO Centro de excelencia Severo Ochoa Program under grant SEV-2012-0249. KR was supported by the National 
Science Centre (Poland) under Grant 2015/19/D/ST2/03136 and the Collaborative Research Center SFB676 of the DFG, ``Particles, Strings, and the Early Universe''.

 \begin{appendices}

 \section{Installation Instructions} 
 \label{sec:Install}

 
 
 
\CheckMATE{} uses a number of external programs and libraries.\footnote{This tutorial has been tested on a Linux machine 
 running under Ubuntu 16.04. The same source files can be used for other operating systems, however 
 some flags might change or some additional system libraries might be required. We refer to the 
 documentation pages of the respective tools and the \Checkmate{} website if problems of that kind occur.} Below we split these into
two categories, those that are always required and those that can be optionally installed to extend 
the functionality of the program. We also note that a step by step interactive online version is available 
at,
\begin{center}
\url{http://checkmate.hepforge.org/tutorial/ver2/start.php}
\end{center}
This tutorial is particularly useful if some of the below steps should be skipped, either because some of the programs have already been installed on the system or if some optional parts are not required. Also, the online tutorial will be continuously updated if the below installation routines especially of the required additional libraries change. 
\subsection{Required Packages}

\CheckMATE{} requires \texttt{Python 2.7.X} where \texttt{X>3}
(note that at the current time, \texttt{Python 3} is NOT supported), the data analysis 
package \texttt{ROOT} (v5.34.36 or later) \cite{Brun:1997pa} and the
detector simulation \texttt{Delphes} (v3.3.3 or later) \cite{deFavereau:2013fsa}. If any of these
packages are already installed, the respective sections of the tutorial can be skipped. The
physics specific programs 
can be downloaded from the relevant project websites,
\begin{center}
\url{https://root.cern.ch}  \\
\url{https://cp3.irmp.ucl.ac.be/projects/delphes}
\end{center}
We begin with the installation\footnote{Note that the \texttt{-j4} flag which we use here 
improves the compilation speed due to paralellisation into four independent processes. The number 
can be changed depending on the number of accessible cores on the computer.} of \texttt{ROOT} and 
we recommend that users do not install a binary
version but rather compile the package from source into a specific installation directory. Here 
and in the following, we use \texttt{[...]} to denote the verbatim output created by the 
respective commands which we enter. These strongly depend on the system setup which is used.
In the following we denote the installation directory by \texttt{\$ROOT},
\begin{bigtextfile}{Terminal}
\begin{Verbatim}[commandchars=\&\"\!]
$ROOT: &userinputcolor mkdir build
$ROOT: &userinputcolor mkdir build/etc
$ROOT: &userinputcolor ./configure --prefix=$ROOT/build --disable-fftw3 --enable-minuit2 --etcdir=$ROOT/build/etc
[...]
$ROOT: &userinputcolor make -j4
[...]
\end{Verbatim}
\begin{Verbatim}[commandchars=\&\"\!]
$ROOT: &userinputcolor make install
[...]
%\end{Verbatim}
\end{bigtextfile}

\bigskip
In order to build the \Delphes{} detector simulation framework in the installation 
directory denoted by \texttt{\$DELPHES}, we 
have to load the above compiled \Root{} libraries
\vspace{0.1cm}
\begin{bigtextfile}{Terminal}
\begin{Verbatim}[commandchars=\\\@\@]
$DELPHES: \userinputcolor source $ROOT/build/bin/thisroot.sh
$DELPHES: \userinputcolor make -j4
[...]
\end{Verbatim}
\end{bigtextfile}
A promising sign of a successful installation is the presence of the 
file \verb@libDelphes.so@ within the \Delphes{} directory. 

\subsection{Optional Packages}

\CheckMATE{} 2 now includes various options to generate events with \texttt{Pythia 8} \cite{Sjostrand:2014zea} and/or 
\Madgraph{}~\cite{Alwall:2014hca}. In addition, if the user wishes to have the possibility to store the generated
events in the \texttt{HepMC} format  \cite{Dobbs:2001ck}, the corresponding library needs to be installed first. These 
programs can be downloaded from the relevant project websites,
\begin{center}
\url{http://hepmc.web.cern.ch/hepmc}  \\
\url{http://home.thep.lu.se/~torbjorn/Pythia.html}  \\
\url{https://launchpad.net/mg5amcnlo}
\end{center}
We start with the installation of the optional \texttt{HepMC} 
library which is only required if the user wishes to save generated events in this 
format. Here the installation directory is denoted by \texttt{\$HEPMC},
\begin{bigtextfile}{Terminal}
\begin{Verbatim}[commandchars=\&\"\!]
$HEPMC: &userinputcolor ./bootstrap
[...]
$HEPMC: &userinputcolor ./configure --with-momentum=GEV --with-length=MM --prefix=$HEPMC/build
[...]
$HEPMC: &userinputcolor make -j4
[...]
$HEPMC: &userinputcolor make install
[...]
\end{Verbatim}
\end{bigtextfile}
If the installation finished successfully the \texttt{build} directory should contain 
the required libraries and header files which are needed by \Pythiaeight{},
\begin{bigtextfile}{Terminal}
\begin{Verbatim}[commandchars=\\\@\@]
$HEPMC: \userinputcolor ls build
include  lib  share
\end{Verbatim}
\end{bigtextfile}
We can continue with the compilation and installation of the \Pythiaeight{} event generator into the
directory denoted by \texttt{\$PYTHIA8}. The command \texttt{--with-hepmc2=\$HEPMC/build}
can be optionally removed if the \texttt{HepMC} library was not installed,
\begin{bigtextfile}{Terminal}
\begin{Verbatim}[commandchars=\&\"\!]
$PYTHIA8: &userinputcolor ./configure --with-hepmc2=$HEPMC/build --prefix=$PYTHIA8/build
[..]
$PYTHIA8: &userinputcolor make -j4
[...]
$PYTHIA8: &userinputcolor make install
[...]
\end{Verbatim}
\end{bigtextfile}
Again, a successful installation procedure should have filled the \verb@build@ directory 
with the necessary library files
\begin{bigtextfile}{Terminal}
\begin{Verbatim}[commandchars=\\\@\@]
$PYTHIA8: \userinputcolor ls build
bin  include  lib  share
\end{Verbatim}
\end{bigtextfile}
The final optional program \CheckMATE{} can link to is \Madgraph{} and this only
requires downloading and unzipping into the directory we denote by \texttt{\$MADGRAPH}.

\subsection{Installing CheckMATE}

With all required libraries being ready, we can finally compile the \Checkmate{} framework
into the directory denoted by \texttt{\$CMMAIN}. In the following, any of the optional \texttt{--with}
commands can be omitted if the relevant program has not been installed (note that \texttt{Delphes} 
and \texttt{Root} are required). The commands \texttt{--with-gzipinc=\$GZIPINC} and \texttt{--with-gziplib=\$GZIPLIB}
are only required if Pythia 8 was compiled with gzip support. In this case, the same source header and library directories
should be used.

\vspace{0.1cm}
\begin{bigtextfile}{Terminal}
\begin{Verbatim}[commandchars=\&\"\!]
$CMMAIN: &userinputcolor ./configure --with-rootsys=$ROOT/build --with-delphes=$DELPHES --with-hepmc=$HEPMC/build \
 &userinputcolor --with-pythia=$PYTHIA8/build --with-gzipinc=$GZIPINC --with-gziplib=$GZIPLIB \
 &userinputcolor --with-madgraph=$MADGRAPH
[...]
&userinputcolor make -j4
[...]
&userinputcolor make install
\end{Verbatim}
\end{bigtextfile}
Let us finish this tutorial with a simple test run, using an example spectrum file which is provided with 
the \Checkmate{} package. It corresponds to a CMSSM scenario 
with $\tan \beta = 10, m_0 = \unit[100]{GeV}, m_{1/2} = \unit[250]{GeV}, A_0 = - \unit[100]{GeV}$ and 
positive $\mu$. This results in a spectrum with all SUSY particles having mass in the range \unit[100--600]{GeV}.
\begin{bigtextfile}{Terminal}
\begin{Verbatim}[commandchars=\&\"\!]
$CMMAIN/bin: &userinputcolor ./CheckMATE -pyp 'p p > go go' -maxev 100 \
 &userinputcolor -slha example_run_cards/auxiliary/testspectrum.slha 
[...]
Is this correct? (y/n) &userinputcolor y
[...]
Evaluating Results
Test: Calculation of r = signal/(95%CL limit on signal)
Result: &badcolor Excluded
Result for r: 1.15744708205
Analysis: atlas_1405_7875
SR: SR03_a.4jl-
\end{Verbatim}
\end{bigtextfile}
Such a SUSY scenario is so constrained by existing LHC searches that even a sample of 
only 100 Monte Carlo events is sufficient to exclude it within seconds. 

\section{Statistical Analysis in \Checkmate{}}
\label{sec:app:statistics}

A standard statistical problem which has to be solved in cut-based collider analyses is the following: What is the $p$-value of observing $N$ events in a certain bin if the Standard Model (= alternative hypothesis) predicts $B \pm \Delta B$ events and the new physics model (= null hypothesis) predicts $S \pm \Delta S$ events in addition to $B$? In this section we explain the exact procedure \Checkmate{} uses, based on the \Cls{} prescription paired with a likelihood ratio discriminator~\cite{Read:2002hq}. For a more detailed description, we refer to Ref.~\cite{Cranmer:2015nia}.
\subsection{1-bin Likelihood and Test Statistics}
\label{sec:app:1bin_like}

If $S$ and $B$ were known with infinite precision, the likelihood of observing $N$ events in a bin where $S + B$ are theoretically expected would be given by the Poisson distribution
\begin{align}
\mathcal{L}(N | S) = \text{Poiss}(N | S+B) \equiv  \frac{(S+B)^N}{N!} e^{-(S+B)}. \label{eq:coll:likelihood}
\end{align}
In reality, we do not know the background for certain but repeated evaluations would result in different values $B_{\text{unc.}}$. This parameter is distributed according to a probability density function $P(B_{\text{unc.}} | B, \Delta_B)$ with fixed $B$, $\Delta_B$. Even though in principle there can be arbitrary many independent $(\Delta B)_i$ from different error sources, \Checkmate{} only considers the combined background error $\Delta B$, cf.~\Cref{sec:resultfolder}. The algorithm described below can however be straightforwardly extended.

 It is a common practice to redefine $B_{\text{unc.}} = B_{\text{unc.}}(\theta)$  in terms of a dimensionless nuisance parameter $\theta$. It is then $\theta$ which is distributed according to a density function $P(\theta | \tilde \theta)$ with $\tilde \theta$ being the \emph{a priori} most probable value. At the beginning, this value is trivially given via $B_{\text{unc.}}(\tilde \theta) = B$, however along the calculation of \Cls{}, the value of $\tilde \theta$ will change as described below. 

In \Checkmate{}, we always assume $B_{\text{unc.}}(\theta) = B \exp{(\theta \Delta B/B )}$ and $\theta$ to be Gaussianly distributed according to $P(\theta | \tilde \theta) \propto \exp (-(\theta - \tilde \theta)^2/2)$.  With this choice of parameters, the \emph{a priori} value for $\tilde{\theta}$ is 0. For small $\Delta B/B$, this \emph{lognormal distribution} will lead to a Gaussianly distributed $B_{\text{unc}}$ with mean $B$ and  standard deviation $\Delta B$. However, for very large uncertainties it prevents $B(\theta)$ from turning to unphysical negative values.  

Following the same approach for $S_{\text{unc}}(\theta)$ and assigning independent uncertainties to signal and background we get the following extended likelihood:\footnote{Note that we can safely ignore normalisation factors for the Gaussian distributions as they will not contribute to the likelihood ratio.}
\begin{align}
\mathcal{L}(N, \tilde \theta_S, \tilde \theta_B | \mu, \theta_{B}, \theta_S) &\equiv \Big(\frac{1}{N!} \Big[\lambda(\mu, \theta_B, \theta_S)\Big]^N e^{-\lambda(\mu, \theta_B, \theta_S)}\Big) \cdot \Big(e^{-(\tilde \theta_B - \theta_B)^2/2} \Big) \cdot \Big(e^{-(\tilde \theta_S - \theta_S)^2/2} \Big), \label{eq:app:likelihood}\\
\lambda(\mu, \theta_B,  \theta_S) &\equiv \mu S e^{\Big(\frac{\Delta S}{S} \theta_S\Big)} + B e^{\Big(\frac{\Delta B}{B} \theta_B\Big)}.
\end{align}
Note that we have introduced the signal strength modifier $\mu$, which will prove more convenient in distinguishing signal and background hypotheses for varying $S_{\text{unc}}(\theta_S)$.

There are different approaches to incorporate the unknown nuisance parameters $\theta_B, \theta_S$ into the test 
statistics. \Checkmate{} uses the \emph{Profile Likelihood Ratio} defined as
\begin{align}
q_\mu(N, \tilde \theta_S, \tilde \theta_B) &\equiv 
     - 2 \log \left( \frac{\mathcal{L}(N, \tilde \theta_S, \tilde \theta_B| \mu, \hat{\theta}^\mu_S, \hat{\theta}^\mu_B)}
                          {\mathcal{L}(N, \tilde \theta_S, \tilde \theta_B | \hat \mu, \hat \theta_S, \hat \theta_B)} \right). \label{eq:coll:likelihood2}
\end{align}
Here, $\hat \mu \in [0, \mu]$\footnote{The lower limit $\hat \mu \geq 0$ leads to a one-sided limit on $S$ while the upper limit $\hat \mu \leq \mu$ ensures that the test statistics is 0 if the global best fit would prefer an even larger signal than the one tested. In other words, we claim perfect compatibility of signal hypothesis and observation also if a larger signal would fit the observation better.   }, $\hat \theta_S$ and $\hat \theta_B$ is the combination of all three parameters which globally maximises $\mathcal{L}$, whereas $\hat \theta_S^\mu, \hat \theta_B^\mu$ are the values which maximise $\mathcal{L}(\mu)$ for fixed $\mu$. $q_\mu(N, \tilde \theta_S, \tilde \theta_B)$ becomes larger for less compatibility of observation and null hypothesis. According to Wilks' theorem \cite{wilks1938}, the maximum likelihood ratio approaches a  $\chi^2$-distribution for large event rates. 

Note that even for our rather simple statistical setup, the numerator cannot be evaluated analytically. As such, \Checkmate{} uses the numerical \texttt{scipy.optimize.root} routine to find the roots of the first derivatives in order to the evaluate the test statistics.


\subsection{Confidence Levels and $p$-values}
\label{sec:app:conflevel}
With the test statistics of Eq.~(\ref{eq:coll:likelihood2}), we can determine the $p$-value of the signal hypothesis $S+B$ after observing $N$ as follows.
If we repeated the experiment infinitely many times we would expect to observe different values $N^\prime$ each time such an experiment is performed due to the statistical nature of the underlying physics. Also, our determination of the distributions for $B_{\text{unc}}(\theta_B), S_{\text{unc}}(\theta_S)$ would have resulted in different values for the expectation values $\tilde \theta_B, \tilde \theta_S$ which we \emph{a priori} assumed to be 0. The values which we would expect depend on the underlying hypothesis we assume. 

If the signal hypothesis $\mu=1$ was correct, according to the likelihood in Eq.~(\ref{eq:app:likelihood}) the most compatible values for the nuisance parameters $\theta_S, \theta_B$ \emph{after} the observation of $N$ would be their best fit values $\hat \theta_S^{\mu=1}$ and $\hat \theta_B^{\mu=1}$. If we thus \emph{a posteriori} assume that these were the true expected values of $\tilde \theta_S, \tilde \theta_B$ and we hypothetically redid the experiment we would expect a random value of $\tilde \theta_B$ according to a Gaussian distribution with expectation value $\hat \theta_B^{\mu=1}$, analogously for $\tilde \theta_S$. Also, the number of observed events $N^\prime$ should be Poisson distributed with the expectation value $\lambda(\mu=1, \hat \theta_S^{\mu=1}, \hat \theta_B^{\mu=1})$. 

For a $p$-value we are interested in the fraction of these hypothetical experiments which would perform in a test statistics at least as bad as the observed one. We call this value for the signal hypothesis $\text{CL}_{S+B}$ and formulate it analytically as follows:
\begin{align}
\text{CL}_{S+B} &\equiv \sum_{N^\prime = 0}^\infty\ \int_{-\infty}^{\infty} \mathrm{d} \tilde \theta^\prime_S \int_{-\infty}^{\infty} \mathrm{d} \tilde \theta^\prime_B\   \Theta\Big(q_{\mu=1}(N^\prime, \tilde \theta^\prime_S, \tilde \theta^\prime_B) - q_{\mu=1}(N, \tilde \theta_S, \tilde \theta_B)\Big)  \times \nonumber\\ 
& \hspace{2cm} \text{Poiss}(N^\prime | \lambda(\mu=1, \hat \theta_S^{\mu=1}, \hat \theta_B^{\mu=1}) \cdot \text{Gauss}(\tilde \theta^\prime_S | \hat \theta^{\mu=1}_S)\cdot \text{Gauss}(\tilde \theta^\prime_B | \hat \theta^{\mu=1}_B), \label{eq:app:clsplusb}
\end{align}
where we define $\Theta(x) = 1$ for $x\geq 0$ and $0$ else.

This approach however has a peculiar property: If, for example, $N$ happens to be much smaller than $B-2\Delta B$, which statistically can happen in a small fraction of experiments even if $B$ describes Nature accurately, $\text{CL}_{S+B}$ will always turn out to be small, regardless of $S$. Therefore, the above interpretation will always claim a tension with the signal hypothesis and could even conclude that a model with $S \ll \Delta B$  is excluded. One should be worried in this case as it intuitively sounds incorrect to conclude anything about a signal which is much smaller than the systematic uncertainty of the experiment. 

A commonly used approach to avoid such a false exclusion is to, in addition to $\text{CL}_{S+B}$, determine the $p$-value for the observation to be compatible with the background-only hypothesis. For that purpose, we simply set $\mu = 0$ in our above explanation and thus evaluate
\begin{align}
1-\text{CL}_{B} &= \sum_{N^\prime = 0}^\infty\ \int_{-\infty}^{\infty} \mathrm{d} \tilde \theta^\prime_S \int_{-\infty}^{\infty} \mathrm{d} \tilde \theta^\prime_B\   \Theta\Big(q_{\mu=1}(N^\prime, \tilde \theta^\prime_S, \tilde \theta^\prime_B) - q_{\mu=1}(N, \tilde \theta_S, \tilde \theta_B)\Big)  \times \nonumber\\ 
& \hspace{2cm} \text{Poiss}(N^\prime | \lambda(\mu=0, \hat \theta_S^{\mu=0}, \hat \theta_B^{\mu=0}) \cdot \text{Gauss}(\tilde \theta^\prime_S | \hat \theta^{\mu=0}_S)\cdot \text{Gauss}(\tilde \theta^\prime_B | \hat \theta^{\mu=0}_B). \label{eq:app:clb}
\end{align}
Note that it is a common misconception to evaluate the test statistics of $1-\text{CL}_{B}$ with $q_{\mu = 0}$ instead of $q_{\mu=1}$. However, through the entire limit setting procedure we are testing the signal hypothesis and at this stage we are estimating what the result of this test would be if the background hypothesis was true. This is why we use $\mu = 0$ for the expectation values of $N^\prime, \tilde \theta_S$ and $\tilde \theta_B$ but still need to evaluate the test statistics for the signal hypothesis $\mu = 1$. 

In the $\text{CL}_S$ prescription, the confidence in the signal hypothesis is calculated by the ratio
\begin{align}
\text{CL}_S \equiv \frac{\text{CL}_{S+B}}{1-\text{CL}_B}
\end{align}
and we interpret this value as the $p$-value for our signal. Consequently, we exclude a signal model if it produces a 
too small $\text{CL}_S$ value. If an experiment shows a perfect agreement with the Standard Model prediction, $1-\text{CL}_B$ 
equals 0.5 and therefore $\text{CL}_S$ is very close to the true $p$-value $\text{CL}_{S+B}$.  For experiments which are in 
tension with the background-only hypothesis, $1-\text{CL}_B$ becomes smaller, $\text{CL}_S$ increases and thus the limit 
weakens. $\text{CL}_S$ 
{$\text{CL}_S$ therefore sets a conservative limit in the presence of under-fluctuations in the data.}

\Checkmate{} evaluates the integrals in Eqs.~(\ref{eq:app:clsplusb}) and (\ref{eq:app:clb}) numerically by generating tuples of random numbers $N^\prime, \tilde \theta_B, \tilde \theta_S$ according to the respective Poisson or Gaussian distribution and counting the relative amount of tuples which yield a value of $q_{\mu=1}$ smaller than $q_{\mu=1}(N, 0, 0)$. 

\subsection{Model Independent Limits $S95$}
In the main text, we explained how \Checkmate{} usually does not calculate $\text{CL}_S$ in each run but normally calculates $r = (S - 1.64 \cdot \Delta S)/S95$ with the model independent upper \unit[95]{\%} confidence limit $S95$. As the value for $\Delta S$ is not known in that case, it is set to 0 for the following evaluation. \Checkmate{} uses simplified versions of the above described formulae with all terms depending on $\theta_S$ removed and using $\lambda(\mu, \theta_B) \equiv \mu S + B \exp(\frac{\Delta B}{B} \theta_B)$.

For the \emph{observed} limit, \texttt{S95\_obs}, one simply needs to find the value of $S$ which yields $\text{CL}_S = 0.05$. \Checkmate{} uses the Pegasus regula-falsi method for this purpose \cite{Dowell1972}.
For the \emph{expected} limit, \texttt{S95\_exp}, \Checkmate{} determines the limit if we observed what we derived as the true background values from our observation, see \Cref{sec:app:conflevel} above. That means we set $N = \lambda(\mu=0, \hat \theta^{\mu=0}_B)$ and the \emph{a priori} $\tilde \theta_B$ to $\hat \theta^{\mu=0}_B$ and follow the same prescription as before. 
 
\subsection{Likelihood}
\label{sec:app:likelihood}
As mentioned in \Cref{sec:app:fulllist}, \Checkmate{} is capable of returning a combined likelihood 
summed over all bins. Here, it calculates the test statistics as in 
Eq.~(\ref{eq:coll:likelihood2}) and sums the result over all bins. For this purpose, we 
remove the restriction $\hat \mu \in [0, \mu]$ from Eq.~(\ref{eq:coll:likelihood2}) as we are not 
trying to perform a one-sided signal test. This calculation can be done for all bins and 
the sum of all likelihoods over all considered signal regions is returned by \Checkmate{}. 

We note that a user should exercise care with the likelihood calculation since it can only be applied
to orthogonal signal regions and some of the analyses included in \CheckMATE{} do not fulfil this condition. If
the user wishes to include such analyses, the likelihood sum should be calculated manually by selecting the 
result of the relevant orthogonal signal regions.

\section{Tuning}
\label{sec:app:tuning}

In this appendix we provide details of the new \Checkmate{} tunings for lepton efficiencies and $b$-tagging. In the case of 
leptons we use recent \Atlas{} parametrisations updated during the 8 and 13~TeV runs.  The new lepton energy 
resolutions have been implemented in the \Delphes{} cards for 13 and 14~TeV analyses. The efficiency is evaluated internally by the respective \texttt{AnalysisHandler}s.

\subsection{Electrons}

\begin{figure}[t]
 \begin{center}
 \includegraphics[width=0.7\textwidth]{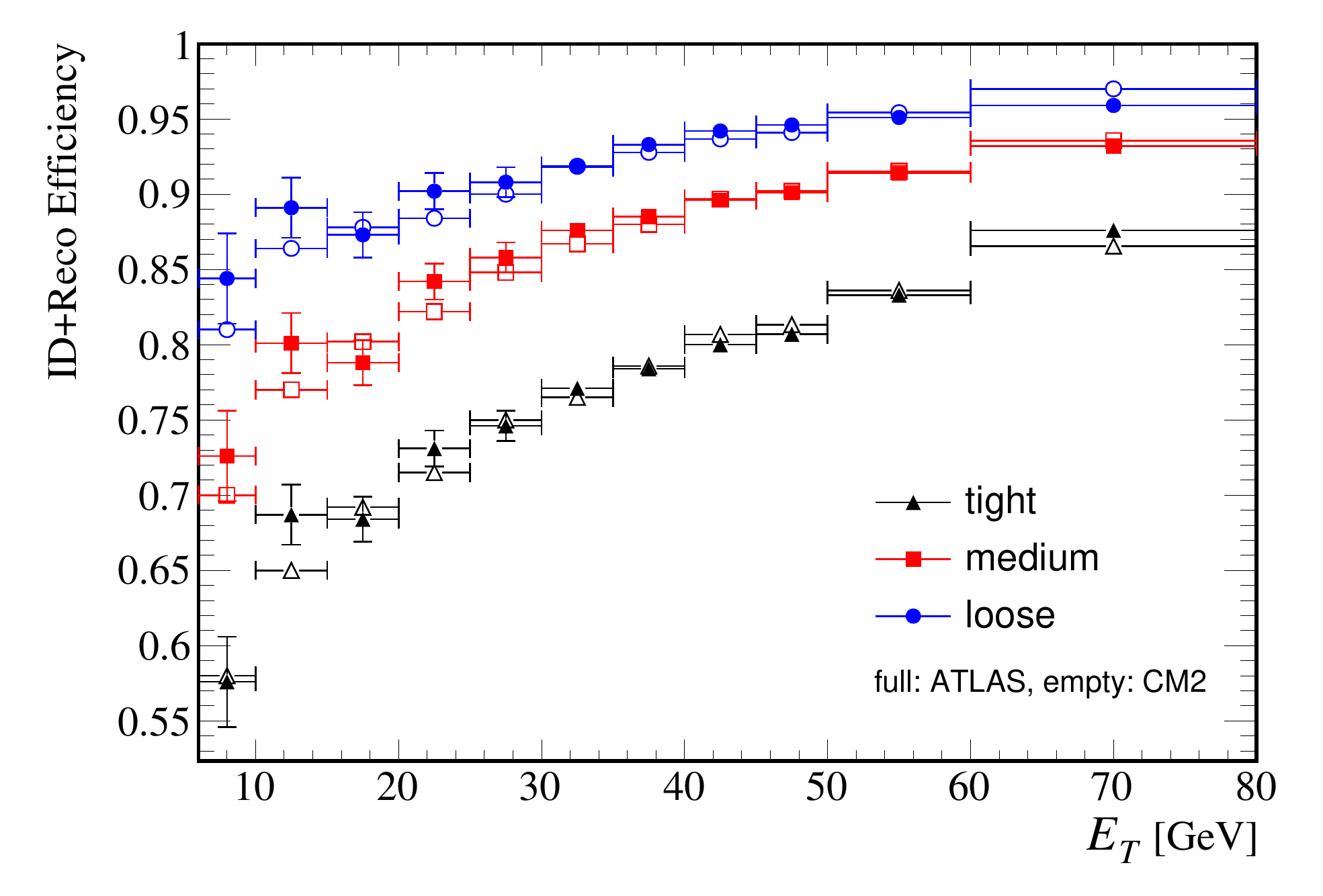}
 \end{center}
 \caption{Comparison of the combined electron reconstruction and identification efficiency reported by \ATLAS{} (full markers; see Figure~14 of Ref.~\cite{ATLAS-CONF-2016-024}) and obtained with \Checkmate{} (empty markers) for tight (black triangle), medium (red square) and loose (blue circle) electrons as a function of electron candidate transverse energy and in the full pseudorapidity range, $|\eta| < 2.47$.\label{fig:electroneff}}
\end{figure}

Electron identification in the \Atlas{} detector is based on the multivariate analysis (MVA) that simultaneously evaluates 
several properties of the electron candidates. Three levels of identification are implemented in \Checkmate{} that 
correspond to the typical \Atlas{} operating points: loose, medium and tight. The efficiency depends on the 
transverse energy of the electron candidate, $E_T$, and to a lesser extent on the rapidity~\cite{ATLAS-CONF-2016-024}. The 
current implementation in the \texttt{AnalysisHandler} takes into account only the former. The following functions are 
used for the 13~TeV setup:
\begin{eqnarray}
 && \epsilon_\mathrm{id}^\mathrm{loose} = 0.976 - 0.0614 \cdot \exp\left(1 - \frac{E_T}{29.1}\right)\,,\\
 && \epsilon_\mathrm{id}^\mathrm{medium} = 0.937 - 0.109 \cdot \exp\left(1 - \frac{E_T}{21.0}\right)\,,\\
 && \epsilon_\mathrm{id}^\mathrm{tight} = 0.8885 - 0.138 \cdot \exp\left(1 - \frac{E_T}{27.5}\right)\,,
\end{eqnarray}
where $E_T$ is the transverse energy of the candidate in GeV. The parameters in the above parametrisation were 
obtained from the fit to the efficiency plots reported by \Atlas{}~\cite{ATLAS-CONF-2016-024,electroneff}. \Cref{fig:electroneff} shows the electron reconstruction and identification efficiency reported by \ATLAS{} and obtained with \Checkmate{} for tight, medium and loose electrons as a function of electron candidate transverse energy.

For the parametrisation of electron energy resolution we follow Ref.~\cite{ATL-PHYS-PUB-2013-004}. The functions describing smearing of energy and momenta are the following:
\begin{eqnarray}
 && \sigma(\mathrm{GeV}) = 0.3 \oplus 0.10\times\sqrt{E(\mathrm{GeV})} \oplus 0.010 \times E(\mathrm{GeV})\quad \mathrm{for}\quad |\eta| < 1.4\,,\\
 && \sigma(\mathrm{GeV}) = 0.3 \oplus 0.15\times\sqrt{E(\mathrm{GeV})} \oplus 0.015 \times E(\mathrm{GeV})\quad \mathrm{for}\quad 1.4 < |\eta| < 2.47\,.
\end{eqnarray}
Here, $a \oplus b \equiv \sqrt{a^2+b^2}$. The functions are implemented in the \texttt{delphes\_skimmed\_ATLAS\_13TeV.tcl} and \texttt{delphes\_skimmed\_ATLAS\_14TeV.tcl} cards.

\subsection{Muons}

Muon reconstruction and identification efficiency shows generally an excellent performance across different rapidities and energy ranges~\cite{Aad:2016jkr}. In \Checkmate{} the efficiency is about $0.99$ for loose muons and $0.97$ for tight muons, except for muon candidates with $|\eta| < 0.1$ where it is set to $0.6$. The dependence on $p_T$, unlike for the electrons, is negligible.  

The combined muons correspond to muons observed in the inner detector (ID) and in the muon spectrometer (MS). The resolution formula combines the information from both systems and is given by~\cite{ATL-PHYS-PUB-2013-009}:
\begin{eqnarray}
  && \sigma_\mathrm{ID} = p_T \times \sqrt{a_1^2+(a_2 \times p_T)^2}\,,\label{eq:sigmaid}\\
  && \sigma_\mathrm{MS} = p_T \times \sqrt{\left(\frac{b_0}{p_T}\right)^2 + b_1^2 + (b_2 \times p_T)^2}\,, \label{eq:sigmams} \\
  && \sigma_\mathrm{CB} = \frac{\sigma_\mathrm{ID} \times \sigma_\mathrm{MS}}{\sqrt{\sigma_\mathrm{ID}^2 +\sigma_\mathrm{MS}^2}}\,,\
\end{eqnarray}
where $p_T$ is the truth transverse momentum in GeV. The coefficients $b_0$, $b_1$, $b_2$ are specified in Table~\ref{tab:muonres} and are the same for 13 and 14~TeV analyses. In the same table we also provide coefficients $a_1$ and $a_2$ for the 13~TeV setup. For the HL option the $a_1$ and $a_2$ coefficients are specified in 15 separate regions in rapidity taking into account planned upgrades to the inner detector. The full list can be found in Ref.~\cite{ATL-PHYS-PUB-2013-009}.  

\begin{table}[t]
\begin{center}\renewcommand*{\arraystretch}{1.2}
\begin{tabularx}{\textwidth}{XXXXXX}
\toprule \midrule
                 & $a_1$     & $a_2$      & $b_0$  & $b_1$     &  $b_2$ \\ \midrule
 $|\eta| < 1.05$ & $0.01607$ & $0.000307$ & $0.24$ & $0.02676$ &  $0.00012$ \\ 
 $|\eta| > 1.05$ & $0.03000$ & $0.000387$ & $0.00$ & $0.03880$ &  $0.00016$ \\ 
\midrule \bottomrule
\end{tabularx}
\caption{The muon resolution coefficients $a_1$, $a_2$, $b_0$, $b_1$, $b_2$ from Eqs.~\eqref{eq:sigmaid} and \eqref{eq:sigmams} for the 13~TeV \Atlas{} setup. \label{tab:muonres}}
\end{center}
\end{table}

\subsection{$B$-Tagger}

\begin{figure}[t]
\begin{subfigure}{0.49\textwidth}
\includegraphics[width=1.0\textwidth]{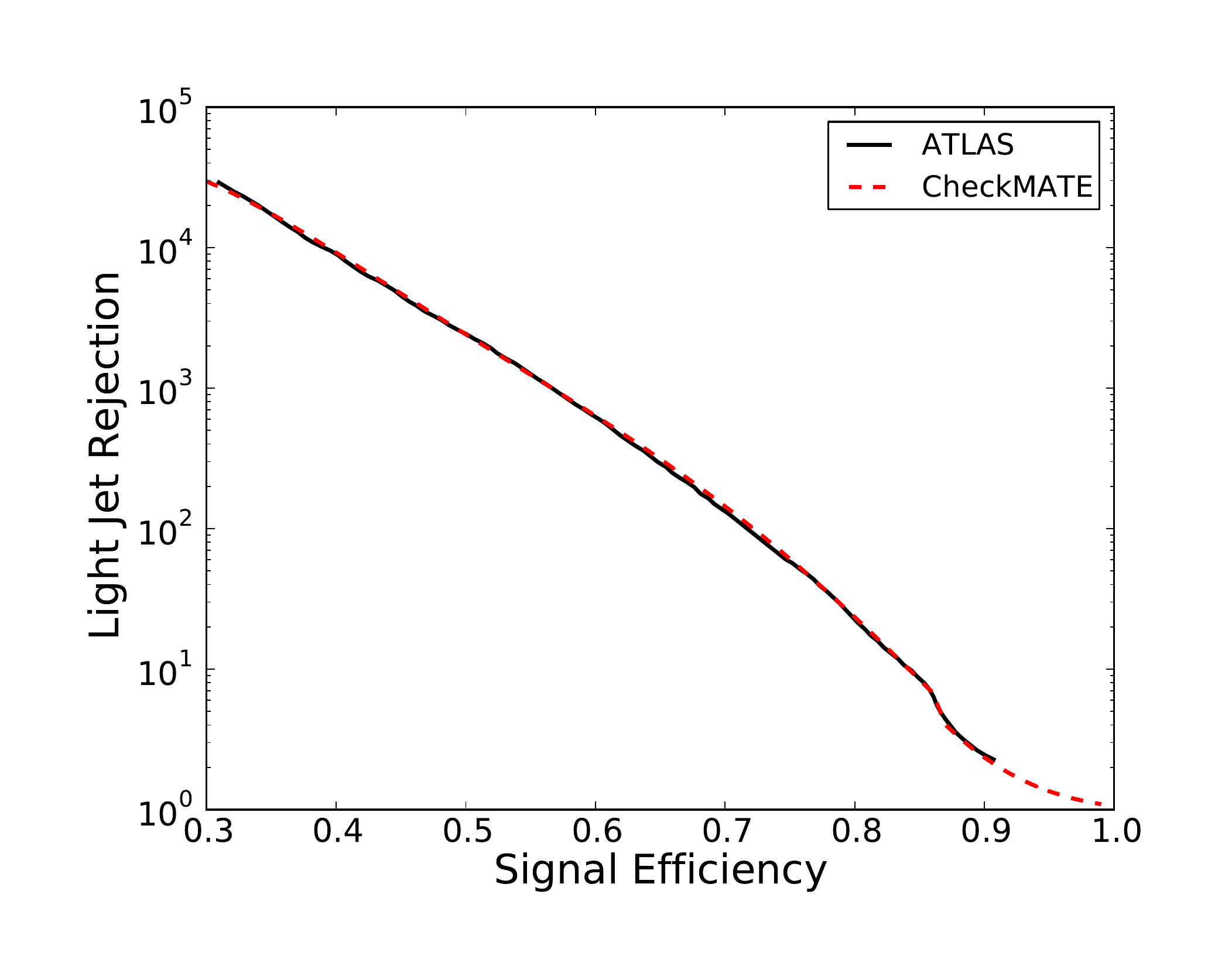}
\caption{Rejection curve for jets containing light quarks only.}
\label{fig:app:tune:btag:roc:a}
\end{subfigure}
\begin{subfigure}{0.49\textwidth}
\includegraphics[width=1.0\textwidth]{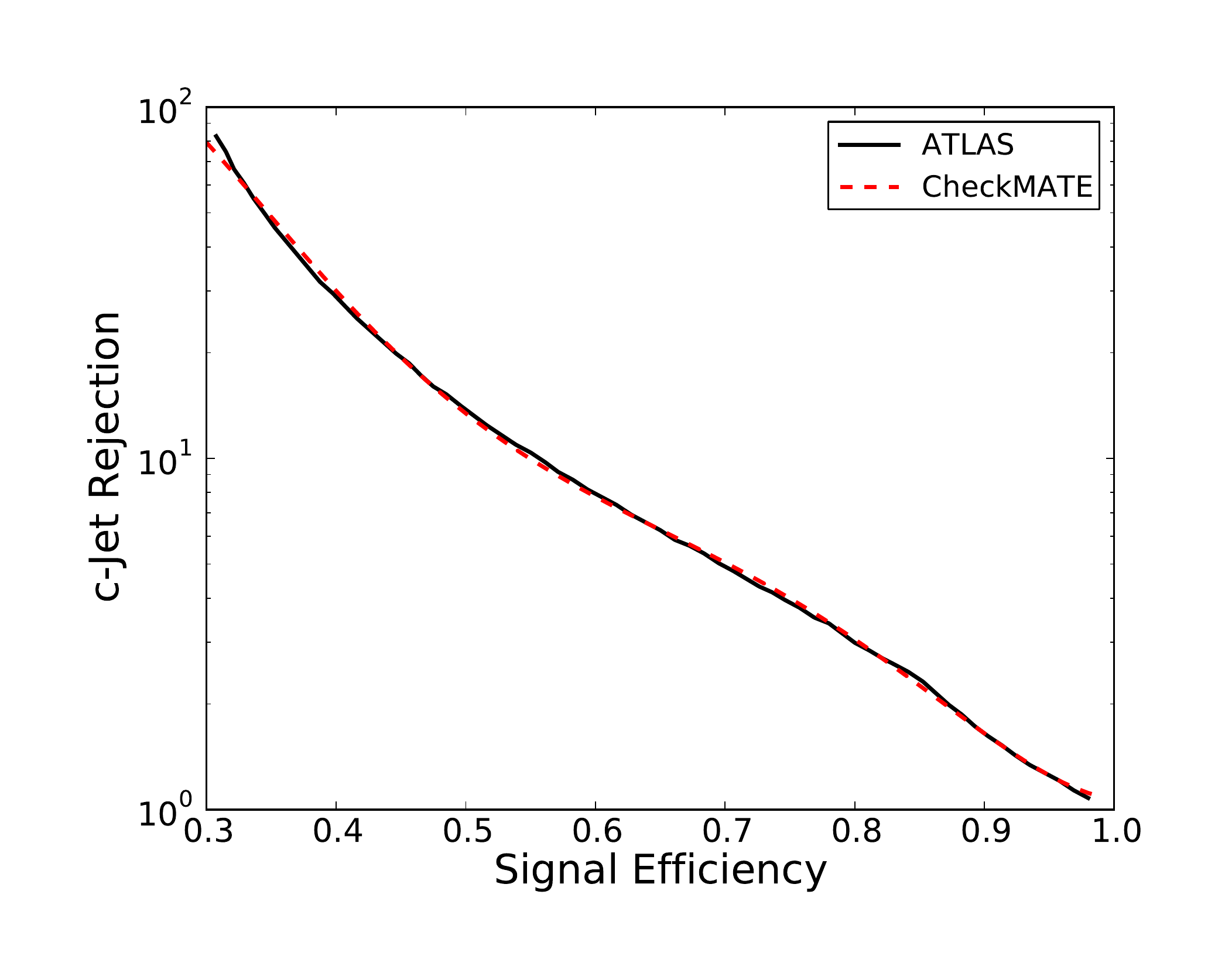}
\caption{Rejection curve for jets with charm content.}
\label{fig:app:tune:btag:roc:b}
\end{subfigure}

\caption{Receiver Operation Characteristic curves for a dependence of the background rejection for jets with different quark contents on a chosen signal efficiency working point of the $b$-tagger~\cite{ATLAS-CONF-2012-043}.}
\label{fig:app:tune:btag:roc}
\end{figure}

The quality of algorithms that try to filter jets containing $b$-quarks from others 
is determined by two main quantities. The signal efficiency describes the probability 
to assign a tag to a jet that actually contains a $b$-quark, whereas the background
efficiency is a measure for the relative amount of jets that are tagged even though 
they do not have any bottom quark content. Since the background efficiency is 
usually small, it is common to use the inverse value, called rejection, for 
illustrative purposes. Also, one usually distinguishes between rejections against 
jets with charm content and other jets that only contain light quarks, as the first 
are harder to distinguish from the signal. 

\begin{figure}[!h]
\begin{subfigure}{0.49\textwidth}
\includegraphics[width=1.0\textwidth]{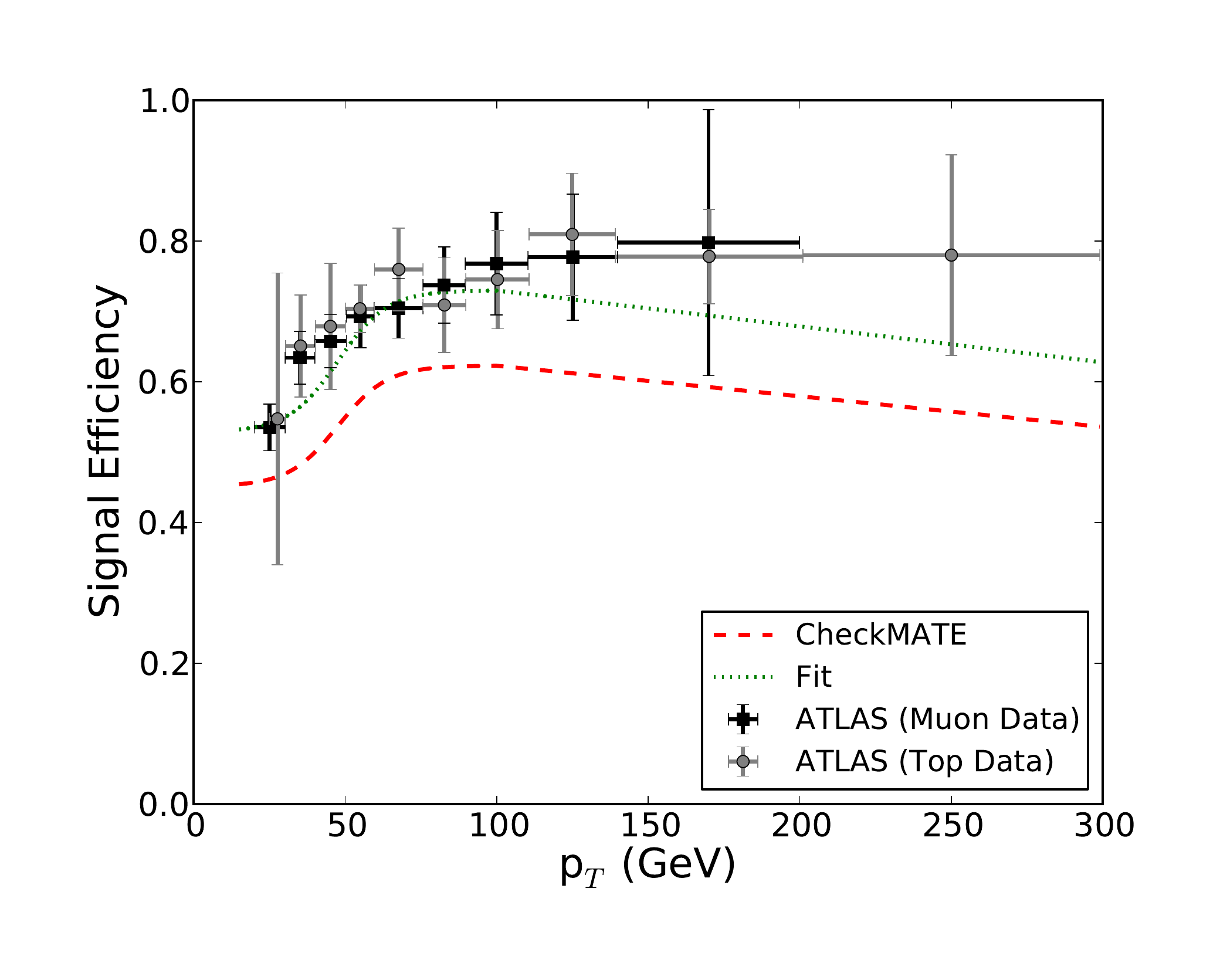}
\caption{Signal efficiencies for $b$-tagging, determined by combining the information from two different search 
channels for $\bar{\epsilon}_\text{S} = 0.7$ \cite{ATLAS-CONF-2012-043,ATLAS-CONF-2012-097}.}
\label{fig:app:tun:btag:pt:a}
\end{subfigure} \hfill\begin{subfigure}{0.49\textwidth}
\includegraphics[width=1.0\textwidth]{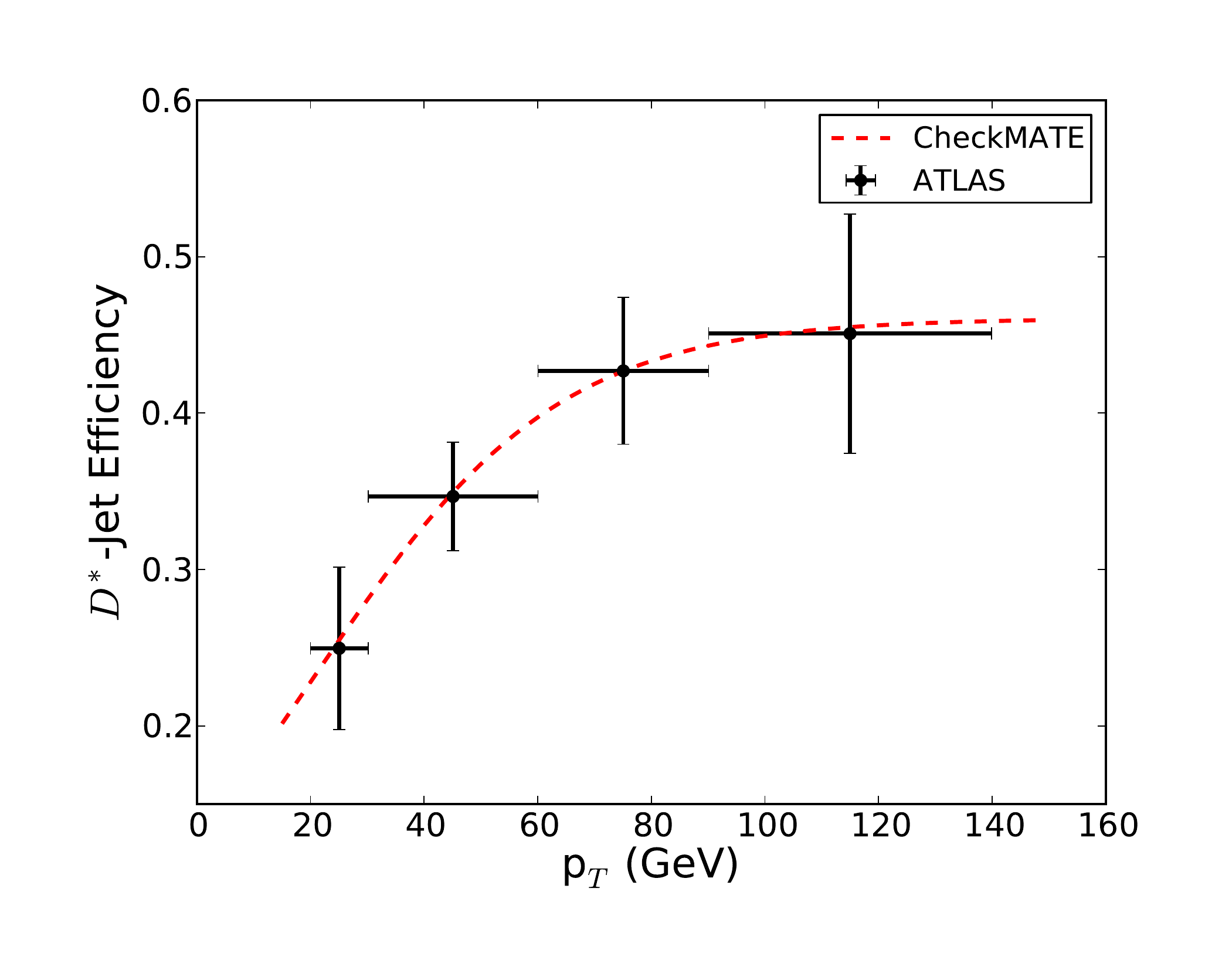} 
\caption{$b$-tagging efficiency of jets containing $D^*$ mesons for $\bar{\epsilon}_\text{S} = 0.7$ \cite{ATLAS-CONF-2012-097}. The inclusive efficiency for $c$-jets is assumed to be \unit[40]{\%} of this $D^*$ efficiency.}
\label{fig:app:tun:btag:pt:b}
\end{subfigure}

\begin{subfigure}{0.49\textwidth}
\includegraphics[width=1.0\textwidth]{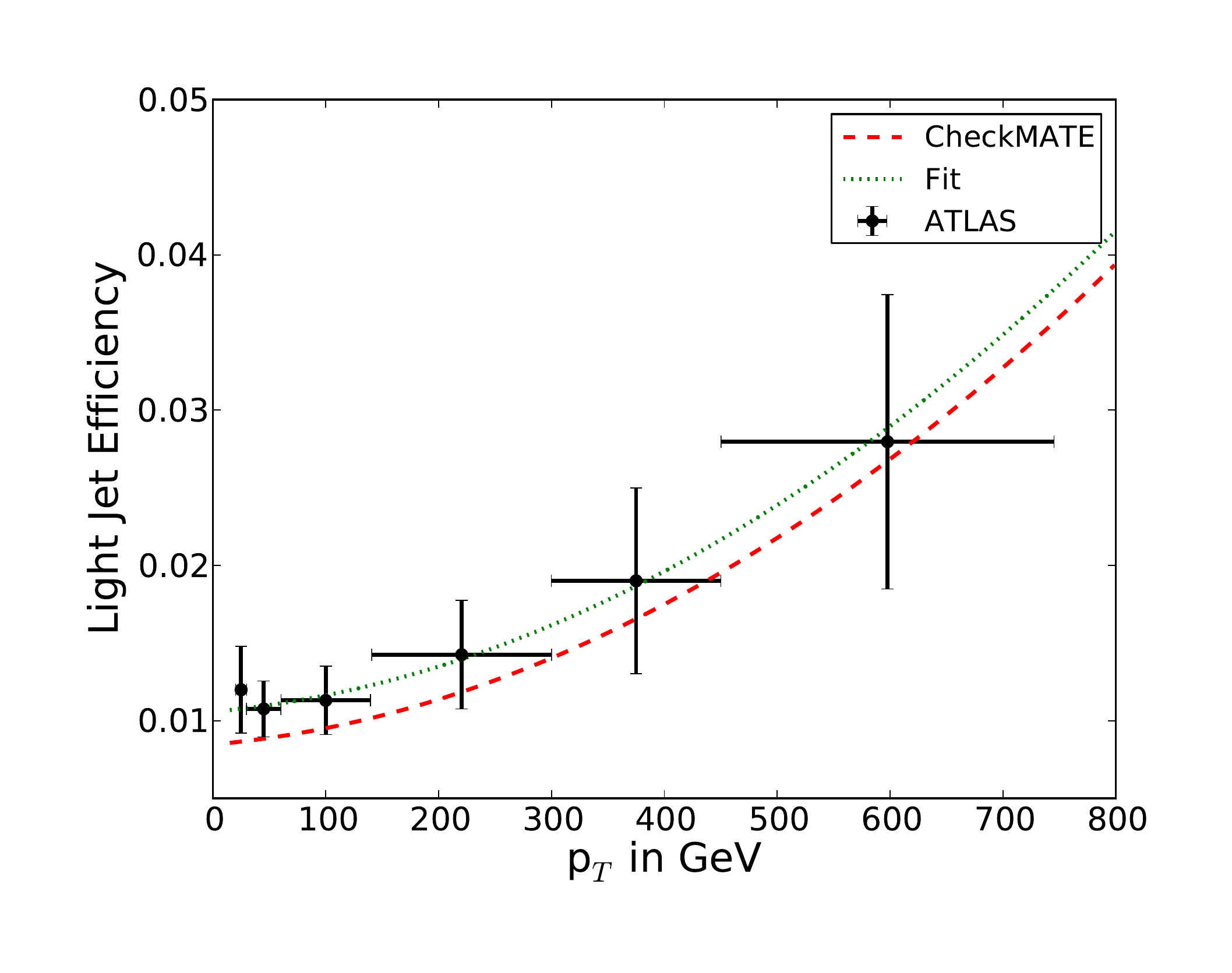}
\caption{$b$-tagging efficiency of jets containing light quarks for $\bar{\epsilon} _\text{S}= 0.7$ and $|\eta| < 1.3$ \cite{ATLAS-CONF-2012-040}.}
\label{fig:app:tun:btag:pt:c}
\end{subfigure}
\begin{subfigure}{0.49\textwidth}
\includegraphics[width=1.0\textwidth]{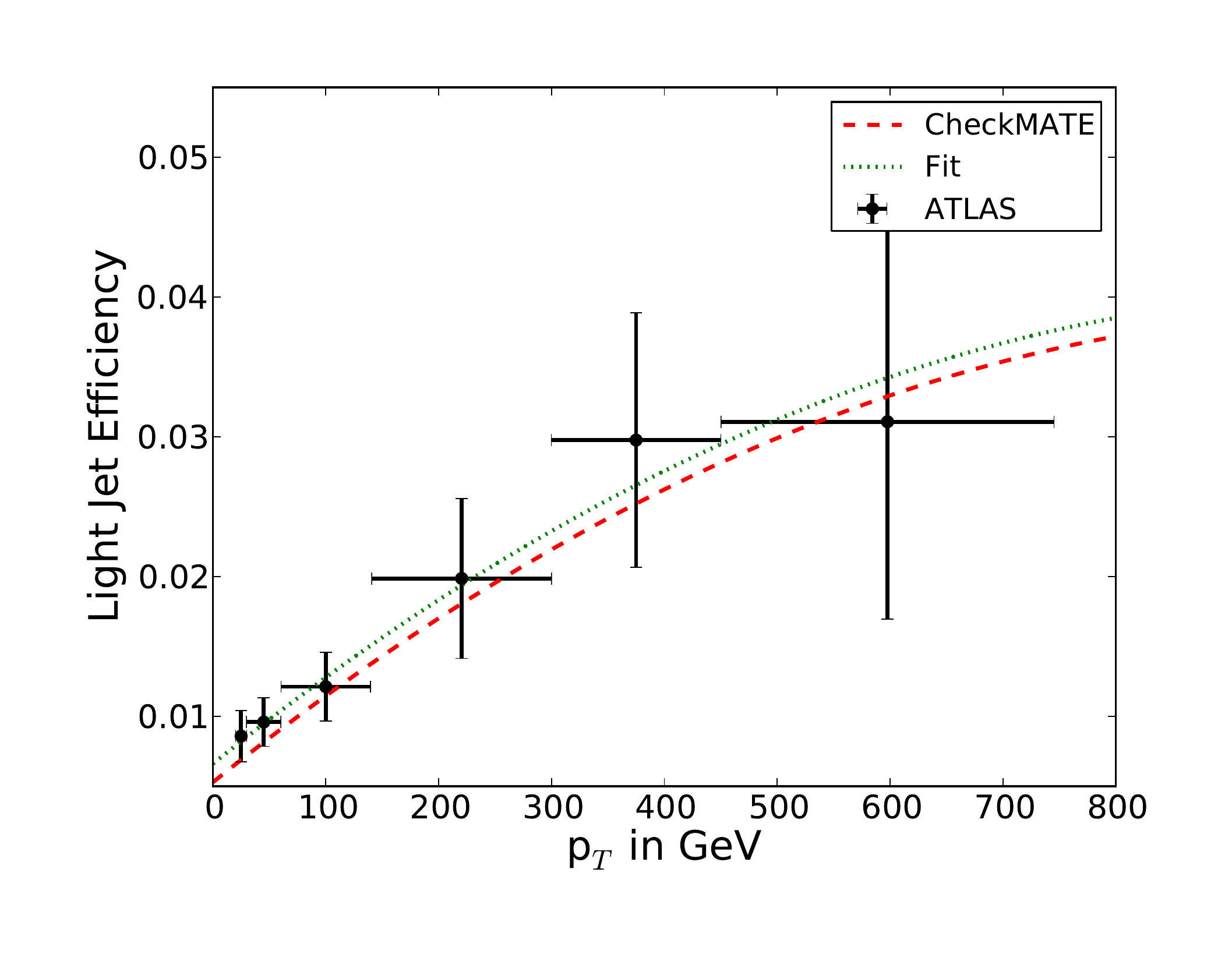}
\caption{Same as (c) for  $1.3 \leq |\eta| < 2.5$. \\ } 
\label{fig:app:tun:btag:pt:d}
\end{subfigure}
\caption{Dependence on the signal and light-jet / $c$-jet background efficiencies for $b$-tagging  on the transverse momentum of the jet candidate.}
\label{fig:app:tun:btag:pt}
\end{figure}

Since the rejection gets weaker with increasing signal efficiency, one has to find a 
balance between signal quantity and signal purity, which depends crucially on the 
details of the respective analysis. For this purpose, one uses the 
ROC (Receiver Operation Characteristic) curve that describes the relation between these 
two quantities.

\subsubsection{8 \TeV{}}
We show the ROC curves for light-jet and $c$-jet 
rejection separately in \Cref{fig:app:tune:btag:roc} and internally parametrise 
these as follows:
\begin{align}
\log_{10}\left[\bar{r}_\text{light}(\bar{\epsilon}_\text{S})\right] &= \left\{
  \begin{array}{lr}
   \displaystyle 52.8\cdot(\bar{\epsilon}_\text{S}-4.045\cdot\bar{\epsilon}_\text{S}^2+7.17\cdot\bar{\epsilon}_\text{S}^3-6.14\cdot\bar{\epsilon}_\text{S}^4+2.01\cdot\bar{\epsilon}_\text{S}^5) &  \quad \text{if } \bar{\epsilon}_\text{S} < 0.87, \\
    -75.4\cdot(\bar{\epsilon}_\text{S}-1.07)^3 &  \quad \text{if }  \bar{\epsilon}_\text{S} \geq 0.87,
  \end{array}
\right. \\
\log_{10}\left[\bar{r}_\text{c}(\bar{\epsilon}_\text{S})\right] &= 29.3\cdot(\bar{\epsilon}_\text{S}-4.572\cdot\bar{\epsilon}_\text{S}^2+8.496\cdot\bar{\epsilon}_\text{S}^3-7.253\cdot\bar{\epsilon}_\text{S}^4+2.33\cdot\bar{\epsilon}_\text{S}^5),
\end{align}
where $\bar{\epsilon}_\text{S}$ is signal efficiency and $\bar{r}_\text{light/c}$ background rejection for light and $c$-jets, 
respectively.

Given a particular working point on the ROC curve, i.e.\ a specific chosen signal 
efficiency, $\bar{\epsilon}_\text{S}$, and the corresponding background 
rejections, $\bar{r}_\text{light/c}$, the actual tagging probabilities depend on the transverse momentum of the considered object. These have been measured 
individually for signal, light-quark and $D^*$ meson\footnote{The tagging probability for
jets containing $D^*$ mesons is roughly 2 times higher than for other $c$-mesons. Using the cutflows 
from various analyses we have tuned this parameter to 0.4 to be in agreement with the
ATLAS results.} jets and we show the results in \Cref{fig:app:tun:btag:pt}. 

For the signal efficiency, we use two different data sets as they have different 
sensitivities at low and high energies; see \Cref{fig:app:tun:btag:pt:a}. In order
to agree with the cutflows of various analyses that require $b$-tagging, a reduction in the 
overall normalisation by 15\% has been applied. In addition, a
significant decrease of the signal efficiency at large energies has been manually 
added in order to get a better agreement with experimental results. 
Furthermore, the light-quark 
jet rejection has been measured for two different $\eta$ regions, which we adapt in 
our parametrisations. We also perform a reduction in the light-quark tagging rates (20\%) in 
order to better agree with experimental cutflows. 

Since the $p_T$ dependent distributions are given for a 
particular working point $\bar{\epsilon}_\text{S} = 0.7$, we linearly rescale the functions to the given 
chosen signal efficiency $\bar{\epsilon}_\text{S}$, or the corresponding background efficiency given 
by the ROC curves ($p_T$ in \GeV):
\begin{align}
\epsilon_\text{S}(\pT) &= \frac{\bar{\epsilon}_\text{S} \cdot 0.85}{0.7}  \left(0.552+ \frac{0.210}{1+\mathrm{e}^{-0.123\cdot(\pT-47.6)}}\right)\nonumber \\
& \qquad \qquad \qquad \qquad \qquad \times\frac{0.7+0.05\cdot \mathrm{e}^{-\frac{\pT}{308}}}{0.75}\left\{
  \begin{array}{ll}
  1 &  \quad \text{if }  \pT \leq 100, \\
\\
   1-7\times 10^{-4} \cdot (\pT-100)  &  \quad \text{if }  \pT > 100,
  \end{array}
\right. \\
\epsilon_{\text{light}}(\pT, \eta) &= \frac{\bar{r}_\text{light}(0.7)\times0.8}{\bar{r}_{\text{light}}(\bar{\epsilon}_\text{S})} \left\{
 \begin{array}{ll}
   1.06 \times 10^{-2} + 6.47 \times 10^{-6} \cdot \pT^2 + 4.03 \times 10^{-8} \cdot \pT^4   & \quad \text{if } |\eta| < 1.3, \\
\\
6.61 \times 10^{-3} + 6.49 \times 10^{-5} \cdot \pT^2 - 3.12 \times 10^{-8} \cdot \pT^4  & \quad \text{if } 1.3 \leq |\eta| < 2.5,
  \end{array}
\right. \\
\epsilon_{\text{c}}(\pT) &= \frac{\bar{r}_\text{c}(0.7)\times0.4}{\bar{r}_{\text{c}}(\bar{\epsilon}_\text{S})} \frac{0.461}{1+\mathrm{e}^{-0.0464\cdot(\pT-20.4)}}.
\end{align}

\subsubsection{13~\TeV{}}

For the \ATLAS{} analyses performed in Run--2, \CheckMATE{} uses efficiencies fitted to the MV2c20 algorithm described in \cite{ATL-PHYS-PUB-2015-022}.
The dependence of the $b$-jet efficiency on $p_T$ is kept the same as in Run--1, but the mistag rates differ from their Run--1 equivalents.

\begin{table}[t]
\begin{center}\renewcommand*{\arraystretch}{1.2}
	\begin{tabularx}{\textwidth}{XXXXXX}
\toprule \midrule
		& $a_{i,0}$ & $a_{i,1}$ & $a_{i,2}$ & $b_{i,1}$ & $b_{i,2}$ \\ \midrule
		light & $\phantom{+}6.77 \cdot 10^{-6}$ & $-9.86 \cdot 10^{-6}$ & $2.79 \cdot 10^{-5}$ & $-1.99 \cdot 10^{2}$ & $2.50 \cdot 10^{2}$ \\
		charm & $-2.97 \cdot 10^{-3}$ & $-2.15 \cdot 10^{-6}$ & $1.29 \cdot 10^{-4}$ & $-1.43 \cdot 10^{2}$ & $1.12 \cdot 10^{2}$ \\
\midrule \bottomrule
	\end{tabularx}
	\captionof{table}{The coefficients for the light- and charm-jet efficiency as a function of $\eta$, Eq.~\eqref{eq:13tev_btagger_fit_eta}.}
	\label{tab:eff_eta}
\end{center}
\end{table}

\begin{figure}[!h]
	\centering
	\captionsetup[subfigure]{aboveskip=0pt}
	\subcaptionbox{Efficiency vs.\ $p_T$. \label{fig:13tev_btagger_ceff_pt}}
		{\includegraphics[width=0.48\textwidth]{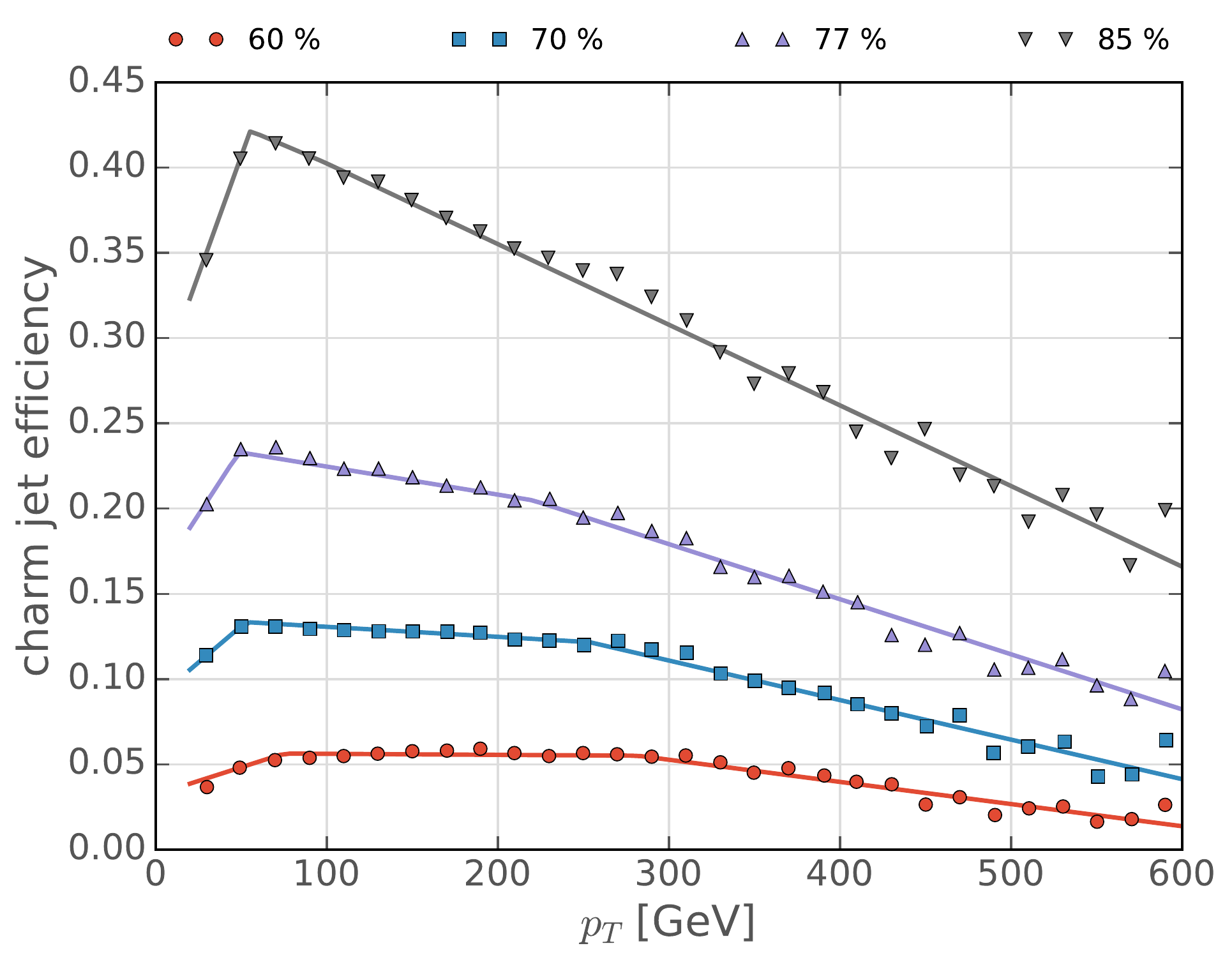}}
	\hfill
	\subcaptionbox{Efficiency vs.\ $\eta$. \label{fig:13tev_btagger_ceff_eta}}
		{\includegraphics[width=0.48\textwidth]{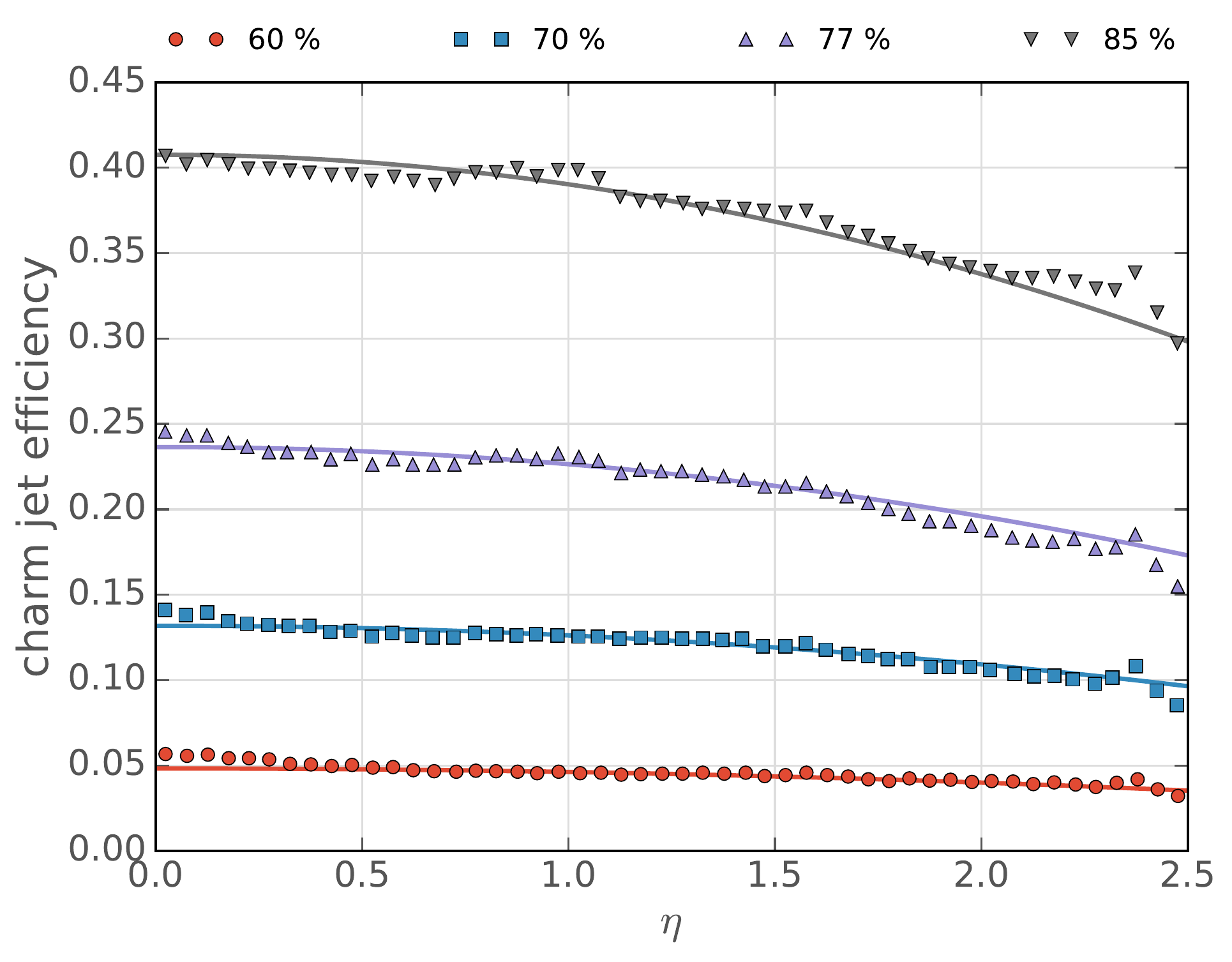}}
	\caption{
		The charm-jet efficiency for the 13~\TeV{} \ATLAS{} $b$-tagger, for four different working points.
		The points show the expected efficiencies as determined by the \ATLAS{} collaboration.
		The solid lines are the functions that \CheckMATETwo{} uses to model the efficiencies.
	}
	\label{fig:13tev_btagger_ceff}
\end{figure}

\begin{figure}[!h]
	\centering
	\captionsetup[subfigure]{aboveskip=0pt}
	\subcaptionbox{Efficiency vs.\ $p_T$. \label{fig:13tev_btagger_leff_pt}}
		{\includegraphics[width=0.48\textwidth]{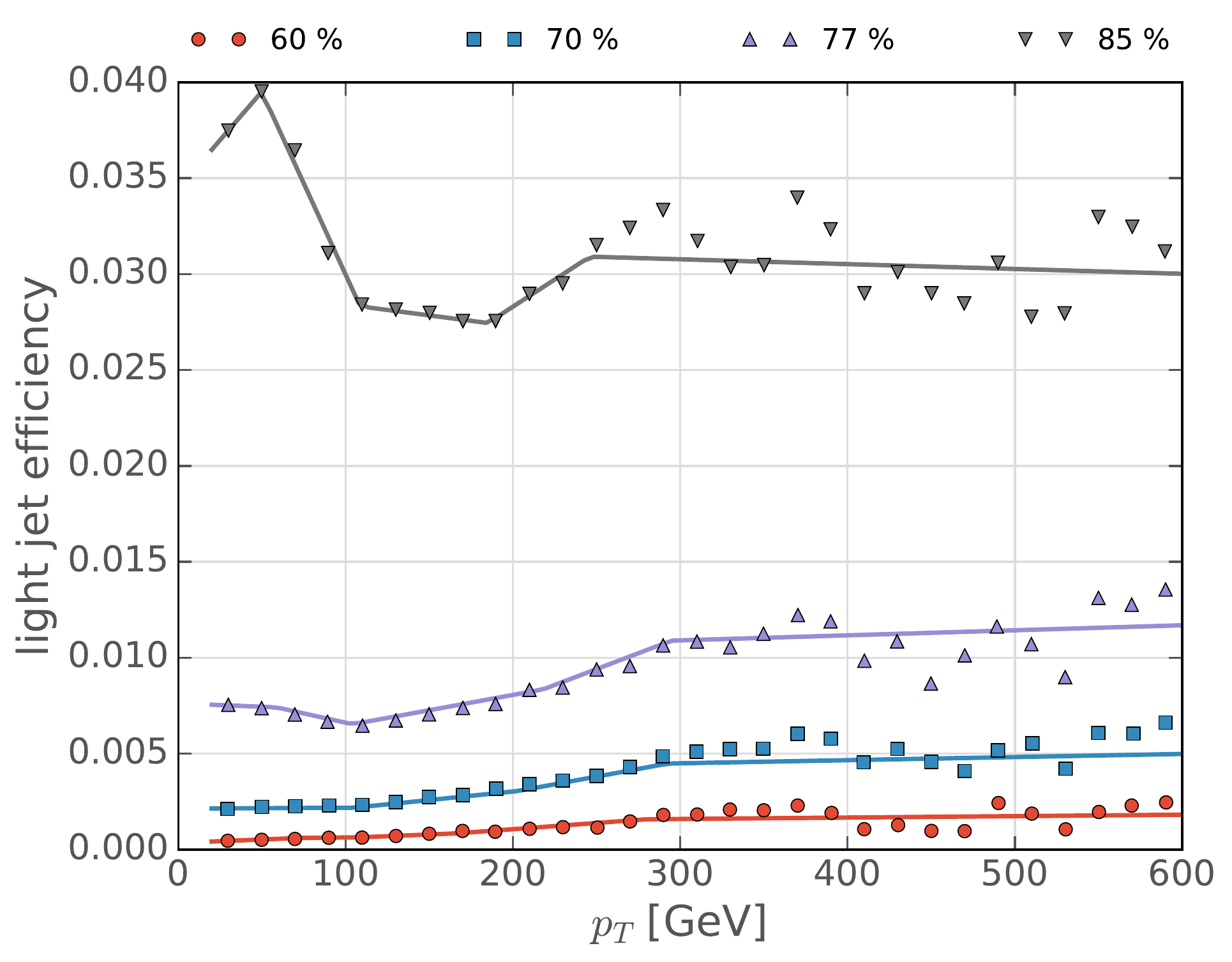}}
	\hfill
	\subcaptionbox{Efficiency vs.\ $\eta$. \label{fig:13tev_btagger_leff_eta}}
		{\includegraphics[width=0.48\textwidth]{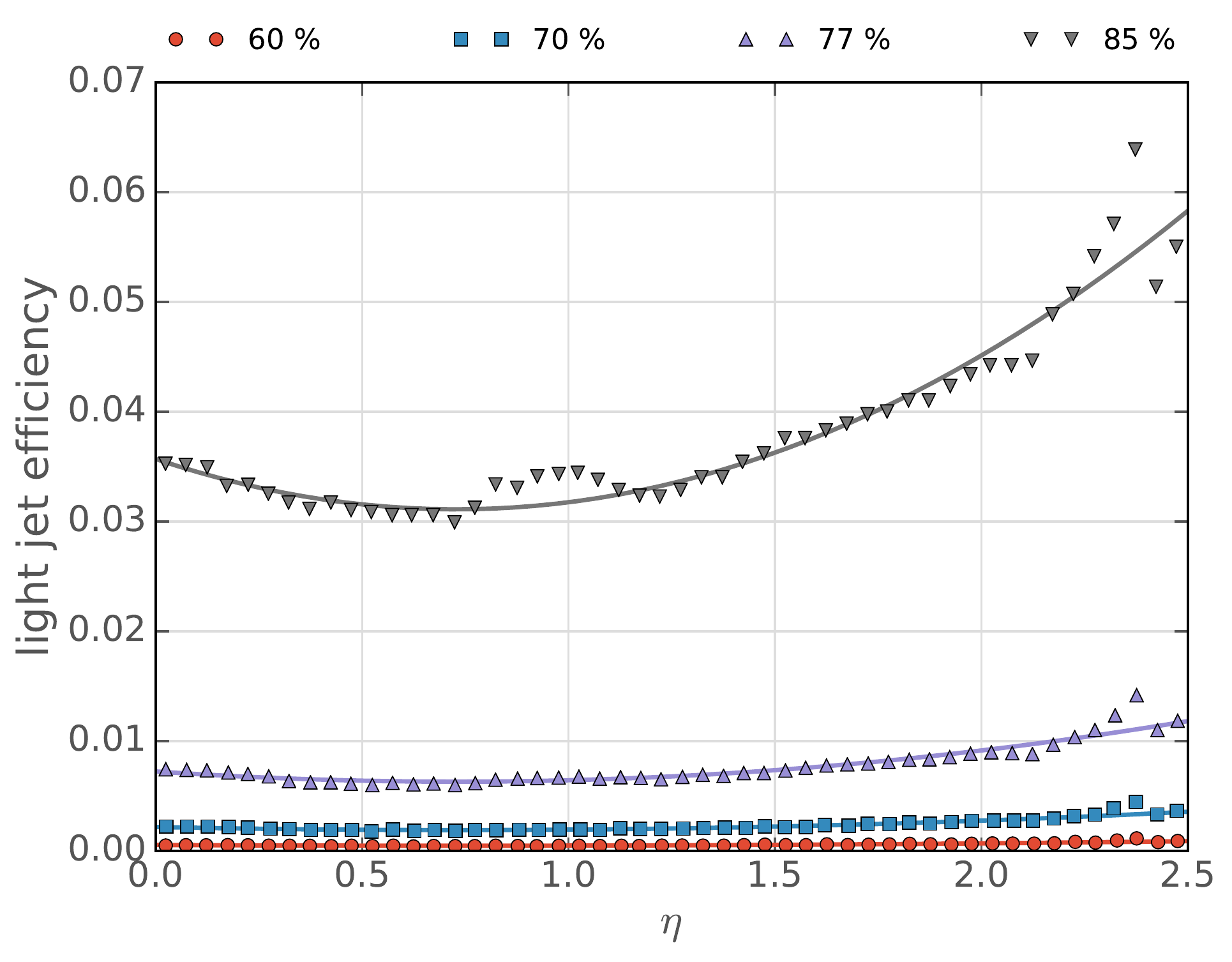}}
	\caption{
		The light-jet efficiency for the 13~\TeV{} \ATLAS{} $b$-tagger, for four different working points.
		The points show the expected efficiencies as determined by the \ATLAS{} collaboration.
		The solid lines are the functions that \CheckMATETwo{} uses to model the efficiencies.
	}
	\label{fig:13tev_btagger_leff}
\end{figure}

The light- and charm-jet efficiencies in \CheckMATE{} depend on both $\eta$ and $p_T$, with the following functional dependencies.
The dependence on the pseudorapidity $\eta$ is as follows:
\begin{equation}
	\epsilon_i(\eta, \bar \epsilon_S) = (a_{i,0} + a_{i,1} \eta + a_{i,2} \eta^2) \cdot (1 + b_{i,1} \bar \epsilon_S + b_{i,2} \bar \epsilon_S^2) \cdot \frac{\bar r_{i}(0.7)}{\bar r_{i}(\bar\epsilon_S)},
	\label{eq:13tev_btagger_fit_eta}
\end{equation}
where $i$ is either $c$ or $l$ for charm and light jets, respectively.
The coefficients of these efficiency functions are shown in \Cref{tab:eff_eta}
and the functional dependence on the $b$-efficiency is included through the ROC curves $\bar r_{i}$, given as
\begin{align}
	\log_{10}\left[ \bar r_\text{light}(\bar \epsilon_S)\right] &= -21.9\,\bar \epsilon_S^2 + 14.5\,\bar\epsilon_S + 7.02, \label{eq:l_roc_run2} \\
	\log_{10} \left[\bar r_{c}(\bar \epsilon_S)\right] &= 7.39\,\bar \epsilon_S^2 - 19.7\,\bar\epsilon_S + 12.4. \label{eq:c_roc_run2}	
\end{align}
The fitted functions given by Eq.~\eqref{eq:13tev_btagger_fit_eta} are shown in \Cref{fig:13tev_btagger_ceff_eta,fig:13tev_btagger_leff_eta} as solid lines, compared to the values from \ATLAS{}, shown as dots.


The $p_T$ dependence is given by a piecewise function, where each piece depends on the $p_T$ of the jet and on the $b$-efficiency.
The form of the pieces is
\begin{equation}
	\epsilon_{i,\alpha}(p_T, \bar \epsilon_S) = \left[(a_{i,\alpha} + b_{i,\alpha} \bar\epsilon_S + c_{i,\alpha} \bar\epsilon_S^2) + (d_{i,\alpha} + e_{i,\alpha} \bar \epsilon_S + f_{i,\alpha} \bar \epsilon_S^2) p_T\right] \cdot \frac{\epsilon_{s,i}(\bar\epsilon_S)}{\epsilon_{s,i}(0.7)},\label{eq:clight_eff_pt}
\end{equation}
where $i$ is either $c$ or $l$ for charm and light jets, respectively, and $\alpha$ enumerates the pieces of the function.
The coefficients for light and charm jets are given in Tables~\ref{tab:l_eff_pt} and \ref{tab:c_eff_pt_const}.

\begin{table}[t]
\begin{center}\renewcommand*{\arraystretch}{1.2}
	\begin{tabularx}{\textwidth}{XXXXXXX}
\toprule \midrule
		$\alpha$                  & 1             & 2               & 3               & 4               & 5              \\ \midrule
		$a_{\text{light},\alpha}$ & $-0.0152$ & $\phantom{+}2.98 \cdot 10^{-3}$  & $\phantom{+}2.40 \cdot 10^{-3}$  & $\phantom{+}2.40 \cdot 10^{-3}$  & $-0.0153$    \\		
		$b_{\text{light},\alpha}$ & $\phantom{+}0.0228$  & $-2.57 \cdot 10^{-3}$ & $-4.31 \cdot 10^{-4}$ & $-4.31 \cdot 10^{-3}$ & $\phantom{+}0.0147$      \\
		$c_{\text{light},\alpha}$ & $-0.00637$  & $\phantom{+}2.70 \cdot 10^{-3}$ & $\phantom{+}0$               & $\phantom{+}0$               & $\phantom{+}0$              \\
		$d_{\text{light},\alpha}$ & $\phantom{+}0.000384$  & $\phantom{+}2.26 \cdot 10^{-5}$  & $\phantom{+}3.53 \cdot 10^{-5}$  & $\phantom{+}1.13 \cdot 10^{-4}$ & $\phantom{+}1.43 \cdot 10^{-5}$  \\		
		$e_{\text{light},\alpha}$ & $-0.00103$ & $-1.64 \cdot 10^{-5}$ & $-1.63 \cdot 10^{-5}$ & $-1.82 \cdot 10^{-4}$ & $-2.32 \cdot 10^{-5}$ \\		
		$f_{\text{light},\alpha}$ & $\phantom{+}0.000684$  & $-2.63 \cdot 10^{-5}$ & $-3.04 \cdot 10^{-5}$ & $\phantom{+}6.20 \cdot 10^{-5}$  & $\phantom{+}0.74 \cdot 10^{-5}$ \\		
\midrule \bottomrule
	\end{tabularx}
	\captionof{table}{Coefficients for the $p_T$-dependent light-jet efficiency function, Eq.~\eqref{eq:clight_eff_pt}.}
	\label{tab:l_eff_pt}
\end{center}
\end{table}
\begin{table}[t]
\begin{center}\renewcommand*{\arraystretch}{1.2}
	\begin{tabularx}{\textwidth}{XXXX}
\toprule \midrule
		$\alpha$       & 1 & 2 & 3 \\ \midrule
		$a_{c,\alpha}$ & $-0.288$        & $\phantom{+}0.576$        &$\phantom{+}1.51$ \\
		$b_{c,\alpha}$ & $\phantom{+}0.375$         & $-0.941$        & $-2.67$ \\
		$c_{c,\alpha}$ & $-0.0329$       & $\phantom{+}0.513$         & $\phantom{+}1.31$   \\
		$d_{c,\alpha}$ & $\phantom{+}1.02 \cdot 10^{-3}$   & $\phantom{+}2.13 \cdot 10^{-4}$  & $-2.13 \cdot 10^{-3}$ \\
		$e_{c,\alpha}$ & $-2.09 \cdot 10^{-4}$   & $-3.52 \cdot 10^{-4}$  & $\phantom{+}4.43 \cdot 10^{-3}$ \\
		$f_{c,\alpha}$ & $-5.00 \cdot 10^{-5}$   & $-5.20 \cdot 10^{-5}$  & $-2.45 \cdot 10^{-3}$ \\
\midrule \bottomrule
	\end{tabularx}
	\captionof{table}{Coefficients for the $p_T$-dependent charm-jet efficiency function, Eq.~\eqref{eq:clight_eff_pt}.}
	\label{tab:c_eff_pt_const}
\end{center}
\end{table}

Outside of the range, $\bar\epsilon_S \in [0.6, 0.85]$, the parameters are frozen to either 60~\% or 85~\% and the scaling is exclusively given by the scale functions $\bar r_{i}$ given by Eqs.~\eqref{eq:l_roc_run2} and \eqref{eq:c_roc_run2}.
The fitted functions are shown in \Cref{fig:13tev_btagger_ceff_pt,fig:13tev_btagger_leff_pt} as solid lines, compared to the points they were fitted to.
The efficiency functions of $p_T$ and $\eta$ are combined into a single function of $p_T$ and $\eta$, and normalised using 12 million $t\bar t$ events,
\begin{align}
	\epsilon_\text{light}(p_T, \eta, \bar \epsilon_S) &= \epsilon_\text{light}(p_T, \bar \epsilon_S) \cdot \epsilon_\text{light}(\eta, \bar \epsilon_S) \cdot 450 \frac{\bar r_\text{light}(\bar \epsilon_S)}{\bar r_\text{light}(0.7)} \left(1+\frac{3 (100 \bar\epsilon_S-60)}{1000}\right), \\
		\epsilon_{c}(p_T, \eta, \bar \epsilon_S) &= \epsilon_{c}(p_T, \bar \epsilon_S) \cdot \epsilon_{c}(\eta, \bar \epsilon_S) \cdot 7.5 \frac{\bar r_{c}(\bar \epsilon_S)}{\bar r_{c}(0.7)} \left(1+\frac{100\bar \epsilon_S-60}{130}\right).
\end{align}
These efficiencies were tested against the subset of the 13~\TeV{} analyses implemented in \CheckMATETwo{} that use $b$-jets.
The validation for this class of analyses was performed using the $b$-tagger described here and the results were found to be in a good agreement with the published results.

\end{appendices}

  \bibliographystyle{elsarticle-num} 
  \bibliography{bibtex}

\begin{thebibliography}{100}
\expandafter\ifx\csname url\endcsname\relax
  \def\url#1{\texttt{#1}}\fi
\expandafter\ifx\csname urlprefix\endcsname\relax\def\urlprefix{URL }\fi
\expandafter\ifx\csname href\endcsname\relax
  \def\href#1#2{#2} \def\path#1{#1}\fi

\bibitem{deFavereau:2013fsa}
J.~de~Favereau, C.~Delaere, P.~Demin, A.~Giammanco, V.~Lema\^{i}tre,
  A.~Mertens, M.~Selvaggi, {DELPHES 3, A modular framework for fast simulation
  of a generic collider experiment}, JHEP 02 (2014) 057.
\newblock \href {http://arxiv.org/abs/1307.6346} {\path{arXiv:1307.6346}}.

\bibitem{Cacciari:2011ma}
M.~Cacciari, G.~P. Salam, G.~Soyez, {FastJet User Manual}, Eur. Phys. J. C72
  (2012) 1896.
\newblock \href {http://arxiv.org/abs/1111.6097} {\path{arXiv:1111.6097}}.

\bibitem{Cacciari:2005hq}
M.~Cacciari, G.~P. Salam, {Dispelling the $N^{3}$ myth for the $k_t$
  jet-finder}, Phys. Lett. B641 (2006) 57--61.
\newblock \href {http://arxiv.org/abs/hep-ph/0512210}
  {\path{arXiv:hep-ph/0512210}}.

\bibitem{Cacciari:2008gp}
M.~Cacciari, G.~P. Salam, G.~Soyez, {The anti-$k_t$ jet clustering algorithm},
  JHEP 0804 (2008) 063.
\newblock \href {http://arxiv.org/abs/0802.1189} {\path{arXiv:0802.1189}}.

\bibitem{Read:2002hq}
A.~L. Read, {Presentation of search results: The $CL_S$ technique}, J. Phys.
  G28 (2002) 2693--2704.

\bibitem{Alwall:2014hca}
J.~Alwall, R.~Frederix, S.~Frixione, V.~Hirschi, F.~Maltoni, O.~Mattelaer,
  H.~S. Shao, T.~Stelzer, P.~Torrielli, M.~Zaro, {The automated computation of
  tree-level and next-to-leading order differential cross sections, and their
  matching to parton shower simulations}, JHEP 07 (2014) 079.
\newblock \href {http://arxiv.org/abs/1405.0301} {\path{arXiv:1405.0301}}.

\bibitem{Sjostrand:2014zea}
T.~Sj{\"o}strand, S.~Ask, J.~R. Christiansen, R.~Corke, N.~Desai, P.~Ilten,
  S.~Mrenna, S.~Prestel, C.~O. Rasmussen, P.~Z. Skands, {An Introduction to
  PYTHIA 8.2}, Comput. Phys. Commun. 191 (2015) 159--177.
\newblock \href {http://arxiv.org/abs/1410.3012} {\path{arXiv:1410.3012}}.

\bibitem{Lester:1999tx}
C.~Lester, D.~Summers, {Measuring masses of semiinvisibly decaying particles
  pair produced at hadron colliders}, Phys. Lett. B463 (1999) 99--103.
\newblock \href {http://arxiv.org/abs/hep-ph/9906349}
  {\path{arXiv:hep-ph/9906349}}.

\bibitem{Barr:2003rg}
A.~Barr, C.~Lester, P.~Stephens, {$m_{T2}$: The Truth behind the glamour}, J.
  Phys. G29 (2003) 2343--2363.
\newblock \href {http://arxiv.org/abs/hep-ph/0304226}
  {\path{arXiv:hep-ph/0304226}}.

\bibitem{Cheng:2008hk}
H.-C. Cheng, Z.~Han, {Minimal Kinematic Constraints and $m_{T2}$}, JHEP 0812
  (2008) 063.
\newblock \href {http://arxiv.org/abs/0810.5178} {\path{arXiv:0810.5178}}.

\bibitem{Bai:2012gs}
Y.~Bai, H.-C. Cheng, J.~Gallicchio, J.~Gu, {Stop the Top Background of the Stop
  Search}, JHEP 1207 (2012) 110.
\newblock \href {http://arxiv.org/abs/1203.4813} {\path{arXiv:1203.4813}}.

\bibitem{Tovey:2008ui}
D.~R. Tovey, {On measuring the masses of pair-produced semi-invisibly decaying
  particles at hadron colliders}, JHEP 0804 (2008) 034.
\newblock \href {http://arxiv.org/abs/0802.2879} {\path{arXiv:0802.2879}}.

\bibitem{Polesello:2009rn}
G.~Polesello, D.~R. Tovey, {Supersymmetric particle mass measurement with the
  boost-corrected contransverse mass}, JHEP 1003 (2010) 030.
\newblock \href {http://arxiv.org/abs/0910.0174} {\path{arXiv:0910.0174}}.

\bibitem{Matchev:2009ad}
K.~T. Matchev, M.~Park, {A General method for determining the masses of
  semi-invisibly decaying particles at hadron colliders}, Phys. Rev. Lett. 107
  (2011) 061801.
\newblock \href {http://arxiv.org/abs/0910.1584} {\path{arXiv:0910.1584}}.

\bibitem{Graesser:2012qy}
M.~L. Graesser, J.~Shelton, {Hunting Mixed Top Squark Decays}, Phys. Rev. Lett.
  111~(12) (2013) 121802.
\newblock \href {http://arxiv.org/abs/1212.4495} {\path{arXiv:1212.4495}}.

\bibitem{Buckley:2013kua}
M.~R. Buckley, J.~D. Lykken, C.~Rogan, M.~Spiropulu, {Super-Razor and Searches
  for Sleptons and Charginos at the LHC}, Phys. Rev. D89~(5) (2014) 055020.
\newblock \href {http://arxiv.org/abs/1310.4827} {\path{arXiv:1310.4827}}.

\bibitem{Aad:2015zhl}
G.~Aad, et~al., {Combined Measurement of the Higgs Boson Mass in $pp$
  Collisions at $\sqrt{s}=7$ and 8 TeV with the ATLAS and CMS Experiments},
  Phys. Rev. Lett. 114 (2015) 191803.
\newblock \href {http://arxiv.org/abs/1503.07589} {\path{arXiv:1503.07589}}.

\bibitem{Khachatryan:2016vau}
G.~Aad, et~al., {Measurements of the Higgs boson production and decay rates and
  constraints on its couplings from a combined ATLAS and CMS analysis of the
  LHC $pp$ collision data at $ \sqrt{s}=7 $ and 8 TeV}, JHEP 08 (2016) 045.
\newblock \href {http://arxiv.org/abs/1606.02266} {\path{arXiv:1606.02266}}.

\bibitem{Alves:2011wf}
D.~Alves, {Simplified Models for LHC New Physics Searches}, J. Phys. G39 (2012)
  105005.
\newblock \href {http://arxiv.org/abs/1105.2838} {\path{arXiv:1105.2838}}.

\bibitem{Kraml:2013mwa}
S.~Kraml, S.~Kulkarni, U.~Laa, A.~Lessa, W.~Magerl, D.~Proschofsky-Spindler,
  W.~Waltenberger, {SModelS: a tool for interpreting simplified-model results
  from the LHC and its application to supersymmetry}, Eur. Phys. J. C74 (2014)
  2868.
\newblock \href {http://arxiv.org/abs/1312.4175} {\path{arXiv:1312.4175}}.

\bibitem{Kraml:2014sna}
S.~Kraml, S.~Kulkarni, U.~Laa, A.~Lessa, V.~Magerl, W.~Magerl,
  D.~Proschofsky-Spindler, M.~Traub, W.~Waltenberger, {SModelS v1.0: a short
  user guide. }\href {http://arxiv.org/abs/1412.1745} {\path{arXiv:1412.1745}}.

\bibitem{Papucci:2014rja}
M.~Papucci, K.~Sakurai, A.~Weiler, L.~Zeune, {Fastlim: a fast LHC limit
  calculator}, Eur. Phys. J. C74~(11) (2014) 3163.
\newblock \href {http://arxiv.org/abs/1402.0492} {\path{arXiv:1402.0492}}.

\bibitem{Barducci:2014gna}
D.~Barducci, A.~Belyaev, M.~Buchkremer, J.~Marrouche, S.~Moretti, L.~Panizzi,
  {XQCAT: eXtra Quark Combined Analysis Tool}, Comput. Phys. Commun. 197 (2015)
  263--275.
\newblock \href {http://arxiv.org/abs/1409.3116} {\path{arXiv:1409.3116}}.

\bibitem{Caron:2016hib}
S.~Caron, J.~S. Kim, K.~Rolbiecki, R.~Ruiz~de Austri, B.~Stienen, {The BSM-AI
  project: SUSY-AI - Generalizing LHC limits on Supersymmetry with Machine
  Learning. }\href {http://arxiv.org/abs/1605.02797} {\path{arXiv:1605.02797}}.

\bibitem{Cranmer:2010hk}
K.~Cranmer, I.~Yavin, {RECAST: Extending the Impact of Existing Analyses}, JHEP
  04 (2011) 038.
\newblock \href {http://arxiv.org/abs/1010.2506} {\path{arXiv:1010.2506}}.

\bibitem{Drees:2013wra}
M.~Drees, H.~Dreiner, D.~Schmeier, J.~Tattersall, J.~S. Kim, {CheckMATE:
  Confronting your Favourite New Physics Model with LHC Data}, Comput. Phys.
  Commun. 187 (2015) 227--265.
\newblock \href {http://arxiv.org/abs/1312.2591} {\path{arXiv:1312.2591}}.

\bibitem{Kim:2015wza}
J.~S. Kim, D.~Schmeier, J.~Tattersall, K.~Rolbiecki, {A framework to create
  customised LHC analyses within CheckMATE}, Comput. Phys. Commun. 196 (2015)
  535--562.
\newblock \href {http://arxiv.org/abs/1503.01123} {\path{arXiv:1503.01123}}.

\bibitem{Conte:2012fm}
E.~Conte, B.~Fuks, G.~Serret, {MadAnalysis 5, A User-Friendly Framework for
  Collider Phenomenology}, Comput. Phys. Commun. 184 (2013) 222--256.
\newblock \href {http://arxiv.org/abs/1206.1599} {\path{arXiv:1206.1599}}.

\bibitem{Dumont:2014tja}
B.~Dumont, B.~Fuks, S.~Kraml, S.~Bein, G.~Chalons, E.~Conte, S.~Kulkarni,
  D.~Sengupta, C.~Wymant, {Toward a public analysis database for LHC new
  physics searches using MADANALYSIS 5}, Eur. Phys. J. C75~(2) (2015) 56.
\newblock \href {http://arxiv.org/abs/1407.3278} {\path{arXiv:1407.3278}}.

\bibitem{Butterworth:2016sqg}
J.~M. Butterworth, D.~Grellscheid, M.~Kr{\"a}mer, D.~Yallup, {Constraining new
  physics with collider measurements of Standard Model signatures. }\href
  {http://arxiv.org/abs/1606.05296} {\path{arXiv:1606.05296}}.

\bibitem{Buckley:2010ar}
A.~Buckley, J.~Butterworth, L.~Lonnblad, D.~Grellscheid, H.~Hoeth, J.~Monk,
  H.~Schulz, F.~Siegert, {Rivet user manual}, Comput. Phys. Commun. 184 (2013)
  2803--2819.
\newblock \href {http://arxiv.org/abs/1003.0694} {\path{arXiv:1003.0694}}.

\bibitem{Skands:2003cj}
P.~Z. Skands, et~al., {SUSY Les Houches accord: Interfacing SUSY spectrum
  calculators, decay packages, and event generators}, JHEP 07 (2004) 036.
\newblock \href {http://arxiv.org/abs/hep-ph/0311123}
  {\path{arXiv:hep-ph/0311123}}.

\bibitem{Allanach:2008qq}
B.~C. Allanach, et~al., {SUSY Les Houches Accord 2}, Comput. Phys. Commun. 180
  (2009) 8--25.
\newblock \href {http://arxiv.org/abs/0801.0045} {\path{arXiv:0801.0045}}.

\bibitem{Degrande:2011ua}
C.~Degrande, C.~Duhr, B.~Fuks, D.~Grellscheid, O.~Mattelaer, T.~Reiter, {UFO -
  The Universal FeynRules Output}, Comput. Phys. Commun. 183 (2012) 1201--1214.
\newblock \href {http://arxiv.org/abs/1108.2040} {\path{arXiv:1108.2040}}.

\bibitem{Christensen:2008py}
N.~D. Christensen, C.~Duhr, {FeynRules - Feynman rules made easy}, Comput.
  Phys. Commun. 180 (2009) 1614--1641.
\newblock \href {http://arxiv.org/abs/0806.4194} {\path{arXiv:0806.4194}}.

\bibitem{Alloul:2013bka}
A.~Alloul, N.~D. Christensen, C.~Degrande, C.~Duhr, B.~Fuks, {FeynRules 2.0 - A
  complete toolbox for tree-level phenomenology}, Comput. Phys. Commun. 185
  (2014) 2250--2300.
\newblock \href {http://arxiv.org/abs/1310.1921} {\path{arXiv:1310.1921}}.

\bibitem{Staub:2015kfa}
F.~Staub, {Exploring new models in all detail with SARAH}, Adv. High Energy
  Phys. 2015 (2015) 840780.
\newblock \href {http://arxiv.org/abs/1503.04200} {\path{arXiv:1503.04200}}.

\bibitem{Staub:2013tta}
F.~Staub, {SARAH 4 : A tool for (not only SUSY) model builders}, Comput. Phys.
  Commun. 185 (2014) 1773--1790.
\newblock \href {http://arxiv.org/abs/1309.7223} {\path{arXiv:1309.7223}}.

\bibitem{Semenov:2014rea}
A.~Semenov, {LanHEP — A package for automatic generation of Feynman rules
  from the Lagrangian. Version 3.2}, Comput. Phys. Commun. 201 (2016) 167--170.
\newblock \href {http://arxiv.org/abs/1412.5016} {\path{arXiv:1412.5016}}.

\bibitem{fritz1}
Wikipedia, \href{https://en.wikipedia.org/wiki/Fritz_(chess)}{{Fritz (chess)}},
  {[Online; accessed May 20, 2016]}.
\newline\urlprefix\url{https://en.wikipedia.org/wiki/Fritz_(chess)}

\bibitem{fritz2}
C.~GmbH, \href{http://en.chessbase.com/}{{ChessBase}}, {[Online; accessed May
  20, 2016]}.
\newline\urlprefix\url{http://en.chessbase.com/}

\bibitem{Beenakker:1996ch}
W.~Beenakker, R.~Hopker, M.~Spira, P.~M. Zerwas, {Squark and gluino production
  at hadron colliders}, Nucl. Phys. B492 (1997) 51--103.
\newblock \href {http://arxiv.org/abs/hep-ph/9610490}
  {\path{arXiv:hep-ph/9610490}}.

\bibitem{Beenakker:1997ut}
W.~Beenakker, M.~Kr{\"a}mer, T.~Plehn, M.~Spira, P.~M. Zerwas, {Stop production
  at hadron colliders}, Nucl. Phys. B515 (1998) 3--14.
\newblock \href {http://arxiv.org/abs/hep-ph/9710451}
  {\path{arXiv:hep-ph/9710451}}.

\bibitem{Beenakker:1999xh}
W.~Beenakker, M.~Klasen, M.~Kr{\"a}mer, T.~Plehn, M.~Spira, P.~M. Zerwas, {The
  Production of charginos/neutralinos and sleptons at hadron colliders}, Phys.
  Rev. Lett. 83 (1999) 3780--3783, [Erratum: Phys. Rev. Lett. 100, 029901
  (2008)].
\newblock \href {http://arxiv.org/abs/hep-ph/9906298}
  {\path{arXiv:hep-ph/9906298}}.

\bibitem{Spira:2002rd}
M.~Spira, {Higgs and SUSY particle production at hadron colliders}, in:
  {Supersymmetry and unification of fundamental interactions. Proceedings, 10th
  International Conference, SUSY'02, Hamburg, Germany, June 17-23}, 2002, pp.
  217--226.
\newblock \href {http://arxiv.org/abs/hep-ph/0211145}
  {\path{arXiv:hep-ph/0211145}}.

\bibitem{Plehn:2004rp}
T.~Plehn, {Measuring the MSSM Lagrangean}, Czech. J. Phys. 55 (2005)
  B213--B220.
\newblock \href {http://arxiv.org/abs/hep-ph/0410063}
  {\path{arXiv:hep-ph/0410063}}.

\bibitem{Kulesza:2008jb}
A.~Kulesza, L.~Motyka, {Threshold resummation for squark-antisquark and
  gluino-pair production at the LHC}, Phys. Rev. Lett. 102 (2009) 111802.
\newblock \href {http://arxiv.org/abs/0807.2405} {\path{arXiv:0807.2405}}.

\bibitem{Kulesza:2009kq}
A.~Kulesza, L.~Motyka, {Soft gluon resummation for the production of
  gluino-gluino and squark-antisquark pairs at the LHC}, Phys. Rev. D80 (2009)
  095004.
\newblock \href {http://arxiv.org/abs/0905.4749} {\path{arXiv:0905.4749}}.

\bibitem{Beenakker:2009ha}
W.~Beenakker, S.~Brensing, M.~Kr{\"a}mer, A.~Kulesza, E.~Laenen, et~al.,
  {Soft-gluon resummation for squark and gluino hadroproduction}, JHEP 0912
  (2009) 041.
\newblock \href {http://arxiv.org/abs/0909.4418} {\path{arXiv:0909.4418}}.

\bibitem{Beenakker:2010nq}
W.~Beenakker, S.~Brensing, M.~Kr{\"a}mer, A.~Kulesza, E.~Laenen, et~al.,
  {Supersymmetric top and bottom squark production at hadron colliders}, JHEP
  1008 (2010) 098.
\newblock \href {http://arxiv.org/abs/1006.4771} {\path{arXiv:1006.4771}}.

\bibitem{Beenakker:2011fu}
W.~Beenakker, S.~Brensing, M.~Kr{\"a}mer, A.~Kulesza, E.~Laenen, et~al.,
  {Squark and Gluino Hadroproduction}, Int. J. Mod. Phys. A26 (2011)
  2637--2664.
\newblock \href {http://arxiv.org/abs/1105.1110} {\path{arXiv:1105.1110}}.

\bibitem{Sjostrand:2007gs}
T.~Sj{\"o}strand, S.~Mrenna, P.~Z. Skands, {A Brief Introduction to PYTHIA
  8.1}, Comput. Phys. Commun. 178 (2008) 852--867.
\newblock \href {http://arxiv.org/abs/0710.3820} {\path{arXiv:0710.3820}}.

\bibitem{pythiawebpage}
{The Pythia 8 Collaboration},
  \href{http://home.thep.lu.se/~torbjorn/pythia82html/Welcome.html}{{PYTHIA 8
  Online Documentation}}, {[Online; accessed May 12, 2016]}.
\newline\urlprefix\url{http://home.thep.lu.se/~torbjorn/pythia82html/Welcome.html}

\bibitem{Kilian:2007gr}
W.~Kilian, T.~Ohl, J.~Reuter, {WHIZARD: Simulating Multi-Particle Processes at
  LHC and ILC}, Eur. Phys. J. C71 (2011) 1742.
\newblock \href {http://arxiv.org/abs/0708.4233} {\path{arXiv:0708.4233}}.

\bibitem{Moretti:2001zz}
M.~Moretti, T.~Ohl, J.~Reuter, {O'Mega: An Optimizing matrix element generator.
  }\href {http://arxiv.org/abs/hep-ph/0102195} {\path{arXiv:hep-ph/0102195}}.

\bibitem{Belyaev:2012qa}
A.~Belyaev, N.~D. Christensen, A.~Pukhov, {CalcHEP 3.4 for collider physics
  within and beyond the Standard Model}, Comput. Phys. Commun. 184 (2013)
  1729--1769.
\newblock \href {http://arxiv.org/abs/1207.6082} {\path{arXiv:1207.6082}}.

\bibitem{miniworkshop}
\href{https://lpsc-indico.in2p3.fr/Indico/event/1085/}{{Mini-workshop on
  recasting ATLAS and CMS new physics searches}}.
\newline\urlprefix\url{https://lpsc-indico.in2p3.fr/Indico/event/1085/}

\bibitem{Agashe:2014kda}
K.~A. Olive, et~al., {Review of Particle Physics}, Chin. Phys. C38 (2014)
  090001.

\bibitem{Kim:2014eva}
J.~S. Kim, K.~Rolbiecki, K.~Sakurai, J.~Tattersall, {'Stop' that ambulance! New
  physics at the LHC?}, JHEP 12 (2014) 010.
\newblock \href {http://arxiv.org/abs/1406.0858} {\path{arXiv:1406.0858}}.

\bibitem{Ball:2012cx}
R.~D. Ball, et~al., {Parton distributions with LHC data}, Nucl. Phys. B867
  (2013) 244--289.
\newblock \href {http://arxiv.org/abs/1207.1303} {\path{arXiv:1207.1303}}.

\bibitem{Aad:2014wea}
G.~Aad, et~al., {Search for squarks and gluinos with the ATLAS detector in
  final states with jets and missing transverse momentum using $\sqrt{s}=8$ TeV
  proton--proton collision data}, JHEP 09 (2014) 176.
\newblock \href {http://arxiv.org/abs/1405.7875} {\path{arXiv:1405.7875}}.

\bibitem{Aad:2015zva}
G.~Aad, et~al., {Search for new phenomena in final states with an energetic jet
  and large missing transverse momentum in pp collisions at $\sqrt{s}=8$ TeV
  with the ATLAS detector}, Eur. Phys. J. C75~(7) (2015) 299, [Erratum: Eur.
  Phys. J. C75, no. 9, 408 (2015)].
\newblock \href {http://arxiv.org/abs/1502.01518} {\path{arXiv:1502.01518}}.

\bibitem{Chatrchyan:2013lya}
S.~Chatrchyan, et~al., {Search for supersymmetry in hadronic final states with
  missing transverse energy using the variables $\alpha_T$ and b-quark
  multiplicity in pp collisions at $\sqrt s=8$ TeV}, Eur. Phys. J. C73~(9)
  (2013) 2568.
\newblock \href {http://arxiv.org/abs/1303.2985} {\path{arXiv:1303.2985}}.

\bibitem{Aad:2013wta}
G.~Aad, et~al., {Search for new phenomena in final states with large jet
  multiplicities and missing transverse momentum at $\sqrt{s}=8$ TeV
  proton-proton collisions using the ATLAS experiment}, JHEP 10 (2013) 130,
  [Erratum: JHEP 01 (2014) 109].
\newblock \href {http://arxiv.org/abs/1308.1841} {\path{arXiv:1308.1841}}.

\bibitem{Cao:2015ara}
J.~Cao, L.~Shang, J.~M. Yang, Y.~Zhang, {Explanation of the ATLAS $Z$-Peaked
  Excess in the NMSSM}, JHEP 06 (2015) 152.
\newblock \href {http://arxiv.org/abs/1504.07869} {\path{arXiv:1504.07869}}.

\bibitem{Aad:2013ija}
G.~Aad, et~al., {Search for direct third-generation squark pair production in
  final states with missing transverse momentum and two $b$-jets in $\sqrt{s} =
  8$ TeV $pp$ collisions with the ATLAS detector}, JHEP 1310 (2013) 189.
\newblock \href {http://arxiv.org/abs/1308.2631} {\path{arXiv:1308.2631}}.

\bibitem{Aad:2014nua}
G.~Aad, et~al., {Search for direct production of charginos and neutralinos in
  events with three leptons and missing transverse momentum in $\sqrt{s} = 8$
  TeV $pp$ collisions with the ATLAS detector}, JHEP 1404 (2014) 169.
\newblock \href {http://arxiv.org/abs/1402.7029} {\path{arXiv:1402.7029}}.

\bibitem{Aad:2014qaa}
G.~Aad, et~al., {Search for direct top-squark pair production in final states
  with two leptons in $pp$ collisions at $\sqrt{s} = 8$ TeV with the ATLAS
  detector}, JHEP 06 (2014) 124.
\newblock \href {http://arxiv.org/abs/1403.4853} {\path{arXiv:1403.4853}}.

\bibitem{Aad:2014mha}
G.~Aad, et~al., {Search for direct top squark pair production in events with a
  $Z$ boson, $b$-jets and missing transverse momentum in $\sqrt{s}=8$ TeV $pp$
  collisions with the ATLAS detector}, Eur. Phys. J. C74~(6) (2014) 2883.
\newblock \href {http://arxiv.org/abs/1403.5222} {\path{arXiv:1403.5222}}.

\bibitem{Aad:2014pda}
G.~Aad, et~al., {Search for supersymmetry at $\sqrt{s}=8$ TeV in final states
  with jets and two same-sign leptons or three leptons with the ATLAS
  detector}, JHEP 06 (2014) 035.
\newblock \href {http://arxiv.org/abs/1404.2500} {\path{arXiv:1404.2500}}.

\bibitem{Aad:2014kra}
G.~Aad, et~al., {Search for top squark pair production in final states with one
  isolated lepton, jets, and missing transverse momentum in $\sqrt s =8$ TeV
  $pp$ collisions with the ATLAS detector}, JHEP 1411 (2014) 118.
\newblock \href {http://arxiv.org/abs/1407.0583} {\path{arXiv:1407.0583}}.

\bibitem{Aad:2014nra}
G.~Aad, et~al., {Search for pair-produced third-generation squarks decaying via
  charm quarks or in compressed supersymmetric scenarios in $pp$ collisions at
  $\sqrt{s}=8~$TeV with the ATLAS detector}, Phys. Rev. D90~(5) (2014) 052008.
\newblock \href {http://arxiv.org/abs/1407.0608} {\path{arXiv:1407.0608}}.

\bibitem{Aad:2014tda}
G.~Aad, et~al., {Search for new phenomena in events with a photon and missing
  transverse momentum in $pp$ collisions at $\sqrt{s}=8$ TeV with the ATLAS
  detector}, Phys. Rev. D91~(1) (2015) 012008, [Erratum: Phys. Rev. D92, no. 5,
  059903 (2015)].
\newblock \href {http://arxiv.org/abs/1411.1559} {\path{arXiv:1411.1559}}.

\bibitem{Aad:2015jqa}
G.~Aad, et~al., {Search for direct pair production of a chargino and a
  neutralino decaying to the 125 GeV Higgs boson in $\sqrt{s} = 8$ TeV ${pp}$
  collisions with the ATLAS detector}, Eur. Phys. J. C75~(5) (2015) 208.
\newblock \href {http://arxiv.org/abs/1501.07110} {\path{arXiv:1501.07110}}.

\bibitem{Aad:2015wqa}
G.~Aad, et~al., {Search for supersymmetry in events containing a same-flavour
  opposite-sign dilepton pair, jets, and large missing transverse momentum in
  $\sqrt{s}=8$ TeV $pp$ collisions with the ATLAS detector}, Eur. Phys. J.
  C75~(7) (2015) 318.
\newblock \href {http://arxiv.org/abs/1503.03290} {\path{arXiv:1503.03290}}.

\bibitem{Aad:2015pfx}
G.~Aad, et~al., {ATLAS Run 1 searches for direct pair production of
  third-generation squarks at the Large Hadron Collider}, Eur. Phys. J.
  C75~(10) (2015) 510, [Erratum: Eur. Phys. J. C76, no. 3, 153 (2016)].
\newblock \href {http://arxiv.org/abs/1506.08616} {\path{arXiv:1506.08616}}.

\bibitem{ATLAS-CONF-2012-104}
{Search for supersymmetry at $\sqrt{s} = 8$ TeV in final states with jets,
  missing transverse momentum and one isolated lepton}, Tech. Rep.
  ATLAS-CONF-2012-104, CERN, Geneva (Aug 2012).

\bibitem{ATLAS-CONF-2012-147}
{Search for New Phenomena in Monojet plus Missing Transverse Momentum Final
  States using 10 fb$^{-1}$ of $pp$ Collisions at $\sqrt{s}=8$ TeV with the
  ATLAS detector at the LHC}, Tech. Rep. ATLAS-CONF-2012-147, CERN, Geneva (Nov
  2012).

\bibitem{ATLAS-CONF-2013-024}
{Search for direct production of the top squark in the all-hadronic $t\bar{t}$
  + $E_T^\text{miss}$ final state in 21 fb$^{-1}$ of $pp$ collisions at
  $\sqrt{s}=8$ TeV with the ATLAS detector}, Tech. Rep. ATLAS-CONF-2013-024,
  CERN, Geneva (Mar 2013).

\bibitem{ATLAS-CONF-2013-049}
{Search for direct-slepton and direct-chargino production in final states with
  two opposite-sign leptons, missing transverse momentum and no jets in 20/fb
  of $pp$ collisions at $\sqrt{s} = 8$ TeV with the ATLAS detector}, Tech. Rep.
  ATLAS-CONF-2013-049, CERN, Geneva (May 2013).

\bibitem{ATLAS-CONF-2013-061}
{Search for strong production of supersymmetric particles in final states with
  missing transverse momentum and at least three $b$-jets using 20.1 fb$^{-1}$
  of $pp$ collisions at $\sqrt{s} = 8$ TeV with the ATLAS Detector}, Tech. Rep.
  ATLAS-CONF-2013-061, CERN, Geneva (Jun 2013).

\bibitem{ATLAS-CONF-2013-089}
{\relax{}The ATLAS Collaboration}, {Search for strongly produced supersymmetric
  particles in decays with two leptons at $\sqrt{s}$ = 8 TeV}, Tech. Rep.
  ATLAS-CONF-2013-089, CERN, Geneva (Aug 2013).

\bibitem{ATLAS-CONF-2015-004}
{\relax{}The ATLAS Collaboration}, {Search for an Invisibly Decaying Higgs
  Boson Produced via Vector Boson Fusion in $pp$ Collisions at $\sqrt{s}=8$ TeV
  using the ATLAS Detector at the LHC}, Tech. Rep. ATLAS-CONF-2015-004, CERN,
  Geneva (Mar 2015).

\bibitem{Khachatryan:2014rra}
V.~Khachatryan, et~al., {Search for dark matter, extra dimensions, and
  unparticles in monojet events in proton–proton collisions at $\sqrt{s} = 8$
  TeV}, Eur. Phys. J. C75~(5) (2015) 235.
\newblock \href {http://arxiv.org/abs/1408.3583} {\path{arXiv:1408.3583}}.

\bibitem{Khachatryan:2015lwa}
V.~Khachatryan, et~al., {Search for Physics Beyond the Standard Model in Events
  with Two Leptons, Jets, and Missing Transverse Momentum in $pp$ Collisions at
  $\sqrt{s} = 8$ TeV}, JHEP 04 (2015) 124.
\newblock \href {http://arxiv.org/abs/1502.06031} {\path{arXiv:1502.06031}}.

\bibitem{Khachatryan:2015nua}
V.~Khachatryan, et~al., {Search for the production of dark matter in
  association with top-quark pairs in the single-lepton final state in
  proton-proton collisions at $\sqrt{s} = 8$ TeV}, JHEP 06 (2015) 121.
\newblock \href {http://arxiv.org/abs/1504.03198} {\path{arXiv:1504.03198}}.

\bibitem{Baek:2016lnv}
S.~Baek, P.~Ko, P.~Wu, {Top-philic Scalar Dark Matter with a Vector-like
  Fermionic Top Partner}, JHEP 10 (2016) 117.
\newblock \href {http://arxiv.org/abs/1606.00072} {\path{arXiv:1606.00072}}.

\bibitem{CMS:2013jea}
{\relax{}The CMS Collaboration}, {Search for new physics in events with
  same-sign dileptons and jets in $pp$ collisions at 8 TeV}, Tech. Rep.
  CMS-PAS-SUS-13-013 (2013).

\bibitem{Aad:2016tuk}
G.~Aad, et~al., {Search for supersymmetry at $\sqrt{s}=13$ TeV in final states
  with jets and two same-sign leptons or three leptons with the ATLAS
  detector}, Eur. Phys. J. C76~(5) (2016) 259.
\newblock \href {http://arxiv.org/abs/1602.09058} {\path{arXiv:1602.09058}}.

\bibitem{Aaboud:2016uro}
M.~Aaboud, et~al., {Search for new phenomena in events with a photon and
  missing transverse momentum in $pp$ collisions at $\sqrt{s}=13$ TeV with the
  ATLAS detector}, JHEP 06 (2016) 059.
\newblock \href {http://arxiv.org/abs/1604.01306} {\path{arXiv:1604.01306}}.

\bibitem{Aaboud:2016tnv}
M.~Aaboud, et~al., {Search for new phenomena in final states with an energetic
  jet and large missing transverse momentum in $pp$ collisions at $\sqrt{s} =
  13$ TeV using the ATLAS detector}, Phys. Rev. D94~(3) (2016) 032005.
\newblock \href {http://arxiv.org/abs/1604.07773} {\path{arXiv:1604.07773}}.

\bibitem{Aaboud:2016zdn}
M.~Aaboud, et~al., {Search for squarks and gluinos in final states with jets
  and missing transverse momentum at $\sqrt{s}$ =13 ${\mathrm{TeV}}$ with the
  ATLAS detector}, Eur. Phys. J. C76~(7) (2016) 392.
\newblock \href {http://arxiv.org/abs/1605.03814} {\path{arXiv:1605.03814}}.

\bibitem{Aad:2016qqk}
G.~Aad, et~al., {Search for gluinos in events with an isolated lepton, jets and
  missing transverse momentum at $\sqrt{s}$ = 13 TeV with the ATLAS detector},
  Eur. Phys. J. C76~(10) (2016) 565.
\newblock \href {http://arxiv.org/abs/1605.04285} {\path{arXiv:1605.04285}}.

\bibitem{Aad:2016eki}
G.~Aad, et~al., {Search for pair production of gluinos decaying via stop and
  sbottom in events with $b$-jets and large missing transverse momentum in $pp$
  collisions at $\sqrt{s} = 13$ TeV with the ATLAS detector}, Phys. Rev.
  D94~(3) (2016) 032003.
\newblock \href {http://arxiv.org/abs/1605.09318} {\path{arXiv:1605.09318}}.

\bibitem{Aaboud:2016lwz}
M.~Aaboud, et~al., {Search for top squarks in final states with one isolated
  lepton, jets, and missing transverse momentum in $\sqrt{s}=13$ TeV $pp$
  collisions with the ATLAS detector}, Phys. Rev. D94~(5) (2016) 052009.
\newblock \href {http://arxiv.org/abs/1606.03903} {\path{arXiv:1606.03903}}.

\bibitem{ATLAS-CONF-2015-082}
\href{http://cds.cern.ch/record/2114854}{{A search for Supersymmetry in events
  containing a leptonically decaying $Z$ boson, jets and missing transverse
  momentum in $\sqrt{s}=13~$TeV $pp$ collisions with the ATLAS detector}},
  Tech. Rep. ATLAS-CONF-2015-082, CERN, Geneva (Dec 2015).
\newline\urlprefix\url{http://cds.cern.ch/record/2114854}

\bibitem{ATLAS-CONF-2016-013}
\href{http://cds.cern.ch/record/2140998}{{Search for production of vector-like
  top quark pairs and of four top quarks in the lepton-plus-jets final state in
  $pp$ collisions at $\sqrt{s}=13$ TeV with the ATLAS detector}}, Tech. Rep.
  ATLAS-CONF-2016-013, CERN, Geneva (Mar 2016).
\newline\urlprefix\url{http://cds.cern.ch/record/2140998}

\bibitem{ATLAS-CONF-2016-050}
\href{https://cds.cern.ch/record/2206132}{{Search for top squarks in final
  states with one isolated lepton, jets, and missing transverse momentum in
  $\sqrt{s} = 13$ TeV $pp$ collisions with the ATLAS detector}}, Tech. Rep.
  ATLAS-CONF-2016-050, CERN, Geneva (Aug 2016).
\newline\urlprefix\url{https://cds.cern.ch/record/2206132}

\bibitem{ATLAS-CONF-2016-076}
\href{http://cds.cern.ch/record/2206249}{{Search for direct top squark pair
  production and dark matter production in final states with two leptons in
  $\sqrt{s} = 13$ TeV $pp$ collisions using 13.3 fb$^{-1}$ of ATLAS data}},
  Tech. Rep. ATLAS-CONF-2016-076, CERN, Geneva (Aug 2016).
\newline\urlprefix\url{http://cds.cern.ch/record/2206249}

\bibitem{CMS-PAS-SUS-15-011}
\href{https://cds.cern.ch/record/2114811}{{Search for new physics in final
  states with two opposite-sign same-flavor leptons, jets and missing
  transverse momentum in $pp$ collisions at $\sqrt{s}=13$ TeV}}, Tech. Rep.
  CMS-PAS-SUS-15-011, CERN, Geneva (2015).
\newline\urlprefix\url{https://cds.cern.ch/record/2114811}

\bibitem{ATL-PHYS-PUB-2014-010}
\href{http://cds.cern.ch/record/1735031}{{Search for Supersymmetry at the high
  luminosity LHC with the ATLAS experiment}}, Tech. Rep. ATL-PHYS-PUB-2014-010,
  CERN, Geneva (Jul 2014).
\newline\urlprefix\url{http://cds.cern.ch/record/1735031}

\bibitem{ATL-PHYS-PUB-2013-011}
\href{http://cds.cern.ch/record/1604505}{{Prospects for benchmark Supersymmetry
  searches at the high luminosity LHC with the ATLAS Detector}}, Tech. Rep.
  ATL-PHYS-PUB-2013-011, CERN, Geneva (Sep 2013).
\newline\urlprefix\url{http://cds.cern.ch/record/1604505}

\bibitem{Barr:2010zj}
A.~J. Barr, C.~G. Lester, {A Review of the Mass Measurement Techniques proposed
  for the Large Hadron Collider}, J. Phys. G37 (2010) 123001.
\newblock \href {http://arxiv.org/abs/1004.2732} {\path{arXiv:1004.2732}}.

\bibitem{vanNeerven:1982mz}
W.~van Neerven, J.~Vermaseren, K.~Gaemers, {Lepton - jet events as a signature
  for $W$ production in p anti-p collisions}, Tech. rep. (1982).

\bibitem{Arnison:1983rp}
G.~Arnison, et~al., {Experimental Observation of Isolated Large Transverse
  Energy Electrons with Associated Missing Energy at $\sqrt{s} = 540$ GeV},
  Phys. Lett. B122 (1983) 103--116.

\bibitem{Banner:1983jy}
M.~Banner, et~al., {Observation of Single Isolated Electrons of High Transverse
  Momentum in Events with Missing Transverse Energy at the CERN anti-p p
  Collider}, Phys. Lett. B122 (1983) 476--485.

\bibitem{Smith:1983aa}
J.~Smith, W.~van Neerven, J.~Vermaseren, {The Transverse Mass and Width of the
  $W$ Boson}, Phys. Rev. Lett. 50 (1983) 1738.

\bibitem{Barger:1987du}
V.~D. Barger, T.~Han, R.~Phillips, {Improved Transverse Mass Variable for
  Detecting Higgs Boson Decays Into $Z$ Pairs}, Phys. Rev. D36 (1987) 295.

\bibitem{Randall:2008rw}
L.~Randall, D.~Tucker-Smith, {Dijet Searches for Supersymmetry at the LHC},
  Phys. Rev. Lett. 101 (2008) 221803.
\newblock \href {http://arxiv.org/abs/0806.1049} {\path{arXiv:0806.1049}}.

\bibitem{Khachatryan:2011tk}
V.~Khachatryan, et~al., {Search for Supersymmetry in $pp$ Collisions at 7 TeV
  in Events with Jets and Missing Transverse Energy}, Phys. Lett. B698 (2011)
  196--218.
\newblock \href {http://arxiv.org/abs/1101.1628} {\path{arXiv:1101.1628}}.

\bibitem{Rogan:2010kb}
C.~Rogan, {Kinematical variables towards new dynamics at the LHC }\href
  {http://arxiv.org/abs/1006.2727} {\path{arXiv:1006.2727}}.

\bibitem{Chatrchyan:2011ek}
S.~Chatrchyan, et~al., {Inclusive search for squarks and gluinos in $pp$
  collisions at $\sqrt{s}=7$ TeV}, Phys. Rev. D85 (2012) 012004.
\newblock \href {http://arxiv.org/abs/1107.1279} {\path{arXiv:1107.1279}}.

\bibitem{Khachatryan:2015pwa}
V.~Khachatryan, et~al., {Search for Supersymmetry Using Razor Variables in
  Events with $b$-Tagged Jets in $pp$ Collisions at $\sqrt{s} = 8$ TeV}, Phys.
  Rev. D91 (2015) 052018.
\newblock \href {http://arxiv.org/abs/1502.00300} {\path{arXiv:1502.00300}}.

\bibitem{Chen:2011ah}
C.~Chen, {New approach to identifying boosted hadronically-decaying particle
  using jet substructure in its center-of-mass frame}, Phys. Rev. D85 (2012)
  034007.
\newblock \href {http://arxiv.org/abs/1112.2567} {\path{arXiv:1112.2567}}.

\bibitem{Brun:1997pa}
R.~Brun, F.~Rademakers, {ROOT: An object oriented data analysis framework},
  Nucl. Instrum. Meth. A389 (1997) 81--86.

\bibitem{Dobbs:2001ck}
M.~Dobbs, J.~B. Hansen, {The HepMC C++ Monte Carlo event record for High Energy
  Physics}, Comput. Phys. Commun. 134 (2001) 41--46.

\bibitem{Cranmer:2015nia}
K.~Cranmer,
  \href{https://inspirehep.net/record/1356277/files/arXiv:1503.07622.pdf}{{Practical
  Statistics for the LHC}}, in: {Proceedings, 2011 European School of
  High-Energy Physics (ESHEP 2011)}, 2015, pp. 267--308, [247(2015)].
\newblock \href {http://arxiv.org/abs/1503.07622} {\path{arXiv:1503.07622}}.
\newline\urlprefix\url{https://inspirehep.net/record/1356277/files/arXiv:1503.07622.pdf}

\bibitem{wilks1938}
S.~S. Wilks, \href{http://dx.doi.org/10.1214/aoms/1177732360}{The large-sample
  distribution of the likelihood ratio for testing composite hypotheses}, Ann.
  Math. Statist. 9~(1) (1938) 60--62.
\newline\urlprefix\url{http://dx.doi.org/10.1214/aoms/1177732360}

\bibitem{Dowell1972}
M.~Dowell, P.~Jarratt, \href{http://dx.doi.org/10.1007/BF01932959}{{The
  ``Pegasus'' method for computing the root of an equation}}, BIT Numerical
  Mathematics 12~(4) (1972) 503--508.
\newline\urlprefix\url{http://dx.doi.org/10.1007/BF01932959}

\bibitem{ATLAS-CONF-2016-024}
\href{http://cds.cern.ch/record/2157687}{{Electron efficiency measurements with
  the ATLAS detector using the 2015 LHC proton-proton collision data}}, Tech.
  Rep. ATLAS-CONF-2016-024, CERN, Geneva (Jun 2016).
\newline\urlprefix\url{http://cds.cern.ch/record/2157687}

\bibitem{electroneff}
\href{https://atlas.web.cern.ch/Atlas/GROUPS/PHYSICS/PLOTS/EGAM-2016-002/}{{Electron
  identification efficiency measured with $Z\to ee$ events using 2016 data}},
  [Online; accessed 12-Nov-2016].
\newline\urlprefix\url{https://atlas.web.cern.ch/Atlas/GROUPS/PHYSICS/PLOTS/EGAM-2016-002/}

\bibitem{ATL-PHYS-PUB-2013-004}
\href{https://cds.cern.ch/record/1527529}{{Performance assumptions for an
  upgraded ATLAS detector at a High-Luminosity LHC}}, Tech. Rep.
  ATL-PHYS-PUB-2013-004, CERN, Geneva (Mar 2013).
\newline\urlprefix\url{https://cds.cern.ch/record/1527529}

\bibitem{Aad:2016jkr}
G.~Aad, et~al., {Muon reconstruction performance of the ATLAS detector in
  proton-proton collision data at $\sqrt{s} =13$ TeV}, Eur. Phys. J. C76~(5)
  (2016) 292.
\newblock \href {http://arxiv.org/abs/1603.05598} {\path{arXiv:1603.05598}}.

\bibitem{ATL-PHYS-PUB-2013-009}
\href{http://cds.cern.ch/record/1604420}{{Performance assumptions based on full
  simulation for an upgraded ATLAS detector at a High-Luminosity LHC}}, Tech.
  Rep. ATL-PHYS-PUB-2013-009, CERN, Geneva (Sep 2013).
\newline\urlprefix\url{http://cds.cern.ch/record/1604420}

\bibitem{ATLAS-CONF-2012-043}
{Measurement of the $b$-tag Efficiency in a Sample of Jets Containing Muons
  with 5 fb$^{-1}$ of Data from the ATLAS Detector}, Tech. Rep.
  ATLAS-CONF-2012-043, CERN, Geneva (Mar 2012).

\bibitem{ATLAS-CONF-2012-097}
{Measuring the $b$-tag efficiency in a top-pair sample with 4.7 fb$^{-1}$ of
  data from the ATLAS detector}, Tech. Rep. ATLAS-CONF-2012-097, CERN, Geneva
  (Jul 2012).

\bibitem{ATLAS-CONF-2012-040}
{Measurement of the Mistag Rate with 5 fb$^{-1}$ of Data Collected by the ATLAS
  Detector}, Tech. Rep. ATLAS-CONF-2012-040, CERN, Geneva (Mar 2012).

\bibitem{ATL-PHYS-PUB-2015-022}
\href{http://cds.cern.ch/record/2037697}{{Expected performance of the ATLAS
  $b$-tagging algorithms in Run-2}}, Tech. Rep. ATL-PHYS-PUB-2015-022, CERN,
  Geneva (Jul 2015).
\newline\urlprefix\url{http://cds.cern.ch/record/2037697}

\end{thebibliography}


\end{document}